\def\spacingNumerator{5}
\def\spacingDenominator{4}

\def\ifundefined#1{\expandafter\ifx\csname#1\endcsname\relax}
\ifundefined{ftmagnification}  \def\ftmagnification{1200} \fi
\ifundefined{spacingNumerator}  \def\spacingNumerator{5} \fi
\ifundefined{spacingDenominator}  \def\spacingDenominator{4} \fi


\magnification\ftmagnification
\tolerance=10000
\hsize=17truecm\vsize=23truecm

\parindent=40pt
\mathsurround=0pt
     \multiply\baselineskip by \spacingNumerator
     \divide \baselineskip by \spacingDenominator 

%
%
\def\today{\ifcase\month\or January\or February\or March\or April\or
     May\or June\or July\or August\or September\or October\or November\or
     December\fi\space\number\day, \number\year}
%
%
\def\dst{\displaystyle}
\def\sst{\scriptstyle}
\def\tst{\textstyle}

%
%
\def\frac#1#2{\dst {#1\over#2}}     
\def\sfrac#1#2{{\tst{#1\over#2}}}   

\def\deqalign#1{\vcenter{\openup1\jot \mathsurround=0pt \ialign{
                \strut\hfil$\displaystyle{##}$&&$\displaystyle{{}##}$\hfil
                \crcr
                #1\crcr}}}         

\def\meqalign#1{\vcenter{\openup1\jot \mathsurround=0pt \ialign{
                &\strut\hfil$\displaystyle{##}$&$\displaystyle{{}##}$\hfil&
                \quad$##$\crcr
                #1\crcr}}}         

%
%
\def\al{\alpha}
\def\be{\beta}
\def\ga{\gamma}
\def\de{\delta}
\def\ep{\epsilon}
\def\ze{\zeta}
\def\et{\eta}

\def\ka{\kappa}
\def\la{\lambda}

\def\si{\sigma}

\def\Si{\Sigma}

\def\Om{\Omega}   
%
%
\def\pmb#1{\setbox0=\hbox{#1}       
     \kern-.025em\copy0\kern-\wd0
     \kern.05em\copy0\kern-\wd0
     \kern-.025em\box0}             
\def\0{{\bf 0}}

\def\k{{\bf k}}

\def\x{{\bf x}}

\def\cG{{\cal G}}
\def\cH{{\cal H}}

\def\cJ{{\cal J}}

\def\cS{{\cal S}}

%
%
\font\tenfrak                 = eufm10
\font\sevenfrak               = eufm7
\font\fivefrak                = eufb5
\newfam\frakfam
     \textfont\frakfam=\tenfrak
     \scriptfont\frakfam=\sevenfrak   
     \scriptscriptfont\frakfam=\fivefrak
\def\frak{\fam\frakfam\tenfrak}
\font \tensans                = cmss10
\font \fivesans               = cmss10 at 5pt
\font \sevensans              = cmss10 at 7pt
\newfam\sansfam
     \textfont\sansfam=\tensans
     \scriptfont\sansfam=\sevensans
     \scriptscriptfont\sansfam=\fivesans
\def\sans{\fam\sansfam\tensans}
%
%
\def\bbbr{{\rm I\!R}}  
\def\bbbn{{\rm I\!N}}

\def\bbbone{{\mathchoice {\rm 1\mskip-4mu l} {\rm 1\mskip-4mu l}    
{\rm 1\mskip-4.5mu l} {\rm 1\mskip-5mu l}}}
\def\bbbc{{\mathchoice {\setbox0=\hbox{$\displaystyle\rm C$}\hbox{\hbox 
to0pt{\kern0.4\wd0\vrule height0.9\ht0\hss}\box0}}
{\setbox0=\hbox{$\textstyle\rm C$}\hbox{\hbox
to0pt{\kern0.4\wd0\vrule height0.9\ht0\hss}\box0}}
{\setbox0=\hbox{$\scriptstyle\rm C$}\hbox{\hbox
to0pt{\kern0.4\wd0\vrule height0.9\ht0\hss}\box0}}
{\setbox0=\hbox{$\scriptscriptstyle\rm C$}\hbox{\hbox
to0pt{\kern0.4\wd0\vrule height0.9\ht0\hss}\box0}}}}
\def\bbbq{{\mathchoice {\setbox0=\hbox{$\displaystyle\rm               
Q$}\hbox{\raise
0.15\ht0\hbox to0pt{\kern0.4\wd0\vrule height0.8\ht0\hss}\box0}}
{\setbox0=\hbox{$\textstyle\rm Q$}\hbox{\raise
0.15\ht0\hbox to0pt{\kern0.4\wd0\vrule height0.8\ht0\hss}\box0}}
{\setbox0=\hbox{$\scriptstyle\rm Q$}\hbox{\raise
0.15\ht0\hbox to0pt{\kern0.4\wd0\vrule height0.7\ht0\hss}\box0}}
{\setbox0=\hbox{$\scriptscriptstyle\rm Q$}\hbox{\raise
0.15\ht0\hbox to0pt{\kern0.4\wd0\vrule height0.7\ht0\hss}\box0}}}}
\def\bbbz{{\mathchoice {\hbox{$\sans\textstyle Z\kern-0.4em Z$}}       
{\hbox{$\sans\textstyle Z\kern-0.4em Z$}}
{\hbox{$\sans\scriptstyle Z\kern-0.3em Z$}}
{\hbox{$\sans\scriptscriptstyle Z\kern-0.2em Z$}}}}
%
%

\def\sgn{{\rm sgn}}
\def\half{\sfrac{1}{2}}

\def\optbar#1{\vbox{\ialign{##\crcr\hfil${\scriptscriptstyle(}\mkern -1mu
         \vrule height 1.2pt width 3pt depth -.8pt
         {\scriptscriptstyle)}$\hfil\crcr
          \noalign{\kern-1pt\nointerlineskip}$\hfil\displaystyle{#1}\hfil$\crcr}}}
\def\<{\left<}
\def\>{\right>}

\def\smprod{\mathop{\textstyle\prod}}
\def\smsum{\mathop{\textstyle\sum}}
\def\set#1#2{\big\{ \ #1\ \big|\ #2\ \big\}}
\def\eval#1{\big|\lower4pt\hbox{$\displaystyle\sst #1$}}
%
%
\font \tafontt                = cmbx10 scaled\magstep2
\font \tbfontt                = cmbx10 scaled\magstep1
\def\titlea#1{\centerline{\tafontt #1 }\vskip.5truein}
\def\titleb#1{\removelastskip\vskip.3truein%
\noindent{\tbfontt #1 }\vskip.25truein}

%
%
\def\newenvironment#1#2#3#4{\long\def#1##1##2{%
\removelastskip\penalty-100\vskip\baselineskip%
\noindent{#3#2\if!##1!.\else\unskip\ \ignorespaces
##1\unskip\fi\ }{#4\ignorespaces##2\vskip\baselineskip}}}
\newenvironment\lemma{Lemma}{\bf}{\it}
\newenvironment\proposition{Proposition}{\bf}{\it}
\newenvironment\theorem{Theorem}{\bf}{\it}
\newenvironment\corollary{Corollary}{\bf}{\it}
\newenvironment\example{Example}{\bf}{\rm}
\newenvironment\problem{Problem}{\bf}{\rm}
\newenvironment\definition{Definition}{\bf}{\rm}
\newenvironment\remark{Remark}{\bf}{\rm}
\newenvironment\hypothesis{Hypothesis}{\bf}{\it}
\newenvironment\convention{Convention}{\bf}{\it}

\def\Item{\vskip.1in\noindent}

%
%
\long\def\proof#1{\removelastskip\penalty-100\vskip\baselineskip\noindent{\bf
            Proof\if!#1!\else\ \ignorespaces#1\fi:\ }\ \ \ignorespaces}
\long\def\prf{\removelastskip\penalty-100\vskip\baselineskip\noindent{\bf
            Proof:\ }\ \ \ignorespaces}
\def\endproof{\hfill\vrule height .6em width .6em depth 0pt\goodbreak\vskip.25in }

\ifundefined{warnForwardRef}  \def\warnForwardRef{n} \fi
\newcount\chapno
\newcount\sectno
\newcount\equano
\newcount\theono
\newcount\probno

\def\IgNoRe#1{}

\chapno=0
\sectno=0
\equano=0
\theono=0
\probno=0
\def\eqhead{}
\def\frefwarning{\if\warnForwardRef y\immediate\write16{   Forward reference on line \the\inputlineno}\fi}
\def\qqqrefwarning{\immediate\write16{   ??? reference on line \the\inputlineno}}

\def\chap#1{\equano=0\sectno=0\theono=0\probno=0\global\advance\chapno by 1%
\def\eqhead{\ifcase\chapno\or I\or II\or III\or IV\or V\or VI\or VII\or
VIII\or IX\or X\or XI\or XII\or XIII\or XIV\or XV\or XVI\or XVII\or XVIII\or
XIX\or XX\or XXI\or XXII\or XXIII\or XXIV\or XXV\or XXVI\or XXVII\or XXVIII\or XXIX\or XXX\or XXXI\or XXXII\or XXXIII\or XXXIV\or XXXV\or XXXVI\or XXXVII\or XXXVIII\or XXXIX\fi.}%
\titlea{\eqhead \hglue 5pt #1}%
}

\def\sect#1{\global\advance\sectno by 1%
\titleb{\eqhead\number\sectno  \hglue 5pt #1}%
}%

\def\appendix#1#2{\equano=0\sectno=0\theono=0\probno=0\def\eqhead{#1.}
\titlea{Appendix #1: #2}%
}

\def\:#1{\def\temp{\expandafter\IgNoRe\string#1}%
\expandafter\ifx\csname\temp\endcsname\relax%
\expandafter\gdef#1{\qqqrefwarning ???}\fi#1}

\def\Eqn{{\hbox{\global\advance\equano by 1}}%
\eqno ({\rm \eqhead\number\equano})}%

\def\Eqno{{\hbox{\global\advance\equano by 1}}%
 ({\rm \eqhead\number\equano})}%

\def\EQN#1{\Eqn\edef\Zwi{\eqhead\number\equano}%
\global\let #1=\Zwi
}

\def\EQNO#1{\Eqno\edef\Zwi{\eqhead\number\equano}%
\global\let #1=\Zwi
}

\def\STM#1{{\global\advance \theono by 1}%
\eqhead\number\theono
\edef\Zwi{\eqhead\number\theono }
\global\let#1=\Zwi
}

\def\PRB#1{{\global\advance \probno by 1}%
\eqhead\number\probno
\edef\Zwi{\eqhead\number\probno }
\global\let#1=\Zwi
}

\def\PG#1{\def\Zwi{\number\pageno }
\global\let#1=\Zwi
}

\def\Stm{{\global\advance \theono by 1}%
\eqhead\number\theono
}

\def\Prb{{\global\advance \probno by 1}%
\eqhead\number\probno
}

\def\EDEF#1#2{
\def\tEmP{#1}\expandafter\gdef\tEmP{#2}
}



\def\suffix{ps}
\newcount\system
\global\system=3   

\def\ifundefined#1{\expandafter\ifx\csname#1\endcsname\relax}
\ifundefined{figdir}\def\figdir{}\fi
%
%
\newcount\firstline
\newdimen\pswidth  \newdimen\xleft
\newdimen\psheight \newdimen\ytop \newdimen\ybot
\newcount\justx \newcount\justy
\global\justx=0 \global\justy=0
\newdimen\vpos \newtoks\labeL 
\newread\labeLfile \newdimen\xcoord \newdimen\ycoord
\newif\ifdoit 
\newbox\labox
\newdimen\xdvikwid 
\newdimen\xdvikht
\newdimen\pspoints
\newdimen\rwi
\pspoints=1bp
\newcount\temp
\def\readdim#1{\global\read\labeLfile to \temp
\global #1=\temp pt}
%
%
%
%
\def\figcrop#1{\par
\openin\labeLfile=\figdir#1.lbl                                              
\global\read\labeLfile to\firstline\message{#1}               
\global\read\labeLfile to\temp
\readdim{\ybot}
\readdim{\xleft}
\readdim{\ytop}
\global\read\labeLfile to\justx
\global\read\labeLfile to\justy
\global\read\labeLfile to\labeL
\readdim{\pswidth}
\global\advance\pswidth by -\xleft
\readdim{\psheight}
\global\advance\ybot by -\psheight
\global\advance\psheight by -\ytop
\global\read\labeLfile to\justx
\global\read\labeLfile to\justy
\global\read\labeLfile to\labeL
\vbox to\psheight{\vfill
\ifnum\system=1
\fi 
\ifnum\system=2
\fi 
\ifnum\system=3
                                                 \fi         
\ifnum\system=4
\fi         
\ifnum\system=1
\hbox to \pswidth{\kern-\xleft\special{postscriptfile \figdir#1.\suffix }\hfil}\fi
\ifnum\system=2
\hbox to \pswidth{\kern-\xleft\special{ps: plotfile \figdir#1.\suffix }\hfil}\fi
\ifnum\system=3
\hbox to \pswidth{\kern-\xleft\includegraphics{\figdir#1.\suffix}\hfil}\fi
\ifnum\system=4
\hbox to \pswidth{\kern-\xleft\includegraphics{\figdir#1.\suffix}\hfil}\fi
\ifnum\system=5
\hbox to \pswidth{\kern-\xleft\includegraphics{\figdir#1.\suffix}\hfil}\fi 
\ifnum\system=6
   \xdvikwid=\pswidth
   \xdvikht=\psheight
   {\global\divide\xdvikwid by \pspoints}
   {\global\divide\xdvikht by \pspoints}
   \rwi=\xdvikwid
    {\global\multiply\rwi by 10}
\hbox to \pswidth{\kern-\xleft\includegraphics{\figdir#1.\suffix\space}\hfil}\fi                   
\vskip -\baselineskip
\vskip -\ybot 
\vskip-\psheight %
\hbox to\pswidth  {\hss}%
\parindent=0pt\offinterlineskip                                       
\vpos=0 pt%
\loop\readdim{\xcoord}                                 
\ifdim \xcoord < -999pt \doitfalse\else\doittrue\fi                        
\ifdoit \advance \xcoord by -\xleft
\readdim{\ycoord}
\advance \ycoord by -\ytop                              
\global\read\labeLfile to\justx                                       
\global\read\labeLfile to\justy                                       
\global\read\labeLfile to\labeL
\global\setbox\labox=\hbox{\labeL\hskip-0.3em}%
\advance\vpos by-\ycoord                                              
\vskip-\vpos \vpos=\ycoord                                         
\hbox to\pswidth{\hskip\xcoord %
\hbox to 0pt{\ifnum\justx>0\hss\fi%
\vbox to0pt{%
\ifnum\justy<2\vss\fi%
\copy\labox\kern0pt%
\ifnum\justy>0\vss\fi}%
\ifnum\justx<2\hss\fi}%
\hss}%
\repeat%
\advance\vpos by-\psheight%
\vskip-\vpos %
}\closein\labeLfile}
%
%
%
\def\figplace#1#2#3{
\openin\labeLfile=\figdir#1.lbl
\ifeof \labeLfile
       \immediate\write16{***Can't find \figdir#1.lbl; Skipping it.***}
\else  \closein\labeLfile
       \null\hskip#2\raise #3 \hbox{\figcrop{#1}}
\fi
}
%
%
%
%
\def\figput#1{
\openin\labeLfile=\figdir#1.lbl
\ifeof \labeLfile
       \immediate\write16{***Can't find \figdir#1.lbl; Skipping it.***}
\else  \closein\labeLfile
       \hbox{\figcrop{#1}}
\fi
}


    \def\squiggle{\raise2pt\hbox{${\scriptstyle\sim}$}}
    \def\stoday{\number\day\space\ifcase\month\or Jan\or Feb\or 
                      Mar\or Apr\or May\or Jun\or Jul\or Aug\or Sep\or 
                      Oct\or Nov\or Dec\fi, \number\year}

    \def\cb{{\frak c}}
    \def\ib{{\rm b}}
    
    \def\cl{;}
    \def\cont#1#2#3{\mathop{{\rm\ \, Con}_{#3}}\limits_{#1\rightarrow#2}}
    \def\Cont#1#2#3{\mathop{{\rm\ \, {\cal C}on}_{#3}}\limits_{#1\rightarrow#2}}
    \def\smchoose#1#2{{\tst {#1\choose #2}}}
    
    \def\dblint{\int\kern-0.7em\int}
    \def\susywedge{\mathop{\raise2ex\hbox{$\mathchar"030E\mkern-5mu\mathchar"030E\mkern-2.5mu\mathchar"030F$}}}

\def\tailind#1#2#3#4#5#6{\lower2pt\hbox{$({\scriptstyle {#1\,#2\,#3\atop #4\,#5\,#6}})$}}

    \def\cC{{\cal C}}
    \def\cG{{\cal G}}
    \def\cJ{{\cal J}}

    \def\cR{{\cal R}}

    \def\veps{{\varepsilon}}

    \def\fN{{\frak N}}

    \def\tn{|\kern-1pt|\kern-1pt|}
    \def\TN{\big|\kern-1.5pt\big|\kern-1.5pt\big|}
    \def\TTN{\Big|\kern-2pt\Big|\kern-2pt\Big|}

    \def\rw{\mathclose{:}}
    \def\lw{\mathopen{:}}
    \def\lW{\mathopen{{\tst{\hbox{.}\atop\raise 2.5pt\hbox{.}}}}}
    \def\rW{\mathclose{{\tst{{.}\atop\raise 2.5pt\hbox{.}}}}}
    \def\lww{\mathopen{{\tst{\raise 1pt\hbox{.}\atop\raise 1pt\hbox{.}}}}}
    \def\rww{\mathclose{{\tst{\raise 1pt\hbox{.}\atop\raise 1pt\hbox{.}}}}}

   \font\sixrm=cmr6   \font\eightrm=cmr8  
   \font\sixi=cmmi6   \font\eighti=cmmi8  
  \font\sixsy=cmsy6  \font\eightsy=cmsy8 
  \font\sixbf=cmbx6  \font\eightbf=cmbx8 
                     \font\eightit=cmti8 
                     \font\eightsl=cmsl8 
                     \font\eighttt=cmtt8 

\font\eightfrak=eufm7 at 8pt

\def\eightpoint{\def\rm{\fam0\eightrm}
 \textfont0=\eightrm \scriptfont0=\sixrm \scriptscriptfont0=\fiverm
 \textfont1=\eighti \scriptfont1=\sixi \scriptscriptfont1=\fivei
 \textfont2=\eightsy \scriptfont2=\sixsy \scriptscriptfont2=\fivesy
 \textfont3=\tenex \scriptfont3=\tenex \scriptscriptfont3=\tenex
 \textfont\itfam=\eightit \def\it{\fam\itfam\eightit}%
 \textfont\slfam=\eightsl \def\sl{\fam\slfam\eightsl}%
 \textfont\ttfam=\eighttt \def\tt{\fam\ttfam\eighttt}%
 \textfont\frakfam=\eightfrak \def\frak{\fam\frakfam\tenfrak}%
 \textfont\bffam=\eightbf \scriptfont\bffam=\sixbf
 \scriptscriptfont\bffam=\fivebf \def\bf{\fam\bffam\eightbf}%
 \normalbaselineskip=9pt
 \setbox\strutbox=\hbox{\vrule height7pt depth2pt width0pt}%
 \let\sc=\sixrm \let\big=\eightbig \normalbaselines\rm}
\catcode`@=11
\def\footnote#1{\edef\@sf{\spacefactor\the\spacefactor}#1\@sf
     \insert\footins\bgroup\eightpoint
     \interlinepenalty100 \let\par=\endgraf
     \leftskip=0pt \rightskip=0pt
     \splittopskip=10pt plus 1pt minus 1pt \floatingpenalty=20000
     \smallskip\item{#1}\bgroup\strut\aftergroup\@foot\let\next}
\skip\footins=12pt plus 2pt minus 4pt
\dimen\footins=30pc
\catcode`@=12


\newcount\CHAPNO
\newcount\APPNO
\CHAPNO=0
\APPNO=1
\def\advCHAPNO{\advance\CHAPNO by 1}
\def\advAPPNO{\advance\APPNO by 1}

\def\caproman#1{\ifcase#1\or I\or II\or III\or IV\or V\or VI\or VII\or
VIII\or IX\or X\or XI\or XII\or XIII\or XIV\or XV\or XVI\or XVII\or XVIII\or
XIX\or XX\or XXI\or XXII\or XXIII\or XXIV\or XXV\or XXVI\or XXVII\or XXVIII\or XXIX\or XXX\or XXXI\or XXXII\or XXXIII\or XXXIV\or XXXV\or XXXVI\or XXXVII\or XXXVIII\or XXXIX\fi}%

\def\capletter#1{\ifcase#1\or A\or B\or C\or D\or E\or F\or G\or
H\or I\or J\or K\or L\or M\or N\or O\or P\or Q\or R\or
S\or T\or U\or V\or W\or X\or Y\or Z\fi}%

\newcount\cHintroI \cHintroI=\CHAPNO \advCHAPNO 
\newcount\cHintroOverview  \cHintroOverview=\CHAPNO \advCHAPNO 
\newcount\cHrenmap \cHrenmap=\CHAPNO \advCHAPNO 

 \advAPPNO

\newcount\cHintroII \cHintroII=\CHAPNO \advCHAPNO 
                              
\newcount\cHfirstscale \cHfirstscale=\CHAPNO \advCHAPNO
                              
\newcount\cHnewsectors \cHnewsectors=\CHAPNO \advCHAPNO
                              
\newcount\cHphladders \cHphladders=\CHAPNO \advCHAPNO
                              
\newcount\cHfinitescale \cHfinitescale=\CHAPNO \advCHAPNO
                              
\newcount\cHstep \cHstep=\CHAPNO \advCHAPNO
                              
\newcount\cHrecurs \cHrecurs=\CHAPNO \advCHAPNO
                              
 \advAPPNO

\newcount\cHintroIII \cHintroIII=\CHAPNO \advCHAPNO
                              
\newcount\cHtildefinitescale \cHtildefinitescale=\CHAPNO \advCHAPNO
                              
\newcount\cHtildenewsectors \cHtildenewsectors=\CHAPNO \advCHAPNO
                              
\newcount\cHtildephladders \cHtildephladders=\CHAPNO \advCHAPNO
                              
\newcount\cHtildestep  \cHtildestep=\CHAPNO \advCHAPNO

 \advAPPNO
 \advAPPNO


  \IgNoRe{PG}
  \IgNoRe{EQN}
  \IgNoRe{STM Assertion }
  \IgNoRe{PG}
  \IgNoRe{STM Assertion }
  \IgNoRe{STM Assertion }
  \IgNoRe{STM Assertion }
  \IgNoRe{STM Assertion }
  \IgNoRe{STM Assertion }
  \IgNoRe{STM Assertion }
  \IgNoRe{EQN}
  \IgNoRe{STM Assertion }
  \IgNoRe{STM Assertion }
  \IgNoRe{STM Assertion }
  \IgNoRe{STM Assertion }
  \IgNoRe{STM Assertion }
  \IgNoRe{STM Assertion }
  \IgNoRe{STM Assertion }
  \IgNoRe{STM Assertion }
  \IgNoRe{PG}
  \IgNoRe{STM Assertion }
  \IgNoRe{STM Assertion }
  \IgNoRe{STM Assertion }
  \IgNoRe{STM Assertion }
  \IgNoRe{STM Assertion }
  \IgNoRe{STM Assertion }
  \IgNoRe{STM Assertion }
  \IgNoRe{STM Assertion }
  \IgNoRe{EQN}
  \IgNoRe{PG}
  \IgNoRe{PG}
  \IgNoRe{STM Assertion }
  \IgNoRe{STM Assertion }
  \IgNoRe{STM Assertion }
  \IgNoRe{STM Assertion }
  \IgNoRe{STM Assertion }
  \IgNoRe{STM Assertion }
  \IgNoRe{PG}
  \IgNoRe{EQN}
  \IgNoRe{EQN}
  \IgNoRe{EQN}
  \IgNoRe{EQN}
  \IgNoRe{EQN}
  \IgNoRe{EQN}
  \IgNoRe{STM Assertion }
  \IgNoRe{STM Assertion }
  \IgNoRe{STM Assertion }
  \IgNoRe{STM Assertion }
  \IgNoRe{PG}
 \def\thmOSinsulators{\frefwarning V.2} \IgNoRe{STM Assertion }
  \IgNoRe{EQN}
  \IgNoRe{STM Assertion }
  \IgNoRe{STM Assertion }
  \IgNoRe{STM Assertion }
  \IgNoRe{STM Assertion }
  \IgNoRe{PG}
  \IgNoRe{STM Assertion }
  \IgNoRe{STM Assertion }
  \IgNoRe{STM Assertion }
  \IgNoRe{STM Assertion }
  \IgNoRe{PG}
  \IgNoRe{EQN}
  \IgNoRe{EQN}
  \IgNoRe{PG}
  \IgNoRe{STM Assertion }
  \IgNoRe{STM Assertion }
  \IgNoRe{STM Assertion }
  \IgNoRe{EQN}
  \IgNoRe{PG}
  \IgNoRe{STM Assertion }
  \IgNoRe{STM Assertion }
  \IgNoRe{EQN}
  \IgNoRe{STM Assertion }
  \IgNoRe{STM Assertion }
  \IgNoRe{STM Assertion }
  \IgNoRe{STM Assertion }
  \IgNoRe{PG}
  \IgNoRe{STM Assertion }
  \IgNoRe{STM Assertion }
  \IgNoRe{STM Assertion }
  \IgNoRe{STM Assertion }
 \def\thmOSfirststep{\frefwarning VIII.6} \IgNoRe{STM Assertion }
  \IgNoRe{STM Assertion }
  \IgNoRe{STM Assertion }
  \IgNoRe{STM Assertion }
  \IgNoRe{PG}
  \IgNoRe{STM Assertion }
  \IgNoRe{STM Assertion }
  \IgNoRe{STM Assertion }
  \IgNoRe{STM Assertion }
  \IgNoRe{STM Assertion }
  \IgNoRe{EQN}
  \IgNoRe{STM Assertion }
  \IgNoRe{STM Assertion }
  \IgNoRe{PG}
  \IgNoRe{STM Assertion }
  \IgNoRe{STM Assertion }
  \IgNoRe{STM Assertion }
  \IgNoRe{STM Assertion }
  \IgNoRe{STM Assertion }
  \IgNoRe{STM Assertion }
  \IgNoRe{STM Assertion }
  \IgNoRe{STM Assertion }
  \IgNoRe{STM Assertion }
  \IgNoRe{EQN}
  \IgNoRe{STM Assertion }
  \IgNoRe{STM Assertion }
  \IgNoRe{PG}
  \IgNoRe{STM Assertion }
  \IgNoRe{STM Assertion }
  \IgNoRe{STM Assertion }
  \IgNoRe{STM Assertion }
  \IgNoRe{STM Assertion }
  \IgNoRe{STM Assertion }
  \IgNoRe{STM Assertion }
  \IgNoRe{PG}
  \IgNoRe{STM Assertion }
  \IgNoRe{PG}
  \IgNoRe{PG}
  \IgNoRe{STM Assertion }
  \IgNoRe{STM Assertion }
  \IgNoRe{PG}
  \IgNoRe{STM Assertion }
  \IgNoRe{STM Assertion }
  \IgNoRe{STM Assertion }
  \IgNoRe{STM Assertion }
  \IgNoRe{STM Assertion }
  \IgNoRe{STM Assertion }
  \IgNoRe{STM Assertion }
  \IgNoRe{STM Assertion }
  \IgNoRe{STM Assertion }
  \IgNoRe{STM Assertion }
  \IgNoRe{STM Assertion }
  \IgNoRe{STM Assertion }
  \IgNoRe{STM Assertion }
  \IgNoRe{STM Assertion }
  \IgNoRe{STM Assertion }
  \IgNoRe{STM Assertion }
  \IgNoRe{EQN}
  \IgNoRe{STM Assertion }
  \IgNoRe{STM Assertion }
  \IgNoRe{PG}
  \IgNoRe{STM Assertion }
  \IgNoRe{EQN}
  \IgNoRe{EQN}
  \IgNoRe{STM Assertion }
  \IgNoRe{EQN}
  \IgNoRe{EQN}
  \IgNoRe{STM Assertion }
  \IgNoRe{EQN}
  \IgNoRe{STM Assertion }
  \IgNoRe{EQN}
  \IgNoRe{STM Assertion }
  \IgNoRe{STM Assertion }
  \IgNoRe{STM Assertion }
  \IgNoRe{STM Assertion }
  \IgNoRe{PG}
  \IgNoRe{STM Assertion }
  \IgNoRe{STM Assertion }
  \IgNoRe{STM Assertion }
  \IgNoRe{STM Assertion }
  \IgNoRe{EQN}
  \IgNoRe{EQN}
  \IgNoRe{STM Assertion }
  \IgNoRe{STM Assertion }
  \IgNoRe{PG}
  \IgNoRe{STM Assertion }
  \IgNoRe{STM Assertion }
 \def\lemOSconcreteintconst{\frefwarning XV.5} \IgNoRe{STM Assertion }
  \IgNoRe{EQN}
  \IgNoRe{STM Assertion }
  \IgNoRe{EQN}
  \IgNoRe{STM Assertion }
  \IgNoRe{STM Assertion }
  \IgNoRe{EQN}
  \IgNoRe{STM Assertion }
  \IgNoRe{EQN}
  \IgNoRe{EQN}
  \IgNoRe{EQN}
  \IgNoRe{EQN}
  \IgNoRe{STM Assertion }
  \IgNoRe{STM Assertion }
  \IgNoRe{STM Assertion }
  \IgNoRe{STM Assertion }
  \IgNoRe{PG}
  \IgNoRe{STM Assertion }
  \IgNoRe{STM Assertion }
  \IgNoRe{STM Assertion }
  \IgNoRe{STM Assertion }
  \IgNoRe{STM Assertion }
  \IgNoRe{STM Assertion }
  \IgNoRe{STM Assertion }
  \IgNoRe{STM Assertion }
  \IgNoRe{STM Assertion }
  \IgNoRe{STM Assertion }
  \IgNoRe{EQN}
  \IgNoRe{EQN}
  \IgNoRe{STM Assertion }
  \IgNoRe{PG}
  \IgNoRe{STM Assertion }
  \IgNoRe{STM Assertion }
  \IgNoRe{EQN}
  \IgNoRe{STM Assertion }
  \IgNoRe{STM Assertion }
  \IgNoRe{EQN}
  \IgNoRe{EQN}
  \IgNoRe{EQN}
  \IgNoRe{EQN}
  \IgNoRe{EQN}
  \IgNoRe{STM Assertion }
  \IgNoRe{STM Assertion }
  \IgNoRe{STM Assertion }
  \IgNoRe{EQN}
  \IgNoRe{EQN}
  \IgNoRe{EQN}
  \IgNoRe{EQN}
  \IgNoRe{EQN}
  \IgNoRe{EQN}
  \IgNoRe{STM Assertion }
  \IgNoRe{STM Assertion }
  \IgNoRe{STM Assertion }
  \IgNoRe{PG}
  \IgNoRe{STM Assertion }
  \IgNoRe{STM Assertion }
  \IgNoRe{EQN}
  \IgNoRe{EQN}
  \IgNoRe{STM Assertion }
  \IgNoRe{EQN}
  \IgNoRe{EQN}
  \IgNoRe{STM Assertion }
  \IgNoRe{EQN}
  \IgNoRe{EQN}
  \IgNoRe{STM Assertion }
  \IgNoRe{STM Assertion }
  \IgNoRe{PG}
  \IgNoRe{STM Assertion }
  \IgNoRe{STM Assertion }
  \IgNoRe{STM Assertion }
  \IgNoRe{PG}
  \IgNoRe{STM Assertion }
  \IgNoRe{STM Assertion }
  \IgNoRe{STM Assertion }
  \IgNoRe{STM Assertion }
  \IgNoRe{PG}
  \IgNoRe{STM Assertion }
  \IgNoRe{STM Assertion }
  \IgNoRe{STM Assertion }
  \IgNoRe{STM Assertion }
  \IgNoRe{STM Assertion }
  \IgNoRe{STM Assertion }
  \IgNoRe{STM Assertion }
  \IgNoRe{STM Assertion }
  \IgNoRe{STM Assertion }
  \IgNoRe{STM Assertion }
  \IgNoRe{STM Assertion }
  \IgNoRe{STM Assertion }
  \IgNoRe{STM Assertion }
  \IgNoRe{STM Assertion }
  \IgNoRe{PG}
  \IgNoRe{STM Assertion }
  \IgNoRe{PG}
  \IgNoRe{STM Assertion }
  \IgNoRe{STM Assertion }
  \IgNoRe{PG}
  \IgNoRe{STM Assertion }
  \IgNoRe{STM Assertion }
  \IgNoRe{EQN}
  \IgNoRe{PG}
  \IgNoRe{EQN}
  \IgNoRe{STM Assertion }
  \IgNoRe{STM Assertion }
  \IgNoRe{PG}
  \IgNoRe{EQN}
  \IgNoRe{EQN}
  \IgNoRe{EQN}
  \IgNoRe{EQN}
  \IgNoRe{EQN}
  \IgNoRe{STM Assertion }
  \IgNoRe{STM Assertion }
  \IgNoRe{EQN}
  \IgNoRe{STM Assertion }
  \IgNoRe{PG}
  \IgNoRe{PG}
  \IgNoRe{PG}
  \IgNoRe{STM Assertion }
  \IgNoRe{EQN}
  \IgNoRe{STM Assertion }
  \IgNoRe{PG}
  \IgNoRe{STM Assertion }
  \IgNoRe{EQN}
  \IgNoRe{STM Assertion }
  \IgNoRe{STM Assertion }
  \IgNoRe{PG}
  \IgNoRe{EQN}
  \IgNoRe{EQN}
  \IgNoRe{EQN}
  \IgNoRe{STM Assertion }
  \IgNoRe{STM Assertion }
  \IgNoRe{STM Assertion }
  \IgNoRe{EQN}
  \IgNoRe{STM Assertion }
  \IgNoRe{STM Assertion }
  \IgNoRe{STM Assertion }
  \IgNoRe{STM Assertion }
  \IgNoRe{STM Assertion }
  \IgNoRe{STM Assertion }
  \IgNoRe{PG}
  \IgNoRe{STM Assertion }
  \IgNoRe{STM Assertion }
  \IgNoRe{STM Assertion }
  \IgNoRe{STM Assertion }
  \IgNoRe{STM Assertion }
  \IgNoRe{STM Assertion }
  \IgNoRe{STM Assertion }
  \IgNoRe{PG}
  \IgNoRe{PG}


 \def\pgRI{\frefwarning 1} \IgNoRe{PG}
 \def\defsuperalgebra{\frefwarning II.1} \IgNoRe{STM Assertion }
 \def\pgRII{\frefwarning 3} \IgNoRe{PG}
 \def\pgRIIa{\frefwarning 3} \IgNoRe{PG}
 \def\exsuper{\frefwarning II.2} \IgNoRe{STM Assertion }
 \def\pgRIIb{\frefwarning 4} \IgNoRe{PG}
 \def\defrengroup{\frefwarning II.3} \IgNoRe{STM Assertion }
 \def\remnormal{\frefwarning II.4} \IgNoRe{STM Assertion }
 \def\eqnpartialanti{\frefwarning II.1} \IgNoRe{EQN}
 \def\defContract{\frefwarning II.5} \IgNoRe{STM Assertion }
 \def\remContractlinearity{\frefwarning II.6} \IgNoRe{STM Assertion }
 \def\pgRIIc{\frefwarning 6} \IgNoRe{PG}
 \def\remkerncontract{\frefwarning II.7} \IgNoRe{STM Assertion }
 \def\lemrestrcontract{\frefwarning II.8} \IgNoRe{STM Assertion }
 \def\eqIIcontr{\frefwarning II.2} \IgNoRe{EQN}
 \def\defGrasscontract{\frefwarning II.9} \IgNoRe{STM Assertion }
 \def\lemGrasscontract{\frefwarning II.10} \IgNoRe{STM Assertion }
 \def\egcontract{\frefwarning II.11} \IgNoRe{STM Assertion }
 \def\remcontract{\frefwarning II.12} \IgNoRe{STM Assertion }
 \def\lemcontract{\frefwarning II.13} \IgNoRe{STM Assertion }
 \def\defFancynormdomain{\frefwarning II.14} \IgNoRe{STM Assertion }
 \def\pgRIId{\frefwarning 11} \IgNoRe{PG}
 \def\defFancynorm{\frefwarning II.15} \IgNoRe{STM Assertion }
 \def\exFancynorm{\frefwarning II.16} \IgNoRe{STM Assertion }
 \def\remFancynorm{\frefwarning II.17} \IgNoRe{STM Assertion }
 \def\defsymnorm{\frefwarning II.18} \IgNoRe{STM Assertion }
 \def\remkernsymnorm{\frefwarning II.19} \IgNoRe{STM Assertion }
 \def\egcompatnorm{\frefwarning II.20} \IgNoRe{STM Assertion }
 \def\defAmnonetonr{\frefwarning II.21} \IgNoRe{STM Assertion }
 \def\lemGrasscompatnorm{\frefwarning II.22} \IgNoRe{STM Assertion }
 \def\deffunctnorm{\frefwarning II.23} \IgNoRe{STM Assertion }
 \def\remfunctnorm{\frefwarning II.24} \IgNoRe{STM Assertion }
 \def\defcontractintbound{\frefwarning II.25} \IgNoRe{STM Assertion }
 \def\egIIcompatnorm{\frefwarning II.26} \IgNoRe{STM Assertion }
 \def\pgRIIe{\frefwarning 16} \IgNoRe{PG}
 \def\eqnIIcompatnorm{\frefwarning II.3} \IgNoRe{EQN}
 \def\defrengroupInfDim{\frefwarning II.27} \IgNoRe{STM Assertion }
 \def\theorII{\frefwarning II.28} \IgNoRe{STM Assertion }
 \def\lemGrasscontractnorm{\frefwarning II.29} \IgNoRe{STM Assertion }
 \def\pgRIIf{\frefwarning 18} \IgNoRe{PG}
 \def\scaleIntConsts{\frefwarning II.30} \IgNoRe{STM Assertion }
 \def\lemwicknorm{\frefwarning II.31} \IgNoRe{STM Assertion }
 \def\corwicknorm{\frefwarning II.32} \IgNoRe{STM Assertion }
 \def\propestCconnect{\frefwarning II.33} \IgNoRe{STM Assertion }
 \def\eqnPropestCconnect{\frefwarning II.4} \IgNoRe{EQN}
 \def\eqnestCconnect{\frefwarning II.5} \IgNoRe{EQN}
 \def\remrenschw{\frefwarning III.1} \IgNoRe{STM Assertion }
 \def\pgRIII{\frefwarning 26} \IgNoRe{PG}
 \def\pgRIIIa{\frefwarning 26} \IgNoRe{PG}
 \def\theoremIII{\frefwarning III.2} \IgNoRe{STM Assertion }
 \def\eqnROPa{\frefwarning III.1} \IgNoRe{EQN}
 \def\remfunctor{\frefwarning III.3} \IgNoRe{STM Assertion }
 \def\defcalR{\frefwarning III.4} \IgNoRe{STM Assertion }
 \def\propcalR{\frefwarning III.5} \IgNoRe{STM Assertion }
 \def\eqRCdef{\frefwarning III.2} \IgNoRe{EQN}
 \def\remcalR{\frefwarning III.6} \IgNoRe{STM Assertion }
 \def\propnormestR{\frefwarning III.7} \IgNoRe{STM Assertion }
 \def\lemnormestcalR{\frefwarning III.8} \IgNoRe{STM Assertion }
 \def\pgRIIIb{\frefwarning 30} \IgNoRe{PG}
 \def\cornormestcalR{\frefwarning III.9} \IgNoRe{STM Assertion }
 \def\propnormestcalS{\frefwarning III.10} \IgNoRe{STM Assertion }
 \def\pgRIIIc{\frefwarning 33} \IgNoRe{PG}
 \def\remjointanalyticity{\frefwarning III.11} \IgNoRe{STM Assertion }
 \def\theoremIVa{\frefwarning IV.1} \IgNoRe{STM Assertion }
 \def\remnormlaCD{\frefwarning IV.2} \IgNoRe{STM Assertion }
 \def\pgRIV{\frefwarning 35} \IgNoRe{PG}
 \def\remjointanalyticitywick{\frefwarning IV.3} \IgNoRe{STM Assertion }
 \def\theoremIVb{\frefwarning IV.4} \IgNoRe{STM Assertion }
 \def\lemprftwoA{\frefwarning IV.5} \IgNoRe{STM Assertion }
 \def\lemfcf{\frefwarning IV.6} \IgNoRe{STM Assertion }
 \def\lemprftwoB{\frefwarning IV.7} \IgNoRe{STM Assertion }
 \def\lemprftwoC{\frefwarning IV.8} \IgNoRe{STM Assertion }
 \def\lemBI{\frefwarning A.1} \IgNoRe{STM Assertion }
 \def\propBII{\frefwarning A.2} \IgNoRe{STM Assertion }
 \def\pgRA{\frefwarning 42} \IgNoRe{PG}
 \def\corBIII{\frefwarning A.3} \IgNoRe{STM Assertion }
 \def\lemBIV{\frefwarning A.4} \IgNoRe{STM Assertion }
 \def\lemBV{\frefwarning A.5} \IgNoRe{STM Assertion }
 \def\corBVIa{\frefwarning A.6} \IgNoRe{STM Assertion }
 \def\lemBVI{\frefwarning A.7} \IgNoRe{STM Assertion }
 \def\lemCovderiv{\frefwarning A.8} \IgNoRe{STM Assertion }
 \def\propGII{\frefwarning B.1} \IgNoRe{STM Assertion }
 \def\exGIII{\frefwarning B.2} \IgNoRe{STM Assertion }
 \def\pgRB{\frefwarning 49} \IgNoRe{PG}
 \def\lemGIII{\frefwarning B.3} \IgNoRe{STM Assertion }
 \def\lemGIV{\frefwarning B.4} \IgNoRe{STM Assertion }
 \def\pgRIref{\frefwarning 54} \IgNoRe{PG}
  \IgNoRe{PG}
  \IgNoRe{STM Assertion }
  \IgNoRe{STM Assertion }
  \IgNoRe{PG}
  \IgNoRe{PG}
  \IgNoRe{STM Assertion }
  \IgNoRe{STM Assertion }
  \IgNoRe{EQN}
  \IgNoRe{STM Assertion }
  \IgNoRe{PG}
  \IgNoRe{STM Assertion }
  \IgNoRe{STM Assertion }
  \IgNoRe{STM Assertion }
  \IgNoRe{PG}
  \IgNoRe{STM Assertion }
  \IgNoRe{STM Assertion }
  \IgNoRe{STM Assertion }
  \IgNoRe{EQN}
  \IgNoRe{STM Assertion }
  \IgNoRe{PG}
  \IgNoRe{EQN}
  \IgNoRe{STM Assertion }
  \IgNoRe{STM Assertion }
  \IgNoRe{STM Assertion }
  \IgNoRe{STM Assertion }
  \IgNoRe{EQN}
  \IgNoRe{EQN}
  \IgNoRe{PG}
  \IgNoRe{PG}
  \IgNoRe{STM Assertion }
  \IgNoRe{STM Assertion }
  \IgNoRe{STM Assertion }
  \IgNoRe{STM Assertion }
  \IgNoRe{STM Assertion }
  \IgNoRe{STM Assertion }
  \IgNoRe{STM Assertion }
  \IgNoRe{STM Assertion }
  \IgNoRe{PG}
  \IgNoRe{STM Assertion }
  \IgNoRe{STM Assertion }
  \IgNoRe{STM Assertion }
  \IgNoRe{STM Assertion }
  \IgNoRe{STM Assertion }
  \IgNoRe{STM Assertion }
  \IgNoRe{STM Assertion }
  \IgNoRe{STM Assertion }
  \IgNoRe{STM Assertion }
  \IgNoRe{STM Assertion }
  \IgNoRe{PG}
  \IgNoRe{STM Assertion }
  \IgNoRe{STM Assertion }
  \IgNoRe{STM Assertion }
  \IgNoRe{PG}
  \IgNoRe{STM Assertion }
  \IgNoRe{STM Assertion }
  \IgNoRe{STM Assertion }
  \IgNoRe{STM Assertion }
  \IgNoRe{PG}
  \IgNoRe{STM Assertion }
  \IgNoRe{STM Assertion }
  \IgNoRe{PG}
  \IgNoRe{STM Assertion }
  \IgNoRe{PG}
  \IgNoRe{STM Assertion }
  \IgNoRe{PG}
  \IgNoRe{STM Assertion }
  \IgNoRe{STM Assertion }
  \IgNoRe{STM Assertion }
  \IgNoRe{STM Assertion }
  \IgNoRe{PG}
  \IgNoRe{PG}
  \IgNoRe{STM Assertion }
  \IgNoRe{STM Assertion }
  \IgNoRe{EQN}
  \IgNoRe{EQN}
  \IgNoRe{STM Assertion }
  \IgNoRe{STM Assertion }
  \IgNoRe{STM Assertion }
  \IgNoRe{STM Assertion }
  \IgNoRe{PG}
  \IgNoRe{STM Assertion }
  \IgNoRe{STM Assertion }
  \IgNoRe{STM Assertion }
  \IgNoRe{STM Assertion }
  \IgNoRe{STM Assertion }
  \IgNoRe{STM Assertion }
  \IgNoRe{STM Assertion }
  \IgNoRe{PG}
  \IgNoRe{STM Assertion }
  \IgNoRe{EQN}
  \IgNoRe{EQN}
  \IgNoRe{PG}
  \IgNoRe{STM Assertion }
  \IgNoRe{EQN}
  \IgNoRe{STM Assertion }
  \IgNoRe{STM Assertion }
  \IgNoRe{STM Assertion }
  \IgNoRe{PG}
  \IgNoRe{STM Assertion }
  \IgNoRe{EQN}
  \IgNoRe{STM Assertion }
  \IgNoRe{PG}
 \def\pgRInot{\frefwarning 56} \IgNoRe{PG}
  \IgNoRe{PG}


{\nopagenumbers
\multiply\baselineskip by \spacingDenominator\divide \baselineskip by\spacingNumerator

\null\vskip3truecm

%
%
\centerline{\tafontt Convergence of Perturbation Expansions}
\centerline{\tafontt  in Fermionic Models}

\vskip0.1in
\centerline{\tbfontt Part 1: Nonperturbative Bounds}

\vskip0.75in
\centerline{Joel Feldman{\parindent=.15in\footnote{$^{*}$}{Research supported 
in part by the
 Natural Sciences and Engineering Research Council of Canada and the Forschungsinstitut f\"ur Mathematik, ETH Z\"urich}}}
\centerline{Department of Mathematics}
\centerline{University of British Columbia}
\centerline{Vancouver, B.C. }
\centerline{CANADA\ \   V6T 1Z2}
\centerline{feldman@math.ubc.ca}
\centerline{http:/\hskip-3pt/www.math.ubc.ca/\squiggle
feldman/}
\vskip0.3in
\centerline{Horst Kn\"orrer, Eugene Trubowitz}
\centerline{Mathematik}
\centerline{ETH-Zentrum}
\centerline{CH-8092 Z\"urich}
\centerline{SWITZERLAND}
\centerline{knoerrer@math.ethz.ch, trub@math.ethz.ch}
\centerline{http:/\hskip-3pt/www.math.ethz.ch/\squiggle
knoerrer/}

\vskip0.75in
\noindent
%
{\bf Abstract.\ \ \ } 
An estimate on the operator norm of an abstract fermionic renormalization 
group map is derived. This abstract estimate is applied in another paper 
to construct the thermodynamic Green's functions of a two dimensional,  
weakly coupled fermion gas with an asymmetric Fermi curve. The estimate
derived here is strong enough to control everything but the 
sum of all quartic contributions to the Green's functions.

\vfill
\eject


\titleb{Table of Contents}
\halign{\hfill#\ &\hfill#\ &#\hfill&\ p\ \hfil#&\ p\ \hfil#\cr
\noalign{\vskip0.05in}
\S I&\omit Introduction                          \span&\:\pgRI&\omit\cr
\noalign{\vskip0.05in}
\S II&\omit The Renormalization Group Map        \span&\:\pgRII\cr
&&Superalgebras                                  &\omit&\:\pgRIIa\cr
&&Grassmann Gaussian Integrals                   &\omit&\:\pgRIIb\cr
&&Tensor algebra and Grassmann algebras          &\omit&\:\pgRIIc\cr
&&Seminorms                                      &\omit&\:\pgRIId\cr
&&An estimate of the renormalization group map   &\omit&\:\pgRIIe\cr
&&Contraction and Integral Bounds                &\omit&\:\pgRIIf\cr
\noalign{\vskip0.05in}
\S III&\omit The Schwinger Functional            \span&\:\pgRIII\cr
&&Description of the Schwinger Functional       &\omit&\:\pgRIIIa\cr
&&Norm estimates on the Schwinger functional    &\omit&\:\pgRIIIb\cr
&&Proof of Theorem \:\theorII                     &\omit&\:\pgRIIIc\cr
\noalign{\vskip0.05in}
\S IV&\omit More Estimates on the Renormalization Group Map\span&\:\pgRIV\cr
\noalign{\vskip0.05in}
{\bf Appendices}\span\cr
\noalign{\vskip0.05in}
\S A&\omit Wick--Ordering                        \span&\:\pgRA\cr
\noalign{\vskip0.05in}
\S B&\omit Gram Bounds                            \span&\:\pgRB\cr
\noalign{\vskip0.05in}
 &\omit References                                    \span&\:\pgRIref \cr
\noalign{\vskip0.05in}
 &\omit Notation                                      \span&\:\pgRInot \cr
}

\vfill\eject
\multiply\baselineskip by \spacingNumerator\divide \baselineskip by\spacingDenominator}
\pageno=1


\chap{Introduction}\PG\pgRI
In a Grassmann algebra ${\bf A}$ with generators $\psi_i$, the renormalization group map with respect to a covariance $C$ is the map that associates to each element $W(\psi)$ of the Grassmann algebra the element
$$
\Om_C(W)(\psi) = \log \sfrac{1}{Z(W)}\int e^{W(\psi +\xi)} d\mu_C(\xi) 
\qquad{\rm where}\quad Z(W)=\int e^{W(\xi)} d\mu_C(\xi)
$$
whenever $Z(W) \ne 0$. Here, $\xi_i$ is a second set of variables that anticommute amongst themselves and with the $\psi_j$'s.
 $\int \cdots d\mu_C(\xi)$ is the Grassmann Gaussian integral with respect to these variables (see \S II). The Schwinger functional with 
 interaction $W$ is the map that associates to a Grassmann function $f(\xi)$ the complex number
$$
{\cal S}(f) = \sfrac{1}{Z(W)}\int f(\xi)\, e^{W(\xi)} d\mu_C(\xi) 
$$
Observe that, if $Z(W)\ne 0$, we may always arrange that $Z(W)=1$ by adding a constant to $W$. In this case
$$\eqalign{
\Om_C(W)(\psi)\ =\ \int_0^1 \sfrac{d\ }{dt} \Om_C(t W)\,dt 
\ =\ \int_0^1 \frac{\int W(\psi+\xi)\, e^{t W(\psi+\xi)}\,d\mu_C(\xi)}
      {\int e^{t W(\psi+\xi)} \,d\mu_C(\xi)}\,dt  
}$$
The $t$--integrand of the right hand side is the Schwinger functional of $W(\psi+\xi)$ with  interaction $tW(\psi+\xi)$, where the Schwinger functional is considered in the Grassmann algebra with generators $\xi_i$ and coefficients in ${\bf A}$. We exploit this observation and the representation of the Schwinger functional in [FKT1] to develop non--perturbative bounds
for the renormalization group map. By ``non--perturbative'' we mean that we bound the sum of the perturbation expansion, not only individual terms in the expansion. 

In a perturbative analysis, one decomposes $W=\smsum_n W_n$, where $W_n$ is homogeneous of degree $n$ in $\psi$. Then $\Om_C(W)$ is the sum of the values of all connected Feynman diagrams with vertices $W_n$ and propagator $C$. In most applications, the kernels $W_n$ are translation invariant. To bound the value of a Feynman diagram, one usually (see [FT]) selects a tree in the diagram (to exploit the connectedness of the diagram) and bounds the lines of the tree differently from the other lines.\footnote{$^{(1)}$}{Typically, the lines of the tree contribute a factor of the $L_1$ norm of the propagator in position space to the bound on the diagram, while the other lines contribute a factor of the $L_\infty$ norm of the propagator to the bound.} The 
non--perturbative analysis we present here is close to the diagrammatic analysis (see the introduction to [FKT1]), but it allows implementation of the Pauli exclusion principle for lines not on the tree. The norms we use in  applications to many fermion systems are quite complicated. Therefore, here, we axiomatize their relevant properties: We assume that there is a system of norms $\|\cdot\|$ on the homogeneous subspaces of the Grassmann algebra. Furthermore, we assume that there is a ``contraction bound'' $\cb$ for $C$,
such that for any two homogeneous elements $f,f'$ in the Grassmann algebra, 
the norm of the diagram

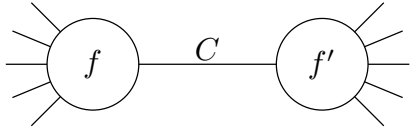
\centerline{\figput{figI1}}

\noindent that is obtained by joining $f$ and $f'$ by one line is bounded by 
$\cb\,\|f\|\,\|f'\|$
and we assume that there is an ``integral bound'' $\ib$ 
that controls the effect of integrating out some of the fields attached to a single vertex. See Definition
\:\defcontractintbound. The contraction bound is analogous to the bound on the tree contribution to a Feynman diagram. The integral bound incorporates the Pauli exclusion principle and is often derived from Gram's estimate for determinants. See Appendix B. It replaces the bound on the non--tree contribution to a Feynman diagram.

When applying a renormalization group transformation, there are often other fields present, that do play no role in the renormalization group transformation. We suppress these fields by allowing a (super)algebra as coefficient ring for the Grassmann algebra on which the renormalization group map is analyzed. See Definition \:\defsuperalgebra. Also, in the analysis of many fermion systems,
we have to control various derivatives (in momentum space) of the effective interactions involved. 
To get a coherent notation for derivative norms, we allow the norms to take values in a formal power series ring, where the powers code the degree of derivatives.
See Definition \:\defFancynormdomain.

In \S II we introduce the concepts discussed above and formulate the 
main estimate on the renormalization group map in these terms (Theorem \:\theorII). \S III discusses the connection with the Schwinger functional and gives the proof of Theorem \:\theorII\footnote{$^{(2)}$}{For other approaches to controlling fermionic renormalization, see [DR] and [SW].}. In \S IV, further estimates of the renormalization group map are discussed, in particular on its derivative with respect to the interaction $W$ and the covariance $C$. Part 2 of the paper
(\S VI--IX) discusses the phenomenon of ``overlapping'' loops, that is
responsible for ``improvements over natural power counting'' and consequently is important for many fermion systems; see [S] and the introduction to 
part 2 of this paper. A notation table is provided at the end of the paper.

\vfill\eject

\chap{ The Renormalization Group Map}\PG\pgRII
\sect{Superalgebras}\PG\pgRIIa

\definition{\STM\defsuperalgebra}{
\Item{(i)} A superalgebra is an associative $\bbbc$-algebra $A$ with unit $1$, together with a decomposition $A=A_+\oplus A_-$ such that
$1\in A_+$ and 
$$\eqalign{
A_+\cdot A_+ \subset A_+  \qquad\qquad & A_-\cdot A_- \subset A_+ \cr
A_+\cdot A_- \subset A_-  \qquad\qquad & A_-\cdot A_+ \subset A_- \cr
}$$
and
$$\meqalign{
ab&=ba && &{\rm if}\quad a\in A_+\quad{\rm or}\quad b\in A_+\cr
ab&=-ba &&\qquad &{\rm if}\quad a,b\in A_-\cr
}$$
The elements of $A_+$ are called even, the elements of $A_-$ odd.
\Item{(ii)} A graded superalgebra is an associative $\bbbc$-algebra $A$ with unit, together with a decomposition $A=\bigoplus_{m=0}^\infty A_m$ such that
$A_0=\bbbc$, $A_m\cdot A_n \subset A_{m+n}$ for all $m,n\ge 0$, and such that
the decomposition $A=A_+\oplus A_-$ with
$$
A_+ = \bigoplus\limits_{m\ {\rm even}} A_m
\qquad\qquad\qquad
A_- = \bigoplus\limits_{m\ {\rm odd}}  A_m
$$
gives $A$ the structure of a superalgebra.
\Item{(iii)} Let $A$ be a graded superalgebra, $f=\sum\limits_m f_m\in A$ with
$f_m\in A_m$. Set $Z(f)=f_0\in A_0=\bbbc$. Clearly, if $f_0\ne 0$,
$$
\sfrac{f}{Z(f)}=1+\sum_{m\ge 1}\sfrac{f_m}{f_0}
$$
In this case, set
$$
\log\sfrac{f}{Z(f)}=\sum_{n=1}^\infty \sfrac{(-1)^{n-1}}{n}\big(\sfrac{f}{Z(f)}-1\big)^n
$$
\Item{(iv)} Let $A$ and $B$ be superalgebras. 
On the tensor product $A\otimes B$ we define multiplication by
$$\eqalign{
\big[ a \otimes (b_+ + b_-) \big]\,\big[ (a_+ + a_-) \otimes b \big]
&= a (a_+ + a_-) \otimes  (b_+ + b_-) b - 2\, a a_- \otimes b_- b\cr
&=  a a_+ \otimes b_+ b+a a_+ \otimes b_- b+a a_- \otimes b_+ b
-a a_- \otimes b_- b\cr
}$$
for $a\in A,\ b\in B\,,\ a_\pm \in A_\pm,\ b_\pm \in B_\pm$.
This multiplication defines an algebra structure on $A\otimes B$. Setting
$(A\otimes B)_+ = (A_+\otimes B_+) \oplus (A_-\otimes B_-)$, \ 
$(A\otimes B)_- = (A_+\otimes B_-) \oplus (A_-\otimes B_+)$ we get a superalgebra. If $A$ and $B$ are graded superalgebras then the decomposition
$A\otimes B = \bigoplus_{m=0}^\infty (A\otimes B)_m$ with
$$
(A\otimes B)_m =  \bigoplus_{m_1+m_2=m} A_{m_1}\otimes B_{m_2}
$$
gives $A\otimes B$ the structure of a graded superalgebra. 
}

\example{\STM\exsuper}{
Let $V$ be a complex vector space.  The Grassmann algebra 
$\bigwedge V=\bigoplus\limits_{m\ge 0} \bigwedge^m V$ over $V$ is a graded superalgebra.
If $A$ is any superalgebra, the Grassmann algebra over $V$ with coefficients in $A$ is the superalgebra 
$$
\bigwedge\nolimits_A V = A \otimes \bigwedge V
$$
where the tensor product is taken as in Definition \defsuperalgebra.iv. If $A$ is a graded superalgebra, so is $\bigwedge\nolimits_A V$.
}

In fact almost all graded superalgebras $A$ that will be used in this paper
will be subalgebras of Grassmann algebras. The one exception is the 
``enlarged'' algebra of section VII.

\sect{ Grassmann Gaussian Integrals}\PG\pgRIIb

Let $A$ be a superalgebra, $V$ be a complex vector space and $C$ an
antisymmetric bilinear form (covariance) on $V$. Then $C$ determines
 an $A$--linear map from $ \bigwedge\nolimits_A V $ to $A$ that is called
 the Grassmann Gaussian integral $d\mu_C$ on $ \bigwedge\nolimits_A V $.
Choose a set $\{\xi_i\}$ of generators for $V$. We write elements of 
$ \bigwedge\nolimits_A V $ as Grassmann functions $f(\xi)$.  A 
Grassmann function $f(\xi)$ is called even if it is an even element of the algebra $ \bigwedge\nolimits_A V $. The Grassmann Gaussian integral of $f(\xi)$ is denoted $\int f(\xi)\, d\mu_C(\xi)$. Then
$$
\int \xi_{i_1}\xi_{i_2}\cdots \xi_{i_n} \, d\mu_C(\xi)\ 
=\ {\rm Pf}\big[ C_{i_k i_\ell} \big]_{1\le k,\ell\le n} 
$$
where $C_{ij} = C(\xi_i,\xi_j)$ and the Pfaffian of an $n\times n$ matrix $M$ is denoted  ${\rm Pf} M$. Observe that, for any even Grassmann function $f(\xi)$,
the Grassmann Gaussian integral $\int f(\xi)\, d\mu_C(\xi)$ is an even element of the coefficient algebra $A$. 

Let $U$ be another vector space. Using the canonical isomorphism
$$
\bigwedge\nolimits_A(U\oplus V) 
= \bigoplus_{r,r'}\, \bigwedge^{r'}\nolimits_A U\, \bigwedge^r\nolimits_A V
\cong \bigwedge\nolimits_A U\, \otimes\, \bigwedge\nolimits_A V
\cong \bigwedge\nolimits_{\bigwedge\nolimits_A U} V
$$
the Grassmann Gaussian integral defines a map $ \int \,\cdot\,d\mu_C(\xi)$ from 
$\bigwedge\nolimits_A(U\oplus V)$ to $\bigwedge\nolimits_A U$. If $\{\ze_i\}$
is any set of vectors in $U$ then
$$
\int e^{\Sigma\, \xi_i\ze_i}\, d\mu_C(\xi) = e^{-1/2\,\Sigma\, \ze_i C_{ij} \ze_j}
$$

\definition{\STM\defrengroup}{
Choose a second copy $V'$ of $V$ and denote the element of $V'$ corresponding to the element $\xi_i$ of $V$ by $\psi_i$. If ${\rm dim}\,V<\infty$, the renormalization group map $\Om_C$ is defined by
$$
\Om_C(W)(\psi) = \log \sfrac{1}{Z}\int e^{W(\psi +\xi)} d\mu_C(\xi) 
\qquad{\rm where}\quad Z= Z\Big(\int e^{W(\xi)} d\mu_C(\xi)\Big)
$$
for all $W\in \bigwedge\nolimits_A V'$ for which 
$Z\Big(\int e^{W(\xi)} d\mu_C(\xi)\Big)\ne 0$. }

\remark{\STM\remnormal}{
\Item i) If ${\rm dim}\,V<\infty$, then $\Om_C(W)$ is a rational function of $W$. 

\Item ii)
By construction $Z\big(\Om_C(W)\big)=0$ for all $W$.

\Item iii) In Definition \:\defrengroupInfDim, below, we extend 
the definition of $\Om_C$ to normed vector spaces.

}

In this paper we state estimates on the renormalization group map for Wick ordered interactions $W$. Recall that Wick ordering with respect to a covariance
$C$
$$
f(\xi) \mapsto \ \lw f(\xi)\rw_{\xi,C} 
$$
is the $A$--linear map on $\bigwedge\nolimits_A V$ characterized by
$$
 \lw e^{\Sigma\, \xi_i\ze_i}\rw_{\xi,C}\ = \ 
e^{1/2\,\Sigma\, \ze_i C_{ij} \ze_j}\,e^{\Sigma\, \xi_i\ze_i}
$$
If the context admits, we delete the Wick ordering covariance $C$ or the variable $\xi$ (or both) from the symbol $\lw \,\cdot\,\rw_{\xi,C}$ for Wick ordering. 

Also recall the integration by parts formula
$$
\int\xi_i\,g(\xi)\ d\mu_C(\xi) =\sum_j C_{ij}\int 
 \sfrac{\de\ \ }{\de \xi_j}g(\xi) \ d\mu_C(\xi)
$$ 
or more generally
$$\eqalign{
\int \lw \xi_{i_1}\cdots\xi_{i_n}\rw\ g(\xi)\ d\mu_C(\xi) 
&=\sum_j C_{i_n\, j}\int \lw \xi_{i_1}\cdots\xi_{i_{n-1}}\rw\,
 \sfrac{\de\ \ }{\de \xi_j}g(\xi) \ d\mu_C(\xi)\cr
&=\sfrac{(-1)^{n-1}}{n}\sum_{i,j} C_{i\, j}\int \lW\sfrac{\de\ \ }{\de \xi_i} \xi_{i_1}\cdots\xi_{i_{n}}\rW\,
 \sfrac{\de\ \ }{\de \xi_j}g(\xi) \ d\mu_C(\xi)\cr
}$$ 

\goodbreak
\sect{ Tensor algebra and Grassmann algebras}\PG\pgRIIc

Again, let $V$ be a complex vector space with a set $\{\xi_i\}$ of generators 
and $A$ a superalgebra. We denote by $V^{\otimes n}$ the $n$--fold tensor product (over the complex numbers) of $V$ with itself. The symmetric group
$S_n$ of all permutations of $\{1,\cdots,n\}$ acts on $V^{\otimes n}$ (from the right) in such a way that
$$
( v_1 \otimes \cdots \otimes v_n )^\pi = 
v_{\pi(1)} \otimes \cdots \otimes v_{\pi(n)}
$$
for all $v_1,\cdots,v_n \in V$ and $\pi \in S_n$. The $n^{\rm th}$ exterior power
$\bigwedge^nV$ of $V$ can be identified with the set of all antisymmetric elements in $V^{\otimes n}$. We have the canonical projection
$$
{\rm Ant}_n:\ V^{\otimes n} \longrightarrow \bigwedge\nolimits^nV \qquad,\qquad
   f \longmapsto \sfrac{1}{n!} \smsum_{\pi \in S_n} \sgn( \pi)\, f^\pi  
$$
By $A$--linearity, ${\rm Ant}_n$ induces a map from $A\otimes V^{\otimes n}$ to
$\bigwedge_A^nV$. The image of $v_1 \otimes \cdots \otimes v_n$ under ${\rm Ant}_n$ is denoted by $v_1\cdots v_n$.

More generally, if $n=n_1+\cdots+n_r$ with nonnegative integers $n_1,\cdots,n_r$ we have the partial antisymmetrization
$$\eqalign{
{\rm Ant}_{n_1,\cdots,n_r}:\ A\otimes V^{\otimes n} & \longrightarrow 
\bigwedge\nolimits_A^{n_1}V \otimes_A \cdots \otimes_A \bigwedge\nolimits_A^{n_r}V \cr
   f & \longrightarrow 
\sfrac{1}{n_1! \cdots n_r!} \smsum_{\pi \in S_{n_1}\times \cdots \times S_{n_r}} \sgn( \pi)\, f^\pi  \cr
}\EQN\eqnpartialanti$$
Here, $S_{n_1}\times \cdots \times S_{n_r}$ is viewed as a subgroup of $S_n$, and we view 
$\bigwedge\nolimits_A^{n_1}V \otimes_A \cdots \otimes_A \bigwedge\nolimits_A^{n_r}V $
as a subspace of $A\otimes V^{\otimes n}$. If the context allows, we delete the subscript $A$ in this tensor product.
Elements of the $r$--fold tensor product
$\bigwedge\nolimits_AV \otimes \cdots \otimes \bigwedge\nolimits_AV$ 
are written as Grassmann functions $f(\xi^{(1)},\cdots ,\xi^{(r)})$, with
$\xi^{(\ell)}$ the variable for the $\ell$--th copy of $\bigwedge\nolimits_AV$.

\definition{\STM\defContract}{
Let $C$ be an antisymmetric bilinear form
on $V$ and let $1 \le i,j\le n$. The contraction of the $i^{\rm th}$ variable to the $j^{\rm th}$ variable is the $A$--linear
map
$$
\Cont{i}{j}{C}:\  A\otimes V^{\otimes n} \longrightarrow A\otimes V^{\otimes (n-2)}
$$
characterized by 
$$
\Cont{i}{j}{C}(v_1 \otimes\cdots\otimes v_n) = 
\epsilon_{ij}\ C(v_i,v_j)\  v_1 \otimes\cdots v_{i-1}\otimes v_{i+1} \otimes\cdots\otimes v_{j-1}\otimes v_{j+1}\otimes\cdots\otimes  v_n
$$
for all $v_1,\cdots,v_n \in V$. Here
$$
\epsilon_{ij} = \cases{ (-1)^{j-i+1} & if\  $j>i$ \cr
                        0            & if\  $j=i$ \cr
                        (-1)^{i-j}   & if\  $j<i$ \cr
}$$
}

\remark{\STM\remContractlinearity}{
Let $C_1,C_2$ be antisymmetric bilinear forms
on $V$, $\la_1,\la_2\in\bbbc$, $1 \le i,j\le n$ and 
$f \in A\otimes V^{\otimes n}$. Then 
$$
{\mathop{{\rm\ \, {\cal C}on}}\limits_{\lower2pt\hbox{$\sst\  i\rightarrow j$}}}_{
\raise6pt\hbox{$\sst(\la_1 C_1 +\la_2 C_2)$}}
\,f
= \la_1 \Cont{i}{j}{C_1}\,f + \la_2\Cont{i}{j}{C_2}\,f
$$

}

\remark{\STM\remkerncontract}{
Assume that $A=\bbbc$ and that $\{\xi_i\},\,i\in {\cal I}$ is a basis of $V$. Then every element $f$ of $V^{\otimes\,n}$ can be uniquely written in the form
$$
f = \smsum_{i_1,\cdots,i_n \in {\cal I}} \varphi(i_1,\cdots,i_n)\
\xi_{i_1} \otimes\cdots\otimes \xi_{i_n}
$$
with a function $\varphi$ on ${\cal I}^n$. Then, for $1\le \mu <\nu \le n$
$$
\Cont{\mu}{\nu}{C} f =  \smsum_{j_1,\cdots,j_{n-2} \in {\cal I}} \varphi'(j_1,\cdots,j_{n-2})\
\xi_{j_1} \otimes\cdots\otimes \xi_{j_{n-2}}
$$
with
$$
\varphi'(j_1,\cdots,j_{n-2}) = 
(-1)^{\nu-\mu+1} \smsum_{i,j\in{\cal I}} C_{ij}\
\varphi(j_1,\cdots,j_{\mu-1},i,j_\mu\cdots,j_{\nu-2},j,j_{\nu-1}\cdots,j_{n-2})
$$
}

\lemma{\STM\lemrestrcontract}{
Let $r\ge 2$, $n=n_1+\cdots+n_r$, $1\le k\ne\ell \le r$ and
$$\eqalign{
n_1+\cdots+n_{k-1}+1 &\le \mu,\mu' \le n_1+\cdots+n_k  \cr
n_1+\cdots+n_{\ell-1}+1 &\le \nu,\nu' \le n_1+\cdots+n_\ell  \cr
}$$
Let $C$ be a covariance (antisymmetric bilinear form) on $V$ and 
$ f \in \bigwedge\nolimits_A^{n_1}V \otimes \cdots \otimes \bigwedge\nolimits_A^{n_r}V $.

\Item{i)} 
$\Cont{\mu}{\nu}{C} f$ is partially antisymmetric, precisely
$$
\Cont{\mu}{\nu}{C} f \in \bigwedge\nolimits_A^{n_1}V \otimes \cdots \otimes
\bigwedge\nolimits_A^{n_k-1}V \otimes \cdots \otimes \bigwedge\nolimits_A^{n_\ell-1}V \otimes \cdots \otimes   \bigwedge\nolimits_A^{n_r}V
$$
\Item{ii)} 
$$
\Cont{\mu'}{\nu'}{C} f = \Cont{\mu}{\nu}{C} f
$$
}

\prf We give the proof in the case $r=2$, $k=1$. The general case is analogous.
\Item ii) Clearly, it suffices to show that $\Cont{\mu}{\nu}{C} f=
\Cont{1}{n_1+1}{C} f$ for all $f\in\bigwedge\nolimits_A^{n_1}V\otimes \bigwedge\nolimits_A^{n_2}V$. Let $n=n_1+n_2$ and
$$\eqalign{
f &= \smsum_{i_1,\cdots,i_n \in {\cal I}} \varphi(i_1,\cdots,i_n)\
\xi_{i_1} \otimes\cdots\otimes \xi_{i_n}\cr
\Cont{\mu}{\nu}{C} f& =  \smsum_{j_1,\cdots,j_{n-2} \in {\cal I}} \varphi'(j_1,\cdots,j_{n-2})\
\xi_{j_1} \otimes\cdots\otimes \xi_{j_{n-2}}\cr
\Cont{1}{n_1+1}{C} f &=  \smsum_{j_1,\cdots,j_{n-2} \in {\cal I}} \varphi''(j_1,\cdots,j_{n-2})\
\xi_{j_1} \otimes\cdots\otimes \xi_{j_{n-2}}\cr
}$$
As $f\in\bigwedge\nolimits_A^{n_1}V\otimes \bigwedge\nolimits_A^{n_2}V$,
$\varphi$ is antisymmetric under permutations of its first $n_1$ arguments
and under permutations of its last $n_2$ arguments. Consequently,
$$\eqalign{
\varphi'(j_1,\cdots,j_{n-2}) &= 
(-1)^{\nu-\mu+1} \smsum_{i,j\in{\cal I}} C_{ij}\
\varphi(j_1,\cdots,j_{\mu-1},i,j_\mu\cdots,j_{\nu-2},j
         ,j_{\nu-1}\cdots,j_{n-2})\cr
&= 
(-1)^{\nu-\mu+1}(-1)^{\mu-1}(-1)^{\nu-n_1-1} \smsum_{i,j\in{\cal I}}\! C_{ij}\
\varphi(i,j_1,\cdots,j_{n_1-1},j,j_{n_1},\cdots,j_{n-2})\cr
&= 
(-1)^{n_1+1} \smsum_{i,j\in{\cal I}}\! C_{ij}\
\varphi(i,j_1,\cdots,j_{n_1-1},j,j_{n_1},\cdots,j_{n-2})\cr
&=\varphi''(j_1,\cdots,j_{n-2})\cr
}$$

\Item i) By part ii, we may assume that $\mu=1$ and $\nu=n_1+1$.
By linearity, may assume that 
$$
f  = v_1\cdots v_{n_1} \otimes w_1\cdots w_{n_2}
$$
with $v_1,\cdots,v_{n_1},\,w_1,\cdots,w_{n_2} \in V$. Using 
$$
v_1\cdots v_{n_1}=\sfrac{1}{n_1}\sum_{i=1}^{n_1}(-1)^{i-1}
v_i\otimes v_1\cdots v_{i-1}v_{i+1}\cdots v_{n_1}
$$
and its analog for $w_1\cdots w_{n_2}$, we have
$$
f=\sfrac{1}{n_1n_2}\sum_{i=1}^{n_1}\sum_{j=1}^{n_2}(-1)^{i-1}(-1)^{j-1}
v_i\otimes v_1\cdots v_{i-1}v_{i+1}\cdots v_{n_1} \otimes 
w_j\otimes w_1\cdots w_{j-1}w_{j+1}\cdots w_{n_2}
$$
Since
$$\eqalign{
&\Cont{1}{n_1+1}{C}v_i\otimes v_1\cdots v_{i-1}v_{i+1}\cdots v_{n_1} \otimes 
w_j\otimes w_1\cdots w_{j-1}w_{j+1}\cdots w_{n_2}\cr
&\hskip.5in=(-1)^{n_1+1}C(v_i,w_j)
 v_1\cdots v_{i-1}v_{i+1}\cdots v_{n_1} \otimes 
 w_1\cdots w_{j-1}w_{j+1}\cdots w_{n_2}\cr
}$$
we have
$$
\cont{\mu}{\nu}{C} f 
 =\sfrac{1}{n_1\,n_2} \!\smsum_{i=1}^{n_1} \smsum_{j=1}^{n_2} 
    (-1)^{n_1+j-i+1} C(v_i,w_j)
 v_1\cdots v_{i-1} v_{i+1} \cdots v_{n_1}\,
  \otimes\, w_1\cdots  w_{j-1} w_{j+1} \cdots  w_{n_2}
 \EQN\eqIIcontr$$
This shows that 
$\cont{\mu}{\nu}{C} f \in \bigwedge\nolimits_A^{n_1-1}V \otimes   
\bigwedge\nolimits_A^{n_2-1}V $.

\endproof

\definition{\STM\defGrasscontract}{
Let $C$ be a covariance on $V$, $r\ge 1$ and $1\le k \ne \ell \le r$.
\Item{i)}
Let $n_1,\cdots,n_r \ge 0$. If $n_k,n_\ell \ge 1$ and 
$ f(\xi^{(1)},\cdots ,\xi^{(r)}) \in 
\bigwedge\nolimits_A^{n_1}V \otimes \cdots \otimes \bigwedge\nolimits_A^{n_r}V$, 
the contraction of the $\xi^{(k)}$--fields to the $\xi^{(\ell)}$--fields is defined
as
$$
\cont{\xi^{(k)}}{\xi^{(\ell)}}{C} f\ =\ n_\ell\,\Cont{\mu}{\nu}{C} f 
$$
where $n_1+\cdots+n_{k-1}+1 \le \mu \le n_1+\cdots+n_k$ and
$n_1+\cdots+n_{\ell-1}+1 \le \nu \le n_1+\cdots+n_\ell$. By Lemma \lemrestrcontract, this definition is independent of the choice of $\mu,\nu$. If $n_k=0$ or $n_\ell=0$, we set $\cont{\xi^{(k)}}{\xi^{(\ell)}}{C}=0$.
Observe that $\cont{\xi^{(k)}}{\xi^{(\ell)}}{C}$ maps
$\bigwedge\nolimits_A^{n_1}V \otimes \cdots \otimes\bigwedge\nolimits_A^{n_r}V$
to $\bigwedge\nolimits_A^{n_1}V \otimes \cdots \bigwedge\nolimits_A^{n_k-1}V \otimes \cdots \bigwedge\nolimits_A^{n_\ell-1}V  \otimes\bigwedge\nolimits_A^{n_r}V$.

\Item{ii)} 
The maps
$\cont{\xi^{(k)}}{\xi^{(\ell)}}{C}$ induce an $A$ linear map from the $r$--fold
tensor product  
$$
\bigwedge\nolimits_AV \otimes \cdots \otimes\bigwedge\nolimits_AV =
\bigoplus_{n_1,\cdots,n_r \ge 0}
\bigwedge\nolimits_A^{n_1}V \otimes \cdots \otimes\bigwedge\nolimits_A^{n_r}V
$$ 
to itself, which is also denoted by $\cont{\xi^{(k)}}{\xi^{(\ell)}}{C}$
}

\lemma{\STM\lemGrasscontract}{
Assume that $\{ \xi_i \}$ is a basis of $V$. 
Then, for every Grassmann function 
$f(\xi^{(1)},\cdots ,\xi^{(r)})$ in 
$\bigwedge\nolimits_A^{n_1}V \otimes \cdots \otimes\bigwedge\nolimits_A^{n_r}V$
$$
\cont{\xi^{(k)}}{\xi^{(\ell)}}{C}(f) = - \sfrac{1}{n_k}\sum_{i,j}\, 
\sfrac{\de\ \ }{\de \xi^{(k)}_i} C_{ij} \sfrac{\de\ \ }{\de \xi^{(\ell)}_j}
\big( f \big)
$$
}

\prf
We give the proof in the case $r=2$, $k=1,\ \ell=2$. The general case is similar. By linearity we may assume that
$$
f = \xi_{i_1}^{(1)} \cdots  \xi_{i_{n_1}}^{(1)} \otimes 
 \xi_{j_1}^{(2)} \cdots  \xi_{j_{n_2}}^{(2)}
$$
The claim then follows directly from (\eqIIcontr).
\endproof

To simplify notation when $r=2$, we write 
$f(\xi,\xi')$ instead of $f(\xi^{(1)},\xi^{(2)})$ for elements of 
$\bigwedge\nolimits_A^{n_1}V \otimes \bigwedge\nolimits_A^{n_2}V$ . Similarly, in the case
$r=3$, we write $f(\xi,\xi',\xi'')$ for $f(\xi^{(1)},\xi^{(2)},\xi^{(3)})$.

\example{\STM\egcontract}{
$$\eqalign{
\cont{\xi}{\xi'}{C} \big(\xi_k\xi'_\ell\big)&=C_{k\ell}\cr
\cont{\xi}{\xi'}{C} \big(\xi_k\xi'_\ell\xi'_m\big)
&=C_{k\ell}\xi'_m-C_{km}\xi'_\ell\cr
\cont{\xi}{\xi'}{C} \big(\xi_j\xi_k\xi'_\ell\big)
&=\half\big[\xi_jC_{k\ell}-\xi_kC_{j\ell}\big]\cr
}$$

} 

\remark{\STM\remcontract}{
Since taking partial derivatives commutes with Wick ordering, for any $f(\xi,\xi') \in \bigwedge\nolimits_A (V\oplus V')$
$$
\cont{\xi}{\xi'}{C} \big( \lw f(\xi,\xi')\rw_{\xi'} \big)
 =\ \lw\!\!\cont{\xi}{\xi'}{C} \big( f(\xi,\xi') \big) \rw_{\xi'} 
$$
}

The main reason for introducing the contraction operator is the following
``integration by parts formula'':

\lemma{\STM\lemcontract}{ Let $f(\xi,\xi',\xi^{\prime\prime})$ be a 
Grassmann function of degree at least one in $\xi'$. Set
$$\eqalign{
\tilde f(\xi,\xi',\xi^{\prime\prime}) 
\ & =\  \cont{\xi'}{\xi''}{C}\,
   f(\xi,\xi',\xi^{\prime\prime}) \cr
\tilde g(\xi,\xi^{\prime\prime})\ &= 
\ \lw \tilde f(\xi,\xi,\xi^{\prime\prime})\rw_\xi  \cr
 g(\xi,\xi^{\prime\prime})\ &=\  \lw f(\xi,\xi,\xi^{\prime\prime})\rw_\xi \cr
}$$
 Then
$$
\int \tilde g(\xi,\xi)\,d\mu_C(\xi)  \ =\ 
\int g(\xi,\xi)\,d\mu_C(\xi)
$$
 That is,
$$
\int \Big[\lw f(\xi,\xi,\xi'')\rw_\xi\Big]_{\xi''=\xi}\,d\mu_C(\xi)  \ =\ 
\int \Big[\lW\big[ \cont{\xi'}{\xi''}{C}f(\xi,\xi',\xi'')\big]_{\xi'=\xi}
\rW_\xi\Big]_{\xi''=\xi}\,d\mu_C(\xi)
$$
}  

\prf It suffices to prove the statement in the case that
$$
f(\xi,\xi',\xi^{\prime\prime}) 
=\xi_{i_1}\cdots \xi_{i_{n-r}}\ \xi'_{i_{n-r+1}}\cdots \xi'_{i_n}\,
\lw \xi^{\prime\prime}_{j_m}\xi^{\prime\prime}_{j_{m-1}}\cdots \xi^{\prime\prime}_{j_1}
\rw_{\xi^{\prime\prime}}
$$
with $1\le r\le n,\ m\ge 1$. Then
$$\eqalign{
\tilde f(\xi,\xi',\xi^{\prime\prime})
& =-\sfrac{1}{r}\,\xi_{i_1}\cdots \xi_{i_{n-r}}\,
  \sum_{k=n-r+1,\cdots n \atop \ell =1,\cdots m} (-1)^{(n+m+\ell)+(k-1)}C_{i_kj_\ell}\cr
& \hskip 4cm
    \xi'_{i_{n-r+1}}\cdots \xi'_{i_{k-1}}\xi'_{i_{k+1}}\cdots \xi'_{i_n}\
\lw \xi^{\prime\prime}_{j_m}\cdots \xi^{\prime\prime}_{j_{\ell+1}}   \xi^{\prime\prime}_{j_{\ell-1}}\cdots \xi^{\prime\prime}_{j_1}\rw  \cr
}$$
and
$$\eqalign{
\int \tilde g(\xi,\xi)\,d\mu_C(\xi)
&= \sfrac{1}{r}\, \sum_{k=n-r+1,\cdots n}\  \sum_{\ell =1,\cdots m}
 (-1)^{n+m+k+\ell}C_{i_kj_\ell} \cr
& \hskip 1.5cm
  \int (\lw \xi_{i_1}\cdots \xi_{i_{k-1}}\xi_{i_{k+1}}\cdots \xi_{i_n}\rw)
(\lw \xi_{j_m}\cdots \xi_{j_{\ell+1}} \xi_{j_{\ell-1}}\cdots \xi_{j_1}\rw)
\,d\mu_C(\xi)
}$$
If $m\ne n$, both $\int \tilde g(\xi,\xi)\,d\mu_C(\xi)$ and 
$\int g(\xi,\xi)\,d\mu_C(\xi)$ are zero by \:\lemBI. In the case $m=n$, again by Lemma \:\lemBI
$$\eqalign{
\int \tilde g(\xi,\xi)\,d\mu_C(\xi)
&= \sfrac{1}{r}\, \smsum_{k=n-r+1,\cdots n} \smsum_{\ell =1,\cdots m}
 (-1)^{k+\ell}C_{i_kj_\ell} \,
\det\big( C_{i_{k'}j_{\ell'}} \big)_{k'\ne k \atop \ell'\ne \ell} \cr
&= \sfrac{1}{r} \smsum_{k=n-r+1,\cdots n} \det\big( C_{i_{k'}j_{\ell'}} \big) \cr
&=  \det\big( C_{i_{k}j_{\ell}} \big) \cr
&=  \int (\lw \xi_{i_1}\cdots \xi_{i_r}\xi_{i_{r+1}}\cdots \xi_{i_n}\rw) \, 
(\lw \xi_{j_m}\xi_{j_{m-1}}\cdots \xi_{j_1}\rw)\,d\mu_C(\xi) \cr
&=  \int  g(\xi,\xi)\,d\mu_C(\xi)}$$
\endproof

\sect{ Seminorms}\PG\pgRIId

We will formulate an estimate on the renormalization group map in terms of norms
for which the contraction maps and Grassmann Gaussian integrals can be controlled. In this subsection we assume that $A$ is a graded superalgebra.

In practice we shall use families of norms on $\bigwedge_A V$ 
that encode information concerning various derivatives of the coefficient functions. To unify such families in a way that incorporates Leibniz's rule
we give  
\definition{\STM\defFancynormdomain}{
\Item{i)}
On $\bbbr_+\cup\{\infty\} = \set{x\in\bbbr}{x\ge0}\cup\{+\infty\}$, addition
 and the total ordering $\le$ are defined in the standard way. With the convention that $0\cdot\infty=\infty$, multiplication is also defined in the standard way.
\Item{ii)}
Let $d\ge 0$. The $d$--dimensional norm domain $\fN_d$ is the set of all formal power series
$$
X = \sum_{\de=(\de_1,\cdots,\de_d)\in\bbbn_0^d} X_\de \
t_1^{\de_1}\cdots t_d^{\de_d}
$$
in the variables $t_1,\cdots,t_d$ with coefficients $X_\de \in \bbbr_+\cup\{\infty\}$. Addition and partial ordering on $\fN_d$ are defined componentwise: If
$$
X = \smsum_{\de\in\bbbn_0^d} X_\de\ t_1^{\de_1}\cdots t_d^{\de_d}
\qquad,\qquad 
X'= \smsum_{\de\in\bbbn_0^d} X'_\de\ t_1^{\de_1}\cdots t_d^{\de_d}
$$
then
$$\eqalign{
X+X'\ \  &= \ \ \smsum_{\de} (X_\de+X'_\de)\ t_1^{\de_1}\cdots t_d^{\de_d} \cr
X \le X' &\iff X_\de \le X'_\de \ \ {\rm for\ all\ } \de \in \bbbn_0^d
}$$
Multiplication is defined by
$$
(X\cdot X')_\de = \smsum_{\be+\ga=\de} X_\be X'_\ga
$$
We identify $\bbbr_+\cup\{\infty\}$ with the set of all $X\in \fN_d$ with
$X_\de=0$ for all $\de\ne \0=(0,\cdots,0)$.

\noindent
If $a>0$, $X_\0\ne\infty$ and $a-X_\0 >0$ then $(a-X)^{-1}$ is defined as
$$
(a-X)^{-1} = \sfrac{1}{a-X_\0} 
\smsum_{n=0}^\infty \big(\sfrac{X-X_\0}{a-X_\0}\big)^n
$$
}

\definition{\STM\defFancynorm}{
Let $E$ be a complex vector space. A ($d$--dimensional) seminorm on $E$ is a 
map $\ \|\cdot\|:\,E\rightarrow \fN_d\ $ such that
$$
\|e_1+e_2\| \le \|e_1\|+\|e_2\| \qquad,\qquad \|\la\,e\| = |\la|\,\|e\|
$$
for all $e,e_1,e_2\in E$ and $\la \in \bbbc$.
}

\example{\STM\exFancynorm}{ Let be the space of all functions $f:\bbbr^d\rightarrow\bbbc$. Define
$$
\| f\|=\sum_{\de\in\bbbn_0^d} \ 
\sup_{x\in\bbbr^d}\big|\partial^\de f(x)\big|\ \ 
t_1^{\de_1}\cdots t_d^{\de_d}
$$
where $\sup_{x\in\bbbr^d}\big|\partial^\de f(x)\big|=\infty$ if
$\partial^\de f(x)$ is not everywhere defined.

}

\remark{\STM\remFancynorm}{ 
\Item i) By convention, $\fN_0=\bbbr_+\cup\{\infty\}$.
\Item ii)
If $E$ is a complex vector space and  $\ \|\cdot\|$ is a 
$0$--dimensional seminorm on $E$ that obeys $\|e\|<\infty$ for all   
$e\in E$, then $\ \|\cdot\|$ is a seminorm on $E$ in the conventional sense.

}

\definition{\STM\defsymnorm}{
Let $m,n\ge 0$.
A seminorm
$\|\,\cdot\,\|$ on the space $A_m\otimes V^{\otimes n}$ is called symmetric, if for all $f \in A_m\otimes V^{\otimes n}$ and all permutations
$\pi \in S_n$
$$
\|f^{\pi}\| = \|f\|
$$
and $\|f\| =0$ if $m=n=0$.
}

\remark{\STM\remkernsymnorm}{
Assume, as in Remark \remkerncontract, that $A=\bbbc$ and that $\{\xi_i\},\,i\in {\cal I}$ is a basis of $V$. Every element $f$ of $V^{\otimes\,n}$ can be uniquely written in the form
$$
f = \smsum_{i_1,\cdots,i_n \in {\cal I}} \varphi(i_1,\cdots,i_n)\
\xi_{i_1} \otimes\cdots\otimes \xi_{i_n}
$$
with a function $\varphi$ on ${\cal I}^n$. Therefore, a symmetric family of 
seminorms corresponds to a family of seminorms $\|\cdot\|$ on the spaces of
functions $\varphi$ on ${\cal I}^n$ such that
$$
\|\varphi(i_1,\cdots,i_n)\| = \|\varphi(i_{\pi(1)},\cdots,i_{\pi(n)})\|
\qquad {\rm for\ all\ } \pi\in S_n
$$
}

\example{\STM\egcompatnorm}{ Let $A=\bbbc$ and 
$V$ be a finite dimensional vector space with basis $\xi_1,\cdots,\xi_D$.
For a function $\varphi$ on $\{1,\cdots,D\}^n$ define the $L_1$--$L_\infty$--norm 
$$
\|\varphi\|_{1,\infty} = \max_{1\le k \le n}\ \sup_{1\le i_k\le D}\ 
\sum_{i_1,\cdots,i_{k-1},i_{k+1},\cdots,,i_n =1}^D  |\varphi(i_1,\cdots,i_n)|
$$
This defines a family of symmetric (0--dimensional) seminorms on the spaces $V^{\otimes n}$. 
}

\definition{\STM\defAmnonetonr}{
By $A_m[n_1,\cdots,n_r]$
we denote the image of $A_m\otimes V^{\otimes (n_1+\cdots+n_r)}$ under the partial antisymmetrization map ${\rm Ant}_{n_1,\cdots,n_r}$
defined in (\eqnpartialanti). It is the
$\bbbc$--linear subspace of 
$$
A[n_1,\cdots,n_r] =
   \bigwedge\nolimits_A^{n_1} V^{(1)} \otimes \cdots \otimes \bigwedge\nolimits_A^{n_r} V^{(r)}  
$$
generated by elements of the form $a_m \,p_1(\xi^{(1)})\cdots p_r(\xi^{(r)})$ with $a_m\in A_m$
and $p_\nu(\xi^{(\nu)}) \in \bigwedge^{n_\nu} V^{(\nu)}$. Elements 
$f(\xi^{(1)},\cdots ,\xi^{(r)})$ of
$A_m[n_1,\cdots,n_r]$ are called homogeneous, and $n_\nu$ is their degree of homogeneity in the variable $\xi_\nu$. Observe that $A_m[n_1,\cdots,n_r]$
is a subspace of $A_m\otimes V^{\otimes (n_1+\cdots+n_r)}$.
}

\lemma{\STM\lemGrasscompatnorm}{ Let $\|\,\cdot\,\|$ be
a family of symmetric seminorms on the spaces $A_m\otimes V^{\otimes n}$. Then for all 
$f(\xi^{(1)},\cdots ,\xi^{(r)})\in A_m[n_1,\cdots,n_r]$
\Item{(i)} for all permutations $\pi\in S_r$  
$$
\|f(\xi^{(1)},\cdots ,\xi^{(r)})\| = 
\|f(\xi^{(\pi(1))},\cdots ,\xi^{(\pi(r))})\|
$$
\Item{(ii)}
$$
\|f(\xi^{(1)},\cdots,\xi^{(r-2)},\xi^{(r-1)},\xi^{(r-1)})\| \le 
\|f(\xi^{(1)},\cdots,\xi^{(r-2)},\xi^{(r-1)},\xi^{(r)})\|
$$
Here, $f(\xi^{(1)},\cdots,\xi^{(r-2)},\xi^{(r-1)},\xi^{(r-1)})$ is an element of $A_m[n_1,\cdots,n_{r-2},n_{r-1}+n_r]$.
\Item{(iii)}
if $\epsilon \in \bbbc$ with $|\epsilon|=1$ and
$$
f(\xi^{(1)},\cdots,\xi^{(r-1)},\xi^{(r)}+ \epsilon \xi^{(r+1)})
= \sum_{k=0}^{n_r}  f_k(\xi^{(1)},\cdots,\xi^{(r-1)},\xi^{(r)},\xi^{(r+1)})
$$
with $\,f_k\in A_m[n_1,\cdots,n_{r-1},n_r-k,k]$
then
$$
\|f_k\| \le  \smchoose{n_r}{ k} \,\|f\|
$$
\Item{(iv)}
$\ 
\|f\|=0
\ $ if $f\in A_0[0,0,\cdots,0]$
}

\prf
Parts (i) and (iv) are trivial. To prove part (ii) set
$$\eqalign{
f'(\xi^{(1)},\cdots,\xi^{(r-2)},\xi^{(r-1)}) 
&= f(\xi^{(1)},\cdots,\xi^{(r-2)},\xi^{(r-1)},\xi^{(r-1)}) \cr
&= {\rm Ant}_{n_1,\cdots,n_{r-2},n_{r-1}+n_r} f \cr
}$$
Then, by Definition \defsymnorm
$$
\|f'\| \ \le\ \sfrac{1}{n_1! \cdots n_{r-2}!\,(n_{r-1}+n_r)!}
\sum_{\pi \in S_{n_1} \times \cdots\times S_{n_{r-2}} \times S_{n_{r-1}+n_r}}
\|f^\pi\| 
\ \le\ \|f\|
$$
We now prove part (iii). For any subset $I$ of 
$\{n_1+\cdots +n_{r-1} +1,\cdots ,n_1+\cdots +n_r\}$ let $\pi_I^{-1}$ be the permutation that brings the sequence $I,\,\{n_1+\cdots +n_{r-1} +1,\cdots ,n_1+\cdots +n_r\} \setminus I$ 
into standard order. Then
$$
f_k = \sum_{I \subset \{n_1+\cdots +n_{r-1} +1,\cdots ,n_1+\cdots +n_r\}
\atop |I| =k} 
\epsilon^{n_r-k}\,\sgn(\pi_I)\ f^{\pi_I}
$$
Again, by Definition \defsymnorm
$$
\|f_k\|\ \le\ \sum_{|I| =k} \|f\| \ =\ \smchoose{n_r}{ k}\, \|f\|
$$
\endproof

\definition{\STM\deffunctnorm}{Let $\|\,\cdot\,\|$ be a family of symmetric d--dimensional seminorms and let
$f(\xi^{(1)},\cdots ,\xi^{(r)})\in \bigwedge\nolimits_A V\otimes\cdots\otimes
\bigwedge\nolimits_A V \cong
\bigwedge\nolimits_A (V\oplus\cdots\oplus V)$. Write
$$
f = \sum_{m, n_1,\cdots, n_r\ge 0} 
f_{m; n_1,\cdots, n_r}
$$ 
with $f_{m; n_1,\cdots, n_r} \in A_m[n_1,\cdots,n_r]$. For any real 
$ \al \ge 1$, $\ib>0$ and $\cb\in\fN_d$ set
$$
N(f\cl \al) = 
\sfrac{1}{\ib^2}\,\cb\!\sum_{m,n_1,\cdots, n_r\ge 0}\,
\al^{|n|}\,\ib^{|n|} \,\|f_{m; n_1,\cdots, n_r}\|
$$
We omit the dependence on $\ib$ and $\cb$ from the symbol $N(f\cl \al)$.
If the context allows, we will also delete the reference to $\al$.
}

In the applications we have in mind (see [FKTo1, Theorem \thmOSinsulators],
[FKTo2, Theorem \thmOSfirststep], [FKTo3, Lemma \lemOSconcreteintconst],
[FKTf1, subsection 7 of \S II]), $\cb$ is a 
bound for various weighted $L^1$--norms of the propagator $C$ (in position space), while $\ib^2$ is a bound for its $L^\infty$--norm. $\sfrac{1}{\al}$ is a (possibly)
fractional power of the coupling constant.

\remark{\STM\remfunctnorm}{Let
$f(\xi^{(1)},\cdots ,\xi^{(r)})\in \bigwedge\nolimits_A V\otimes\cdots\otimes
\bigwedge\nolimits_A V \cong
\bigwedge\nolimits_A (V\oplus\cdots\oplus V)$. Then
\Item{(i)}
for $1\le s\le r$
$$
N\big(f(\xi^{(1)}+\xi^{(r+1)},\cdots ,\xi^{(s)}+\xi^{(r+s)},
\xi^{(s+1)},\cdots,\xi^{(r)})\cl \al\big)
\le \ N\big(f(\xi^{(1)},\cdots ,\xi^{(r)})\cl 2\al\big)
$$
\Item{(ii)} for all $a\ge 1$
$$\eqalign{
N(f\cl \al) &\le\ N(f\cl a\al)\cr
}$$

}

\prf 
\Item{(i)} Since
$$
f(\xi^{(1)}+\xi^{(r+1)},\cdots ,\xi^{(s)}+\xi^{(r+s)},\cdots,\xi^{(r)})
=f(\xi^{(1)}+\xi^{(r+1)},\cdots ,\xi^{(r)}+\xi^{(2r)})
 \Big|_{\xi^{(r+1)}=\cdots=\xi^{(2r)}=0}
$$
it suffices to prove the claim in the case $s=r$.
Write
$$
f = \sum_{m,n_1,\cdots, n_r\ge 0} 
f_{m; n_1,\cdots, n_r}
$$ 
with $f_{m; n_1,\cdots, n_r} \in A_m[n_1,\cdots,n_r]$. By part (iii) of 
Lemma \lemGrasscompatnorm
$$\eqalign{
f_{m; n_1,\cdots, n_r}&(\xi^{(1)}+\xi^{(r+1)},\cdots,\xi^{(r)}+\xi^{(2r)}) \cr
&= \sum_{k_i=0,\cdots,n_i \atop {\rm for\ } i=1,\cdots,r}  
f_{m;\, n_1-k_1,\cdots,n_r-k_r,k_1,\cdots,k_r}
(\xi^{(1)},\cdots,\xi^{(r)},\xi^{(r+1)},\cdots,\xi^{(2r)})
}$$
with $\,f_{m;\, n_1-k_1,\cdots,n_r-k_r,k_1,\cdots,k_r}
\in A_m[n_1-k_1,\cdots,n_r-k_r,k_1,\cdots,k_r]$
and
$$
\big\|f_{m;\, n_1-k_1,\cdots,n_r-k_r,k_1,\cdots,k_r}\big\| 
\le  \,\|f_{m; n_1,\cdots, n_r}\|\,\smprod_{i=1}^r \smchoose{n_i}{k_i}
$$
Consequently
$$\eqalign{
N\big(f(\xi^{(1)}+\xi^{(r+1)},\cdots &,\xi^{(r)}+\xi^{(2r)})\big) \cr
 & = \sfrac{1}{\ib^2}\,\cb\sum_{m,n_1,\cdots, n_r\ge 0}
\sum_{k_i=0,\cdots,n_i \atop {\rm for\ } i=1,\cdots,r} 
\al^{|n|}\,\ib^{|n|}\,
\big\|f_{m;\, n_1-k_1,\cdots,n_r-k_r,k_1,\cdots,k_r}\big\|  \cr
& \le \sfrac{1}{\ib^2}\,\cb\sum_{m;\,  n_1,\cdots n_r}\,
\al^{|n|}\,\ib^{|n|}\,
\|f_{m; n_1,\cdots, n_r}\|\,
\sum_{k_i=0,\cdots,n_i \atop {\rm for\ } i=1,\cdots,r} 
\smprod_{i=1}^r\smchoose{n_i}{k_i}
\cr
& = \sfrac{1}{\ib^2}\,\cb\sum_{m;\,  n_1,\cdots n_r}\,
\al^{|n|}\,\ib^{|n|}\,
2^{n_1+\cdots+n_r} \,\|f_{m; n_1,\cdots, n_r}\| \cr
&= \ N\big(f(\xi^{(1)},\cdots ,\xi^{(r)})\cl 2\al\big) \cr
}$$
\Item{(ii)} is trivial
\endproof

\sect{ An estimate of the renormalization group map}\PG\pgRIIe

Let $A$ be a graded superalgebra, $\|\,\cdot\,\|$ be a family of 
symmetric $d$--dimensional seminorms on the spaces $A_m\otimes V^{\otimes n}$
 and $C$ be a covariance on $V$.

\definition{\STM\defcontractintbound}{
\Item{(i)} 
We say that $\cb\in\fN_d$, with $\cb_\0\ne 0,\infty$,
 is a contraction bound for $C$ with respect to these seminorms if for all $f\in A_m\otimes V^{\otimes n}$, 
$f'\in A_{m'}\otimes V^{\otimes n'}$ and all $1\le i \le n,\ 1\le j \le n'$
$$
\| \Cont{i}{n+j}{C} (f\otimes f') \| \le \cb\, \|f\|\,\|f'\|
$$
Observe that 
$\Cont{i}{n+j}{C} (f\otimes f') \in A_{m+m'}\otimes V^{\otimes (n+n'-2)}$.

\Item{ii)}
For $n'\le n$ define the partial antisymmetrization
$$
Ant_{n'} = {\rm Ant}_{n',1,1,\cdots,1}
$$
It is characterized by
$$
Ant_{n'}(v_1\otimes\cdots\otimes v_{n'}\,\otimes\,
w_1\otimes\cdots\otimes w_{n-n'})
\ =\  
v_1 \cdots v_{n'} \, \otimes\, w_1\otimes \cdots \otimes w_{n-n'}
\ \in\ \bigwedge\nolimits_A^{n'}V \otimes_A  V^{\otimes (n-n')}
$$ 
for all $ v_1,\cdots ,v_{n'},w_1\cdots,w_{n-n'}\in V$. 
We say that the real number $\ib\ge 0$ is a (Grassmann) integral bound for $C$ 
with respect to the family of seminorms if for every $f\in A_m\otimes V^{\otimes n}$ and every $n'\le n$
$$
\Big\| \int Ant_{n'}(f) d\mu_C \Big\| \le (\ib/2)^{n'} \|f\|
$$
Here, the Grassmann Gaussian integral maps 
$\bigwedge\nolimits_A^{n'}V\otimes_A V^{\otimes (n-n')} $ to 
$A \otimes V^{\otimes(n-n')}$. 
}

\example{\STM\egIIcompatnorm}{ In the example \egcompatnorm, a contraction bound
for a covariance $C=\big(C_{ij}\big)_{1\le i,j\le D}$ with respect to the
$L_1$--$L_\infty$--norm is given by
$$
\cb=\|C\|_{1,\infty} =\max_{1\le i\le D}\sum_{1\le j\le D}\big|C_{ij}\big|
$$
Let $\ib>0$ be such that 
$$
\Big| \int \xi_{i_1} \cdots \xi_{i_n}\,    
d\mu_C(\xi)  \Big| 
\le (\ib/2)^{n}
\EQN\eqnIIcompatnorm$$
for all\ $n\ge 0$ and all 
$i_1,\cdots,i_n$.
Then $\ib$ is an integral bound for the covariance $C$.
In Appendix B we give criteria under which (\eqnIIcompatnorm) is fulfilled. 
}

When ${\rm dim}\,V=\infty$, it is not a priori clear whether or not the renormalization group map
$$
\Om_C(W)(\psi) = \log \sfrac{1}{Z}\int e^{W(\psi +\xi)} d\mu_C(\xi) 
$$
of Definition \defrengroup\ makes sense. However, in the case of interest, we
can define it as a formal power series in $W$.  The Taylor expansion of 
$\int e^{W(\psi +\xi)} d\mu_C(\xi) $
is $\sum_{n=1}^\infty\cG_n(W,\cdots,W)$ where the 
$n^{\rm th}$ term is the $n$--linear map
$$
\cG_n(W_1,\cdots,W_n)= \sfrac{1}{n!}
\int  W_1(\psi +\xi)\cdots W_n(\psi +\xi)\ d\mu_C(\xi)
$$
from 
$\bigwedge_A V'\times\cdots\times \bigwedge_A V'$ to $\bigwedge_A V'$, 
restricted to the diagonal. Explicit evaluation of the Grassmann 
integral expresses $\cG_n$ as the sum of all graphs with vertices 
$W_1,\ \cdots,\ W_n$ and lines $C$. The (formal) Taylor coefficient 
$\sfrac{d\hfill}{dt_1}\cdots\sfrac{d\hfill}{dt_n}
\Om_C(t_1 W_1+\cdots +t_n W_n)\Big|_{t_1=\cdots=t_n=0}$ 
of $\Om_C(W)$ is similar, but with only connected graphs.

\definition{\STM\defrengroupInfDim}{ 
Let $\|\,\cdot\,\|$ be a family of symmetric seminorms and let
$C$ be a covariance on $V$ with a finite integral
bound and a contraction bound $\cb$ obeying  $\cb_0<\infty$.
Then 
$$
\sfrac{d\hfill}{dt_1}\cdots\sfrac{d\hfill}{dt_n}\Om_C(t_1W_1+\cdots +t_n W_n)
\Big|_{t_1=\cdots=t_n=0}
$$
interpreted as a sum of graphs, as above, is a bounded $n$--linear map .
Then $\Om_C$ is defined to be the formal Taylor series associated to the sequence of these multilinear maps.

}

\theorem{\STM\theorII}{ Let $\|\,\cdot\,\|$ be a family of symmetric 
seminorms and let $C$ be a covariance on $V$ with contraction bound 
$\cb$ and integral bound $\ib$. Then the formal Taylor series $\Om_C(\lw W\rw)$
converges to an analytic map\footnote{$^{(1)}$}{For an elementary discussion of analytic maps between Banach spaces see, for example, Appendix A of [PT].} on $\set{W\in\bigwedge_A V'}{W\hbox{ even},\  
N\big(W\cl 8\al \big)_\0 <\sfrac{\al^2}{4}}$.
Furthermore, if $W(\psi) \in \bigwedge_A V'$ is an even Grassmann function 
such that
$$
N\big(W\cl 8\al \big)_\0 <\sfrac{\al^2}{4} 
$$
then
$$
N\big(\Om_C(\lw W\rw)-W\cl \al\big)\ \le\ 
\sfrac{2}{\al^2}\,
     \sfrac{N( W\cl 8\al)^2}{1-{4\over\al^2}N(W\cl 8\al)}
$$
}
\noindent
This Theorem is proven in the \S III.3.

\sect{ Contraction and Integral Bounds}\PG\pgRIIf
In this subsection we investigate properties of contraction and 
integral bounds as introduced in Definition \defcontractintbound. These 
properties will be used in \S III to prove Theorem \theorII.

\lemma{\STM\lemGrasscontractnorm}{ Assume that $\cb$ is a
contraction bound and $\ib$ an integral bound for $C$.
\Item{(i)} If $n_r, n_{r+1}'\ge 1$, then for 
$$\eqalign{
f_1(\xi^{(1)},\cdots,\xi^{(r-1)},\xi^{(r)})&\in A_m[n_1,\cdots,n_{r-1},n_r,0]\cr
f_2(\xi^{(1)},\cdots,\xi^{(r-1)},\xi^{(r+1)})&\in 
A_{m'}[n_1',\cdots n_{r-1}',0,n_{r+1}']\cr
}$$
one has
$$\eqalign{
\big\| \cont{\xi^{(r)}}{\xi^{(r+1)}}{C}&\,
f_1(\xi^{(1)},\cdots,\xi^{(r-1)},\xi^{(r)})\,
f_2(\xi^{(1)},\cdots,\xi^{(r-1)},\xi^{(r+1)}) \big\| \cr
&\le n_{r+1}' \cb\,\|f_1(\xi^{(1)},\cdots,\xi^{(r-1)},\xi^{(r)})\|\,
\|f_2(\xi^{(1)},\cdots,\xi^{(r-1)},\xi^{(r+1)}) \| \cr
}$$
Observe that the contraction $\cont{\xi^{(r)}}{\xi^{(r+1)}}{C}\,
f_1(\xi^{(1)},\cdots,\xi^{(r-1)},\xi^{(r)})\,
f_2(\xi^{(1)},\cdots,\xi^{(r-1)},\xi^{(r+1)})$ lies in 
$A_{m+m'}[n_1+n_1',\cdots,n_{r-1}+n_{r-1}',n_r-1,n_{r+1}'-1]$.
\Item{(ii)} 
Let  $0\le s\le t\le r$ and let
$$ 
f(\xi^{(1)},\cdots,\xi^{(s)},\xi^{(s+1)},\cdots,\xi^{(t)},\xi^{(t+1)},
\cdots,\xi^{(r)})\in A_m[n_1,\cdots, n_s,n_{s+1},\cdots, n_t,n_{t+1},\cdots,n_r]
$$
Set
$$
\tilde f(\xi^{(1)},\cdots,\xi^{(s)},\xi^{(s+1)},\cdots,\xi^{(t)},
\cdots,\xi^{(r)})
=\lw  f(\xi^{(1)},\cdots,\xi^{(s)},\xi^{(s+1)},\cdots,\xi^{(t)},
\cdots,\xi^{(r)}) \rw_{\xi^{(1)},\cdots,\xi^{(s)}} 
$$
where $\lw \,\cdot\,\rw_{\xi^{(1)},\cdots,\xi^{(s)}} $ denotes Wick ordering, 
with respect to the covariance $C$, in the variables $\xi^{(1)},\cdots,\xi^{(s)}$ separately.
Then 
$$
\Big\| \int \tilde f(\xi,\cdots,\xi,\xi^{(t+1)}\cdots,\xi^{(r)}) 
\,d\mu_C(\xi)\,
\Big\| 
\le \ib^{n_1+\cdots+n_t}\,
\| f(\xi^{(1)},\cdots,\xi^{(r)})\|
$$
Observe that
$\int \tilde f(\xi,\cdots,\xi,\xi^{(t+1)}\cdots,\xi^{(r)}) \,d\mu_C(\xi)\,
\in A_m[0,\cdots, 0,n_{t+1},\cdots,n_r]$. 
}

\prf
\Item{i)}
By definition
$$\eqalign{
\cont{\xi^{(r)}}{\xi^{(r+1)}}{C}&\,
f_1(\xi^{(1)},\cdots,\xi^{(r-1)},\xi^{(r)})\,
f_2(\xi^{(1)},\cdots,\xi^{(r-1)},\xi^{(r+1)}) \cr
& \hskip 2cm= \pm\,n'_{r+1}\, {\rm Ant}_{n_1+n'_1,\cdots,n_{r-1}+n'_{r-1},n_r-1,n_{r+1}-1}
\big(\Cont{\mu}{\nu}{C} f_1 \otimes f_2\big)^\pi
}$$
where $1+\sum_{i=1}^{r-1}n_i \le \mu \le \sum_{i=1}^{r}n_i$, 
$1+\sum_{i=1}^{r}n_i+\sum_{i=1}^{r-1}n'_i \le \nu \le n'_{r+1}+\sum_{i=1}^{r}n_i+\sum_{i=1}^{r-1}n'_i$ and $\pi$ is the
permutation that maps the sequence 
$$
1,\cdots,n_1,\cdots,n_1+\cdots+n_r-1,\cdots ,n_1+\cdots +n_r + n'_1+ \cdots
+n'_{r+1} -2
$$ 
to the sequence 
$$
1,\cdots,n_1,n_1+\cdots +n_r-1 +1,\cdots,n_1+\cdots +n_r-1+n'_1, n_1+1,\cdots
$$
The desired estimate now follows from Definitions \defcontractintbound,
\defsymnorm\ and \defGrasscontract.

\Item{(ii)} Set $g(\xi,\xi^{(t+1)}\cdots,\xi^{(r)})=\tilde f(\xi,\cdots,\xi,\xi^{(t+1)}\cdots,\xi^{(r)})$.
 In the case $s=0$ one has $\tilde f=f$ and hence $\|g\| \le \|\tilde f\|
\le \|f\| $ 
by part (ii) of Lemma \lemGrasscompatnorm. Therefore, by Definition \defcontractintbound,
$$
\Big\| \int \tilde f(\xi,\cdots,\xi,\xi^{(t+1)}\cdots,\xi^{(r)}) 
\,d\mu_C(\xi)\,\Big\| 
\le (\ib/2)^{n_1+\cdots+n_t}\,\| g\|
\le (\ib/2)^{n_1+\cdots+n_t}\,\| f\|
$$
In the general case, we have by part (i) of Proposition \:\propBII\ and part (iii) of Lemma
\lemGrasscompatnorm
$$\eqalign{
\tilde f &= \int  f(\xi^{(1)}+\imath\ze^{(1)},\cdots,\xi^{(s)}+\imath\ze^{(s)},
\xi^{(s+1)},\cdots,\xi^{(t)},\cdots,\xi^{(r)})\,
d\mu_C(\ze^{(1)},\cdots,\ze^{(s)}) \cr
&= \sum_{k_i = 1,\cdots,n_i \atop i=1,\cdots,s} 
 \int  f_{k_1,\cdots,k_s}(\xi^{(1)},\ze^{(1)},\cdots,\xi^{(s)},\ze^{(s)},
\xi^{(s+1)},\cdots,\xi^{(t)},\cdots,\xi^{(r)})\,
d\mu_C(\ze^{(1)},\cdots,\ze^{(s)}) \cr
}$$
with
$$
f_{k_1,\cdots,k_s} \in A_m[n_1-k_1,k_1,\cdots,n_s-k_s,k_s,\,n_{s+1}, \cdots,n_r]
$$
fulfilling
$$
\| f_{k_1,\cdots,k_s}\| \le  \|f\|\,\smprod_{i=1}^s \smchoose{n_i}{k_i}
$$
By the special case discussed above
$$\eqalign{
\Big\| &\int   f_{k_1,\cdots,k_s}(\xi^{(1)},\ze^{(1)},\cdots,\xi^{(s)},\ze^{(s)},
\xi^{(s+1)},\cdots,\xi^{(t)},\xi^{(t+1)},\cdots,\xi^{(r)})\,
d\mu_C(\ze^{(1)},\cdots,\ze^{(s)}) \Big\| \cr
& \hskip 9cm \le (\ib/2)^{k_1+\cdots+k_s}  \|f_{k_1,\cdots,k_s}\|
}$$
Therefore, again by the special case discussed above,
$$\eqalign{
\Big\| &\int\Big[ \int f_{k_1,\cdots,k_s}(\xi,\ze^{(1)},\cdots,\xi,\ze^{(s)},
\xi,\cdots,\xi,\xi^{(t+1)},\cdots,\xi^{(r)})\,
d\mu_C(\ze^{(1)},\cdots,\ze^{(s)})\Big] \,d\mu_C(\xi)\Big\| \cr
& \hskip 9cm \le (\ib/2)^{n_1+\cdots+n_t}  \|f_{k_1,\cdots,k_s}\|
}$$
Consequently
$$\eqalign{
\Big\| \int \tilde f(\xi,\cdots,\xi,\xi^{(t+1)}\cdots,\xi^{(r)}) 
\,d\mu_C(\xi)\,\Big\| 
&\le \ib^{n_1+\cdots+n_t} \| f\|\,
\sum_{k_i = 1,\cdots,n_i \atop i=1,\cdots,s}
\smprod_{i=1}^s \sfrac{1}{2^{n_i}}\smchoose{n_i}{k_i} \cr
&\le \ib^{n_1+\cdots+n_t} \| f\|
}$$
\endproof

\remark{\STM\scaleIntConsts}{ 
Let $C_1,C_2$ be  covariances on $V$ and $\la_1,\la_2\in\bbbc$.
\Item{i)} 
If $\cb_1$ is a contraction bound for $C_1$ and $\cb_2$ is a contraction bound for $C_2$ then $|\la_1|\cb_1+|\la_2|\cb_2$ is a contraction
bound for $\la_1 C_1 +\la_2 C_2$.
\Item{ii)} 
If $\ib_1$ and $\ib_2$ are integral bounds for $C_1$ and $C_2$, then 
$\ \sqrt{|\la_1|}\, \ib_1 + \sqrt{|\la_2|}\, \ib_2\ $ 
is an integral
bound for $\la_1 C_1 +\la_2 C_2$.
}
\prf Part (i) follows from Remark \remContractlinearity. To prove part (ii), let $n'\le n$. For $I\subset \{1,\cdots,n'\}$ let $Ant_I$ be the map from $A\otimes V^{\otimes n}$ to
$\bigwedge\nolimits_A^{|I|}V \otimes_A\bigwedge\nolimits_A^{n'-|I|}V \otimes_A   V^{\otimes (n-n')}$
characterized by
$$
Ant_{I} (v_1\otimes\cdots \otimes v_{n'}\, \otimes\,  w_1\otimes \cdots \otimes w_{n-n'})\ =\  
\ep_I\ \big( \smprod_{i\in I}v_i \big) \otimes 
\big(\smprod_{j\in \{1,\cdots,n'\}\setminus I}v_i\big) \,
 \otimes\, w_1\otimes \cdots \otimes w_{n-n'}
$$ 
where $\ep_I$ is the sign of the permutation that puts the sequence 
$I,\,\{1,\cdots,n'\}\setminus I$ back in increasing order. Then
$$
\int Ant_{n'}(f)\, d\mu_{\la_1 C_1 +\la_2 C_2} 
= \sum_{I\subset \{1,\cdots,n'\} \atop |I| \ {\rm even}}
   \int Ant_I(f)(\xi,\xi')\,d\mu_{\la_1C_1}(\xi)\,d\mu_{\la_2C_2}(\xi')
$$
and consequently
$$\eqalign{
\Big\| \int Ant_{n'}(f) d\mu_C \Big\| 
& \le \smsum_{r=0}^{n'}\ \smsum_{I\subset \{1,\cdots,n'\} \atop |I|=r} 
   \Big\| \int Ant_I(f)(\xi,\xi')\,d\mu_{\la_1C_1}(\xi)\,d\mu_{\la_2C_2}(\xi')      \Big\|  \cr
& \le \smsum_{r=0}^{n'} \smchoose{n'}{r}\ 
  \Big(\sfrac{\sqrt{|\la_1|}\ \ib_1}{2}\Big)^r \,
  \Big(\sfrac{\sqrt{|\la_2|}\ \ib_2}{2}\Big)^{n'-r} \|f\|\cr
& \le \big( \sfrac{1}{2}\big)^{n'}\,
   \Big( \sqrt{|\la_1|}\  \ib_1 + \sqrt{|\la_2|}\  \ib_2 \Big)^{n'}\ \|f\| \cr
}$$
\endproof

\lemma{\STM\lemwicknorm}{
Let
$ f(\xi^{(1)},\cdots,\xi^{(s)},\xi^{(s+1)},\cdots,\xi^{(t)},\xi^{(t+1)},
\cdots,\xi^{(r)})$ be a Grassmann function in 
$\bigwedge\nolimits_A V\otimes\cdots\otimes
\bigwedge\nolimits_A V \cong
\bigwedge\nolimits_A (V\oplus\cdots\oplus V)$. Set
$$
\tilde f(\xi^{(1)},\cdots,\xi^{(s)},\xi^{(s+1)},\cdots,\xi^{(t)},
\cdots,\xi^{(r)})
=\ \lw  f(\xi^{(1)},\cdots,\xi^{(s)},\xi^{(s+1)},\cdots,\xi^{(t)},
\cdots,\xi^{(r)}) \rw_{\xi^{(1)},\cdots,\xi^{(s)}} 
$$
Let $0<\veps\le \al$.
If $\veps\ib$ is is an integral bound for the covariance $C$, then
$$
N\Big( \int \tilde f(\xi,\cdots,\xi,\xi^{(t+1)},\cdots,\xi^{(r)})\,d\mu_C(\xi)
\ ;\ \al\Big)  \le N(f;\al)
$$
If $f(0,\cdots,0,\xi^{(t+1)},\cdots,\xi^{(r)})=0$ then
$$
N\Big( \int \tilde f(\xi,\cdots,\xi,\xi^{(t+1)},\cdots,\xi^{(r)})\,d\mu_C(\xi)
\ ;\ \al\Big)  \le \sfrac{\veps^2}{\al^2}\,N(f;\al)
$$

}

\prf Set 
$$
g(\xi^{(t+1)},\cdots,\xi^{(r)}) = 
 \int \tilde f(\xi,\cdots,\xi,\xi^{(t+1)},\cdots,\xi^{(r)})\,d\mu_C(\xi)
$$
Write
$$
f = \sum_{m,n_1,\cdots, n_r\ge 0} 
f_{m; n_1,\cdots, n_r}
$$ 
with $f_{m; n_1,\cdots, n_r} \in A_m[n_1,\cdots,n_r]$ and set
$$\eqalign{
\tilde f_{m; n_1,\cdots, n_r}(\xi^{(1)},\cdots,\xi^{(t)},\cdots,\xi^{(r)})
&=\ \lw  f_{m; n_1,\cdots, n_r}
(\xi^{(1)},\cdots,\xi^{(s)},\cdots,\xi^{(t)},\cdots,\xi^{(r)}) 
\rw_{\xi^{(1)},\cdots,\xi^{(s)}}  \cr
g_{m; n_1,\cdots, n_r}(\xi^{(t+1)},\cdots,\xi^{(r)}) 
&=  \int \tilde f_{m; n_1,\cdots, n_r}
(\xi,\cdots,\xi,\xi^{(t+1)},\cdots,\xi^{(r)})\,d\mu_C(\xi) \cr
}$$
Then $g = \sum_{m; n_1,\cdots, n_r\ge 0} g_{m; n_1,\cdots, n_r}$ and therefore,
by part (ii) of Lemma \lemGrasscontractnorm,
$$\eqalign{
N(g;\al) & \le \sfrac{1}{\ib^2}\,\cb 
  \sum_{m;n_{t+1},\cdots,n_r} \big(\al\ib\big)^{n_{t+1}+\cdots+n_r}
  \sum_{n_1,\cdots,n_t} (\veps\ib)^{n_1+\cdots+n_t}
    \|f_{m; n_1,\cdots, n_r}\|
\cr
& = \sfrac{1}{\ib^2}\,\cb 
  \sum_{m;n_1,\cdots,n_{t+1},\cdots,n_r} \Big(
     \sfrac{\veps}{\al}\Big)^{n_1+\cdots+n_t}
     \al^{|n|} \ib^{|n|} \|f_{m; n_1,\cdots, n_r}\| \cr
}$$
Since $ \sfrac{\veps}{\al} \le 1$, this implies that 
$N(g;\al)\le N(f;\al)$. If
$f(0,\cdots,0,\xi^{(t+1)},\cdots,\xi^{(r)})=0$ then
$f_{m; n_1,\cdots, n_r}=0$ for $n_1=\cdots=n_t=0$ and
$g_{m; n_1,\cdots, n_r}=0$ for $n_1+\cdots+n_t \le 1$,
since $\int\xi_i\,d\mu_C(\xi)=0$. 
Consequently
$$
N(g;\al)\ \le\  \sfrac{1}{\ib^2}\,\cb 
  \sum_{m;n_1+\cdots+n_t \ge 2 \atop n_{t+1},\cdots,n_r} 
  \Big( \sfrac{\veps}{\al}\Big)^{n_1+\cdots+n_t}
     \al^{|n|} \ib^{|n|} \|f_{m; n_1,\cdots, n_r}\|
\ \le \ \sfrac{\veps^2}{\al^2}\,N(f;\al)
$$
\endproof

\corollary{\STM\corwicknorm}{
Let $f(\xi) \in \bigwedge_A V$. 
\Item{(i)}
If $\al\ib$ is an integral bound for the covariance $C$, then
$$\eqalign{
N(\lw f\rw_C\,\cl \al) &\le N(f\cl 2\al) \cr
N(f\cl \al)  &\le N(\lw f\rw_C\,\cl 2\al) \cr
}$$
\Item{(ii)}
Let $C_1,C_2$ be two covariances and $0<\veps \le \sfrac{\al}{\sqrt{2}}$. 
Assume that $\veps\ib$ is an integral bound for $C_1-C_2$. Set
$$
\lw f(\xi)\rw_{C_2}\ =\ \lw f'(\xi)\rw_{C_1}
$$
Then
$$
N(f'-f\cl \al) \le \sfrac{2\veps^2}{\al^2}N(f\cl 2\al)
$$
\Item{(iii)}
Let, for $\ka$ in a neighbourhood of zero, $C_\ka$ be a covariance on $V$. 
Assume that $C_0$ and $\sfrac{d\hfill}{d\ka}C_\ka\big|_{\ka=0}$ have integral
bounds $\ib$ and $\ib'$, respectively. Define $f_\ka$ by
$\lW f_\ka\rW_{C_\ka}=f$. Then
$$
N\big(\sfrac{d\hfill}{d\ka}f_\ka\big|_{\ka=0}\,\cl \al\big) 
\le \sfrac{1}{(\al-1)^2}\big(\sfrac{\ib'}{\ib}\big)^2N(f\cl 2\al)
$$

}

\prf
\Item{(i)}
By part (i) of Proposition \:\propBII,  Remark \scaleIntConsts,
the Lemma above with $s=0$ and $\veps=\al$, and part (i) of Remark \remfunctnorm
$$\eqalign{
N(\lw f\rw_C\,\cl \al)
&= N\Big( \int f(\xi+\xi')\,d\mu_{-C}(\xi')\cl \al\Big) \cr
& \le  N\big( f(\xi+\xi')\cl \al\big) \cr
& \le  N(f\cl 2\al) \cr
}$$
The proof of the other bound is similar.
\Item{(ii)} 
Define $f_z$
by
$\ 
\lw f(\xi)\rw_{C_2}\ =\ \lw f_z(\xi)\rw_{C_2+z(C_1-C_2)}
\ $.
By part (i) of Lemma \:\lemBIV
$$
f_z(\xi)\ =\ \lW f(\xi)\rW_{-z(C_1-C_2)}
$$
By Remark  \:\scaleIntConsts.ii, $-z(C_1-C_2)$
has integral bound $\sqrt{|z|}\,\veps\,\ib$, which is bounded by $\al\ib$ for 
all $|z|\le\big(\sfrac{\al}{\veps}\big)^2$. Hence, by part (i),
$$
N(f_z\cl \al)\ \le \ N(f\cl 2\al)
$$
for all $|z|\le\big(\sfrac{\al}{\veps}\big)^2$. Let $\cC$ be the contour
$|\ze|=\big(\sfrac{\al}{\veps}\big)^2$ in the complex plane, with standard
orientation. Then
$$
f'-f=f_1-f_0
=\sfrac{1}{2\pi \imath}\int_{\cC}\big(\sfrac{1}{\ze-1}
-\sfrac{1}{\ze}\big)\ f_\ze\ d\ze
=\sfrac{1}{2\pi \imath}\int_{\cC}\sfrac{1}{\ze(\ze-1)}\ f_\ze\ d\ze
$$
Hence, 
$$
N\big(f'-f\cl\al\big)
\ \le \ \sfrac{1}{({\al\over \veps})^2-1}\sup_{|\ze|=({\al\over \veps})^2}
N\big(f_\ze\cl \al\big) 
\ \le\ 2\sfrac{\veps^2}{\al^2}\,N(f\cl 2\al)
$$

\Item{(iii)} Set $D_z=C_0+z\sfrac{d\hfill}{d\ka}C_\ka\big|_{\ka=0}$
and define $g_z$ by $\lW g_z\rW_{D_z}=f$.
Since $C_\kappa$ and $D_\kappa$ agree to order $\kappa$, 
$\sfrac{d\hfill}{d\ka}f_\ka\big|_{\ka=0}
=\sfrac{d\hfill}{dz}g_z\big|_{z=0}$.
By Remark \:\scaleIntConsts.ii, $D_z$
has integral bound $\ib+\sqrt{|z|}\,\ib'$. For all 
$|z|\le(\al-1)^2\big(\sfrac{\ib}{\ib'}\big)^2$
this is bounded by $\al\ib$. By part (i),
$$
N(g_z\cl \al)\ \le \ N(f\cl 2\al)
$$
for all $|z|\le(\al-1)^2\big(\sfrac{\ib}{\ib'}\big)^2$. Hence, by the 
Cauchy integral formula
$$\eqalign{
N\big(\sfrac{d\hfill}{d\ka}f_\ka\big|_{\ka=0}\cl \al\big) 
&=N\big(\sfrac{d\hfill}{dz}g_z\big|_{z=0}\cl \al\big) 
\le \sfrac{1}{(\al-1)^2}\big(\sfrac{\ib'}{\ib}\big)^2
\sup_{|z|=(\al-1)^2({\ib\over\ib'})^2}N\big(g_z\cl \al\big) \cr
&\le \sfrac{1}{(\al-1)^2}\big(\sfrac{\ib'}{\ib}\big)^2N(f\cl 2\al)\cr
}$$
\endproof

The following Proposition is the key to our estimate on the renormalization 
group map and will be used in the proof of Proposition \:\propnormestR.

\proposition{\STM\propestCconnect}{
Assume that $\cb$ is a contraction bound and $\ib$ 
an integral bound for $C$.
Let $\ell\ge 1$ and $r\ge s\ge 1$. Let $f_i(\xi^{(1)},\cdots,\xi^{(s)},\xi^{(s+1)},\cdots,\xi^{(r)})$, $1\le i\le \ell$ and $f(\xi^{(1)},\cdots,\xi^{(r)})$ be Grassmann functions. Assume that
for each $1\le i\le \ell$ there is a $1\le j_i\le s$ such that $f_i$ vanishes
when $\xi^{(j_i)}=0$.
Set
$$\eqalign{
g(\xi^{(s+1)},\cdots,\xi^{(r)}) 
&= \int
\lW\smprod_{i=1}^\ell f_i(\xi^{(1)},\cdots,\xi^{(s)},\cdots,\xi^{(r)})
  \rW_{\xi^{(1)},\cdots,\xi^{(s)}} \cr
&\hskip3cm
 \lW f(\xi^{(1)},\cdots,\xi^{(s)},\cdots,\xi^{(r)})
  \rW_{\xi^{(1)},\cdots,\xi^{(s)}}
  \smprod_{i=1}^s d\mu_C(\xi^{(i)})\cr
}$$
Then, for $\al \ge 2$, we have the two bounds
$$\eqalign{
N(g\cl \al) \ 
& \le  \sfrac{\ell!}{\al^\ell}  
N(f\cl \al)\,\smprod_{i=1}^\ell N(f_i\cl \al) \cr
N(g\cl \al) \ 
& \le  \sfrac{\ell^\ell}{\al^{2\ell}}  
N(f\cl \al)\,\smprod_{i=1}^\ell N(f_i\cl \al) \cr
}$$

}
\prf
We first prove the statement in the case that $f$ and all the $f_i$'s are  
homogeneous. That is
$$\eqalign{
f & \in A_m[n^{(1)},\cdots,n^{(s)},\cdots,n^{(r)}] \cr
f_i & \in A_{m_i}[n_i^{(1)},\cdots,n_i^{(r)}]\qquad 1\le i\le \ell \cr
}$$
 By hypothesis
$n_i^{(j_i)}\ge 1$ for each $1\le i\le \ell$. Set, for each $1\le i\le \ell$,
${\rm Con}_i
=\cont{\xi^{(j_i)}_i}{\ze^{(j_i)}}{C}$ and
$$\eqalign{
g'(\xi^{(1)},\cdots,\xi^{(r)};\ze^{(1)},\cdots,\ze^{(r)}) 
&= \smprod_{i=1}^\ell{\rm Con}_i\,
   \smprod_{i=1}^\ell f_i(\xi^{(1)}_i,\cdots,\xi^{(r)}_i)\  
    f(\ze^{(1)},\cdots,\ze^{(r)})\Big|_{\xi_i^{(j)}=\xi^{(j)}}\cr
g'_w(\xi^{(1)},\cdots,\xi^{(r)};\ze^{(1)},\cdots,\ze^{(r)})
&=\lW g'(\xi^{(1)},\cdots,\xi^{(r)};\ze^{(1)},\cdots,\ze^{(r)})
 \rW_{\xi^{(1)},\cdots,\xi^{(s)}\atop \ze^{(1)},\cdots,\ze^{(s)} }
}$$
The $\Big|_{\xi_i^{(j)}=\xi^{(j)}}$ signifies that  $\xi_i^{(j)}$ is to 
be evaluated at $\xi^{(j)}$ for all for each $1\le j\le r$ and 
$1\le i\le \ell$. 
By Lemma \lemcontract, $\ell$ times, 
$$
g = \int g'_w(\xi^{(1)},\cdots,\xi^{(r)};\xi^{(1)},\cdots,\xi^{(r)}) \,
  \smprod_{i=1}^s d\mu_C(\xi^{(i)})
$$

Observe that $g=0$ unless $n^{(j)}=\smsum_{i=1}^\ell n_i^{(j)}$ 
for all $1\le j \le s$ and then $g'$ is of degree 
$$ 
\smsum_{j=1}^s \Big[n^{(j)}+\smsum_{i=1}^\ell n_i^{(j)}\Big]-2\ell
\ =\ \smsum_{j=1}^s 2n^{(j)}\ \ -\ 2\ell
$$ 
in $\xi^{(1)},\cdots,\xi^{(s)}$. Hence 
$$\eqalign{
\|g\|
\ &\le\ \ib^{2(n^{(1)}+\cdots+n^{(s)}-\ell)}\,
\|g'(\xi^{(1)},\cdots,\xi^{(r)};\xi^{(1)},\cdots,\xi^{(r)})\| \cr
&\hskip8cm\hbox{by Lemma \lemGrasscontractnorm.ii}\cr
\ &\le\ \ib^{2(n^{(1)}+\cdots+n^{(s)}-\ell)}\,
\| \smprod_{i=1}^\ell{\rm Con}_i\,
   \smprod_{i=1}^\ell f_i(\xi^{(1)}_i,\cdots,\xi^{(r)}_i)\  
    f(\ze^{(1)},\cdots,\ze^{(r)})\|\cr 
&\hskip8cm\hbox{by Lemma \lemGrasscompatnorm.ii}\cr
}$$
Set, for each
$1\le j\le s$
$$
p_j=\#\set{i}{1\le i\le \ell,\ j_i=j}
$$
Note that $p_j\le n^{(j)}$ and $\smsum_{j=1}^sp_j=\ell$.  Therefore, by Lemma \lemGrasscontractnorm.i,
$$\eqalignno{
\|g\|
\ &\le\ \smprod_{j=1}^s p_j!\smchoose{n^{(j)}}{p_j}\ 
 \cb^\ell\ib^{2(n^{(1)}+\cdots+n^{(s)}-\ell)}\,
\|f\|\, \smprod_{i=1}^\ell\| f_i\|&\EQNO\eqnPropestCconnect\cr 
\ &=\ \smprod_{j=1}^s\Big[ p_j!\smchoose{n^{(j)}}{p_j}\sfrac{1}{\al^{2n^{(j)}}}\Big]\ 
 \big(\al\ib\big)^{2(n^{(1)}+\cdots+n^{(s)})}\,
\|f\|\, \smprod_{i=1}^\ell \ib^{-2}\cb\| f_i\|\cr 
\ &=\ \smprod_{j=1}^s\Big[ p_j!\smchoose{n^{(j)}}{p_j}\sfrac{1}{\al^{2n^{(j)}}}\Big]\ 
 \big(\al\ib\big)^{n^{(1)}+\cdots+n^{(s)}}\,
\|f\|\, \smprod_{i=1}^\ell \ib^{-2}\cb
\big(\al\ib\big)^{n_i^{(1)}+\cdots+n_i^{(s)}}\| f_i\|\cr 
}$$
As $g$ is of degree 
$\smsum_{j=s+1}^r\big[n^{(j)}+\smsum_{i=1}^\ell n_i^{(j)}\Big]$
in $\xi^{(s+1)},\cdots,\xi^{(r)}$,
Definition \deffunctnorm\ implies that
$$\eqalignno{
N(g;\al) &= \sfrac{1}{\ib^2}\,\cb\,
\big(\al\ib\big)^{\Si_j(n^{(j)}+\Si_i n_i^{(j)})} \,\|g\| \cr
&\le \smprod_{j=1}^s\Big[ p_j!\smchoose{n^{(j)}}{p_j}\sfrac{1}{\al^{2n^{(j)}}}\Big]\ 
N(f;\al)\, \smprod_{i=1}^\ell N(f_i;\al) &\EQNO\eqnestCconnect\cr
}$$
To complete the proof in the homogeneous case, we derive two bounds on
$p!\smchoose{n}{p}\sfrac{1}{\al^{2n}}$ for all $p\le n$. The first is
$$
p!\smchoose{n}{p}\sfrac{1}{\al^{2n}}
\le p! \sfrac{2^n}{\al^{2n}}\le p! \sfrac{1}{\al^{n}}\le p! \sfrac{1}{\al^{p}}
$$
since $\al\ge 2$. It yields the first bound of the Proposition, since 
$$
\smprod_{j=1}^s\Big[ p_j!\frac{1}{\al^{p_j}}\Big]
\le \big(\Si_j p_j\big)!\ \frac{1}{\al^{\Si_j p_j}}
=\ell!\frac{1}{\al^\ell}
$$
The second is, setting $n=p+m$,
$$\eqalign{
p!\smchoose{p+m}{p}\sfrac{1}{\al^{2(p+m)}}
&=\sfrac{1}{\al^{2p}}\sfrac{(p+m)!}{m!}\sfrac{1}{\al^{2m}}
\le \sfrac{1}{\al^{2p}}(p+m)^p\sfrac{1}{\al^{2m}}
= \sfrac{p^p}{\al^{2p}}\big(1+\sfrac{m}{p}\big)^p\sfrac{1}{\al^{2m}}\cr
&\le\sfrac{p^p}{\al^{2p}}\big(e^{m/p}\big)^p\sfrac{1}{\al^{2m}}
=\sfrac{p^p}{\al^{2p}}\big(\sfrac{e}{\al^{2}}\big)^m
\le \sfrac{p^p}{\al^{2p}}
}$$
since $\al^2>e$. It yields the second bound of the Proposition, since 
$$
\smprod_{j=1}^s\Big[ \sfrac{p_j^{p_j}}{\al^{2{p_j}}}\Big]
\le \big(\Si_j p_j\big)^{\Si_j p_j}\ \frac{1}{\al^{2\Si_j p_j}}
=\ell^\ell\frac{1}{\al^{2\ell}}
$$

The general case now follows by decomposing $f$ and the $f_i$'s into homogeneous pieces.
\endproof

\vfill\eject

\chap{ The Schwinger Functional}\PG\pgRIII
\sect{ Description of the Schwinger Functional}\PG\pgRIIIa

Let $A$ be a superalgebra and $V$ be a complex vector space with  
generators $\{\xi_i\}$. Furthermore let $C$ be an
antisymmetric bilinear form (covariance) on $V$. 
First, suppose that $V$ is finite dimensional. Let 
$U(\xi) \in \bigwedge\nolimits_A V $ be an even Grassmann function such that 
$$
Z(U,C) = \int e^{U(\xi)}\, d\mu_C(\xi)
$$
is invertible in $A$. For any $f(\xi)\in \bigwedge\nolimits_A V$ set
$$
\cS(f) = \cS_{U,C}(f) = \sfrac{1}{Z(U,C)}\, \int f(\xi)\, e^{U(\xi)}\, d\mu_C(\xi)
$$ 
$\cS$ is called the Schwinger functional. If $V$ is infinite dimensional,
we define $\cS$ as a formal power series in $U$, as in Definition
\defrengroupInfDim. The coefficient that is of order $n$ in $U$ is a sum
of connected graphs that has $n$ vertices $U$, one vertex $f$ and lines
$C$.

\remark{\STM\remrenschw}{
\Item{(i)}
The renormalization group map can be expressed in terms of the Schwinger functional. Recall that
$$
\Om_C(W)(\psi) = \log \sfrac{1}{Z}\int e^{W(\psi +\xi)} d\mu_C(\xi)
\qquad\qquad \hbox{where}\qquad
Z=Z\Big(\int e^{W(\xi)} d\mu_C(\xi)\Big) 
$$
where the $\psi_i$ are the generators of a vector space $V'$,
 which is a second copy of $V$. As $\Om_C(0)=0$, for even Grassmann 
functions $W(\psi)$,
$$\eqalign{
\Om_C(W)(\psi) & = \int_0^1 \sfrac{d\ }{dt} \Om_C(t W)\,dt \cr
 &=\int_0^1
\frac{\int W(\psi+\xi)\, e^{t W(\psi+\xi)}\,d\mu_C(\xi)}
      {\int e^{t W(\psi+\xi)} \,d\mu_C(\xi)}\,dt-\log Z \cr 
 &=\int_0^1 \cS_{t U, C}(U)\,dt-\log Z \cr
}$$
where in the integral $U(\psi;\xi) = W(\psi+\xi) \in \bigwedge\nolimits_A(V'\oplus V) \cong 
 \bigwedge\nolimits_{\bigwedge\nolimits_A V'} V $
and the Schwinger functional is taken in the Grassmann algebra over $V$ with coefficients in the algebra $\bigwedge\nolimits_A V'$.
\Item{(ii)} More generally, if $W_1$ and $W_2$ are even Grassmann functions and $W_2=W_1+W$, then
$$\eqalign{
\Om_C(W_2)(\psi)-\Om_C(W_1)(\psi) & 
= \int_0^1 \sfrac{d\ }{dt} \Om_C(W_1+t W)\,dt \cr
 &=\int_0^1 \cS_{U_1+t U, C}(U)\,dt
 -  \log\sfrac{Z_1}{Z_2} \cr
}$$
where $U_1(\psi;\xi) = W_1(\psi+\xi)$ and $U(\psi;\xi) = W(\psi+\xi)$. 
}

In [FKT1] we gave a representation for Schwinger functionals which we repeat in the present context. Choose an additional copy $V^{\prime\prime}$ of the vector space $V$. We denote the canonical isomorphism from $V$ to $V^{\prime\prime}$ by $\sigma$ and set $\eta_i=\sigma(\xi_i)$. The antisymmetric bilinear form $C^{\prime\prime}$ on $V''$ corresponding to $C$ is given by $C^{\prime\prime}(v,w) = C(\si^{-1}v,\si^{-1}w)$.
Using the canonical isomorphisms
$$
\bigwedge\nolimits_A(V\oplus V^{\prime\prime}) 
= \bigoplus_{r,r'}\, \bigwedge^r\nolimits_A V\, \otimes\, \bigwedge^{r'}\nolimits_A V^{\prime\prime}
\cong \bigwedge\nolimits_A V\, \otimes\, \bigwedge\nolimits_A V^{\prime\prime}
\cong \bigwedge\nolimits_{\bigwedge\nolimits_A V} V^{\prime\prime}
$$
$C''$ defines the Grassmann Gaussian integral $d\mu_{C''}(\eta)$ 
as a map from
$\bigwedge\nolimits_A(V\oplus V^{\prime\prime})$ to $\bigwedge\nolimits_A V$. The diagonal embedding 
$ {V\rightarrow V\oplus V'\atop v\mapsto v\oplus\si(v)}$
induces an embedding
${\bigwedge\nolimits_A V \rightarrow \bigwedge_A(V\oplus V'')\atop
f(\xi)\mapsto f(\xi+\eta)\strut}$
and the isomorphism ${V\rightarrow V''\atop v\mapsto \si(v)}$ induces an isomorphism
${\bigwedge\nolimits_A V \rightarrow \bigwedge\nolimits_A V''\atop f(\xi)\mapsto f(\eta)\strut}$. 
With this notation we define the map
$$
R=R_{U,C}: \bigwedge\nolimits_A V \longrightarrow \bigwedge\nolimits_A V
$$
by
$$
f \longmapsto \int \lw e^{U(\xi+\eta)-U(\xi)} -1\rw_{\eta} \,f(\eta)\,d\mu_{C}(\eta)
$$
Once again, if ${\rm dim} V=\infty$, $R$, is, a priori, defined as a
formal power series in $U$, i.e. as a sequence of multilinear maps. 
In this case, it is easy to explicitly find maps. The $n^{\rm th}$ map
is
$$
(U_1,U_2,\cdots,U_n,f)\mapsto 
\sfrac{1}{n!}
\int \lW\smprod_{i=1}^n[U_i(\xi+\eta)-U_i(\xi)]\rW_{\eta} \,f(\eta)\,d\mu_{C}(\eta)
$$   
As in [FKT1] we have

\theorem{\STM\theoremIII}{ For all Grassmann functions 
$f\in \bigwedge\nolimits_A V $
$$
\cS_{U,C}(f) = \int  \big(\bbbone - R_{U,C} \big)^{-1}(f)\, d\mu_C
$$
}
\prf
We first prove 
$$
\int\! f(\xi)\,  e^{U(\xi)}\, d\mu_C(\xi)
 = \int\! f(\et)\,d\mu_C(\et)\,\int e^{U(\xi)}\,   d\mu_C(\xi)
\ +\ \int\! R_{U,C}(f) (\xi)\,  e^{U(\xi)}\,   d\mu_C(\xi)
\EQN\eqnROPa$$
Inserting the definition of $R_{U,C}(f)$ into the right hand side,
$$\deqalign{
&\int\! f(\et)\,d\mu_C(\et)\,\int e^{U(\xi)}\,   d\mu_C(\xi)
\ +\ \int\! R_{U,C}(f) (\xi)\,  e^{U(\xi)}\,   d\mu_C(\xi)\cr
\ &\hskip1in=\ 
\int\!\bigg[\int \lW e^{U(\xi+\et)-U(\xi)}\rW_\et\, f(\et)\, d\mu_C(\et)\bigg]\, e^
{U(\xi)}\   d\mu_C(\xi)\hidewidth\cr
\ &\hskip1in=\ 
\int\!\! \int\, \lW e^{U(\xi+\et)}\rW_\et\, f(\et) \ \, d\mu_C(\et)\  \, d\mu_C(\xi)
\cr
}$$
since $\lW e^{U(\xi+\et)-U(\xi)}\rW_\et=\lW e^{U(\xi+\et)}\rW_\et\ e^{-U(\xi)}$.
Continuing,
$$\deqalign{
&\int\! f(\et)\,d\mu_C(\et)\,\int e^{U(\xi)}\,   d\mu_C(\xi)
\ +\ \int\! R_{U,C}(f) (\xi)\,  e^{U(\xi)}\,   d\mu_C(\xi)\cr
\ &\hskip1in=\ 
\int\!\! \int\, \lW e^{U(\xi+\et)}\rW_\xi\, f(\et) \ \, d\mu_C(\et)\  \, d\mu_C(\xi)
&\hbox{by Proposition \propBII.ii}\cr
\ &\hskip1in=\ 
\int\, f(\et)\ e^{U(\et)}\  \, d\mu_C(\et)
&\hbox{by Proposition \propBII.i}\cr
\ &\hskip1in=\ 
\int\, f(\xi)\ e^{U(\xi)}\  \, d\mu_C(\xi)\cr
}$$
This completes the proof of (\eqnROPa). Now we prove the Theorem itself.
For all $g(\xi)\in\bigwedge\nolimits_A V$
$$
\int (\bbbone-R_{U,C})(g)\, e^{U(\xi)}\   d\mu_C(\xi)
\, =\,Z(U,C) \int g(\xi)\ d\mu_C(\xi)
$$
by (\eqnROPa). Since $R_{0,C}=0$, the map $\bbbone-R_{U,C}$ trivially has
a formal power series inverse and we may choose 
$g=(\bbbone-R_{U,C})^{-1}(f)$. So 
$$
\int f(\xi)\, e^{U(\xi)}\   d\mu_C(\xi)
\, =\,Z(U,C) \int (\bbbone-R_{U,C})^{-1}(f)(\xi)\ d\mu_C(\xi)
$$
The left hand side does not vanish for all $f\in\bigwedge\nolimits_A V$ (for example, for
$f=e^{-U}$) so $Z(U,C)$ is nonzero and
$$
\sfrac{1}{Z(U,C)}\int f(\xi)\, e^{U(\xi)}\   d\mu_C(\xi)
\ =\ \int (\bbbone-R_{U,C})^{-1}(f)(\xi)\ d\mu_C(\xi)
$$

\endproof

This construction is functorial in the following sense:

\remark{\STM\remfunctor}{ Let $\pi_A: \tilde A \rightarrow A$ be a homomorphism of superalgebras, and $\pi_V: \tilde V \rightarrow V$ a linear map
of complex vector spaces. Define the antisymmetric bilinear $\tilde C$
form on $\tilde V$ by $\ \tilde C (v,w) = C(\pi_V(v),\pi_V(w))\ $.
$\pi_A$ and $\pi_V$ induce an algebra homomorphism 
$\ \pi_*: \bigwedge\nolimits_{\tilde A} \tilde V \rightarrow \bigwedge\nolimits_A V\ $. Let
$\tilde U \in \bigwedge\nolimits_{\tilde A} \tilde V$ with $\pi_*(\tilde U) = U$. Then for all even $\tilde f \in \bigwedge\nolimits_{\tilde A} \tilde V$
$$
\pi_A \big( \cS_{\tilde U, \tilde C} (\tilde f) \big)
= \cS_{U,C}(\pi_*\,f)
$$ 
and
$$
\pi_*\,R_{\tilde U, \tilde C} (\tilde f) = R_{U,C} ( \pi_*\,\tilde f )
$$
}

In our context the Grassmann functions will all be Wick ordered with respect to the covariance $C$. We give a description of the map $R$ of Theorem 
\theoremIII\ adapted to Wick ordering. We use a further copy of the 
vector space $V$ with generators $\{\xi_i'\}$ corresponding to the generators 
$\{\xi_i\}$ of $V$.

\definition{\STM\defcalR}{
For any Grassmann function $K(\xi,\xi',\eta)$ define the map
$\cR_{K,C}: \bigwedge\nolimits_A V \rightarrow \bigwedge\nolimits_A V$ by
$$
\cR_{K,C}(f) = \lww  \int \int \lW  e^{\lw K(\xi,\xi',\eta)\rw_{\xi'}} -1 \rW_\eta\,f(\eta)\,
d\mu_C(\xi')\,d\mu_C(\eta) \rww_\xi
$$
Yet again, when ${\dim}\,V=\infty$, $\cR_{K,C}$ is a formal power series
in $K$.
}

\proposition{\STM\propcalR}{
Assume that $U(\xi) = \lw \hat U(\xi)\rw$. Set 
$K(\xi,\xi',\eta)=\hat U(\xi+\xi'+\eta)-\hat U(\xi+\xi')$.  Then
$$
R_{U,C} =\ \cR_{K,C}
$$
}

\prf By part (ii) of Proposition \:\propBII
$$
U(\xi+\eta)-U(\xi)=\  \lw \hat U(\xi+\eta)-\hat U(\xi)\rw_\xi
$$
Hence by part (iv) of  Proposition \:\propBII
$$
e^{U(\xi+\eta)-U(\xi)} = \ 
\lW \int e^{\lw \hat U(\xi+\xi'+\eta)-\hat U(\xi+\xi')\rw_{\xi'}} d\mu_C(\xi') \rW_\xi
$$
Consequently
$$\eqalign{
R_{U,C}(f) &=  \int \lW  e^{U(\xi+\eta)-U(\xi)} -1\rW_{\eta} \,f(\eta)\,d\mu_{C}(\eta) \cr
&=  \int \lww \big( \lW \int e^{\lw \hat U(\xi+\xi'+\eta)-\hat U(\xi+\xi')\rw_{\xi'}} d\mu_C(\xi') \rW_\xi -1\big)\rww_{\eta} \,f(\eta)\,d\mu_{C}(\eta) \cr
&=  \int \lww  \lW \int\big( e^{\lw K(\xi,\xi',\eta)\rw_{\xi'}}-1\big) d\mu_C(\xi') \rW_\xi \rww_{\eta} \,f(\eta)\,d\mu_{C}(\eta) \cr
& = \  \lww\int \int \lW  e^{\lw K(\xi,\xi',\eta)\rw_{\xi'}} -1 \rW_\eta\,f(\eta)\,d\mu_C(\xi')\,d\mu_C(\eta) \rww_\xi
}$$
\endproof

To perform estimates we expand the exponential on the right hand side of Definition \defcalR. For even Grassmann functions 
$K_1(\xi,\xi',\eta),\cdots,K_\ell(\xi,\xi',\eta)$ define the map
$$
R_C(K_1,\cdots,K_\ell): \bigwedge\nolimits_A V \longrightarrow \bigwedge\nolimits_A V
$$
by
$$
f \longmapsto \ 
\lww  \int \int \lW  \big( \smprod_{i=1}^\ell \lw K_i(\xi,\xi',\eta)\rw_{\xi'} \big) \rW_{\eta} \,f(\eta)\,d\mu_{C}(\xi')\,d\mu_{C}(\eta) \rww_{\xi}
\EQN\eqRCdef$$
Observe that $R_C(K_1,\cdots,K_\ell)$ is multilinear and symmetric 
in $K_1,\cdots,K_\ell$.

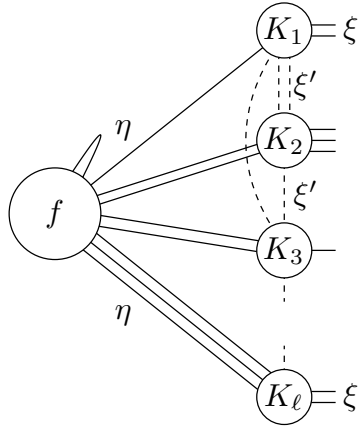
\centerline{\figput{figIII1}}

\noindent
Expanding the exponential gives

\remark{\STM\remcalR}{
For any even Grassmann function $K(\xi,\xi',\eta)$ 
$$
\cR_{K,C} =\ \smsum_{\ell=1}^\infty \sfrac{1}{\ell !}\, R_C(K,\cdots,K)
$$
That is, the $\ell^{\rm th}$ order term in the formal Taylor expansion
of $\cR_{K,C}$ is the $\ell$--linear map 
$\sfrac{1}{\ell!}R_C(K_1,\cdots,K_\ell)$.
}

\sect{ Norm estimates on the Schwinger functional}\PG\pgRIIIb

Again we assume that $A$ is a graded superalgebra and that we are given a  family of symmetric seminorms on the spaces $A_m\otimes V^{\otimes n}$. 
Assume that $\cb$ is a contraction bound and $\ib$ an integral bound for
the covariance $C$. (See Definition \defcontractintbound.) 
Fix $\al>1$. We write $N(f)$ for the
$N(f\cl \al)$ of Definition \deffunctnorm.

\proposition{\STM\propnormestR}{Let $K^{(1)}(\xi,\xi',\eta),\cdots,K^{(\ell)}(\xi,\xi',\eta) $ be  even Grassmann functions that
obey $K^{(i)}(\xi,\xi',0)=0$. Furthermore let $f(\xi)$ be a Grassmann function and $\ell\ge 1$. Set
$$
\sfrac{1}{\ell !}\, R_C(K^{(1)},\cdots,K^{(\ell)})(\lw f\rw)(\xi) = \ \lw f'(\xi)\rw
$$
\Item{(i)} Assume that $ f\in A_m[n]$ for some index $m$ and some $n\ge0$.
Then $f'=0$ if $\ell >n$, and
$$
N(f') \le \sfrac{1}{\al^{2n}}\,\smchoose{n}{\ell}\, N(f)\,
\smprod_{i=1}^\ell\, N(K^{(i)})
$$
\Item{(ii)}
In general, if $\al\ge 2$, then
$$
N(f') \le \sfrac{1}{\al^{\ell}}\,N(f)\,\smprod_{i=1}^\ell\, N(K^{(i)})
$$
}
\prf
Set
$$\eqalign{
G_1(\xi;\xi^{\prime(1)},\cdots,\xi^{\prime(\ell)};\eta) 
&= \lW \smprod_{i=1}^\ell K^{(i)}(\xi,{\xi'}^{(i)},\eta)\rW_\et 
\lw f(\eta)\rw_\et\cr
G_2(\xi;\xi^{\prime(1)},\cdots,\xi^{\prime(\ell)}) 
&=\int G_1(\xi;\xi^{\prime(1)},\cdots,\xi^{\prime(\ell)};\eta)\, d\mu_C(\eta)\cr 
G_3(\xi;\xi^{\prime(1)},\cdots,\xi^{\prime(\ell)})
&= \ \lw  G_2(\xi;\xi^{\prime(1)},\cdots,\xi^{\prime(\ell)}) 
\rw_{\xi^{\prime(1)},\cdots,\xi^{\prime(\ell)}} \cr
}$$
Then
$$
f'(\xi)  = \sfrac{1}{\ell!}\int G_3(\xi;\xi',\cdots,\xi') 
\,d\mu_C(\xi')
$$
By Lemma \lemwicknorm
$$
N(f')\le \sfrac{1}{\ell!}N(G_2)
$$
By Proposition \propestCconnect, with $s=1$ and $r=\ell+2$,
$$
N(G_2)\le \sfrac{\ell!}{\al^{\ell}}\,N(f)\,\smprod_{i=1}^\ell\, N(K^{(i)})
$$
This proves part ii. Part i follows from (\eqnestCconnect) with $s=1$, $p_1=\ell$ and
$n^{(1)}=n$
\endproof

\lemma{\STM\lemnormestcalR}{Let $f(\xi)$ be a Grassmann function over $A$.
The formal Taylor series $\cR_{K,C}(\lw f\rw)$
converges to an analytic map on $\set{K(\xi,\xi',\eta)}{K\hbox{ even},\  
K(\xi,\xi',0)=0,\ N(K)_\0 < \sfrac{\al^2}{2}}$.
Furthermore, if $K(\xi,\xi',\eta)$ is an even Grassmann function over $A$ with 
$K(\xi,\xi',0)=0$ and  $N(K)_\0 < \sfrac{\al^2}{2}$ and if 
 $\lw f'\rw\ =\ \cR_{K,C}(\lw f\rw)$ then
$$
N(f') \le  \ \sfrac{2}{\al^2}\,\sfrac{N(K)}{1-{2\over\al^2}N(K)}\, N(f)
$$
}

\prf Write
$$
f(\xi) =\sum_{m, n\ge 0} f_{m;n}(\xi)
$$
with $f_{m;n} \in A_m[n]$ and set
$$
\lw f'_{m;n}\rw \ = \smsum_{\ell=1}^\infty \sfrac{1}{\ell !}\, R_C(K,\cdots,K)(\lw f_{m;n}\rw)
$$
By part (i) of Proposition \propnormestR, 
$\sfrac{1}{\ell !}\, R_C(K,\cdots,K)(\lw f_{m;n}\rw) =0$ for $\ell>n$ and
$$\eqalign{
N\big(f'_{m;n}\big) \ 
&\le  \ \sfrac{1}{\al^{2n}}\,N(f_{m;n})\,
\smsum_{\ell=1}^n \smchoose{n}{\ell} \,N(K)^\ell\cr
&\le  \ N(f_{m;n})\,
\smsum_{\ell=1}^n \sfrac{2^\ell}{\al^{2\ell}} \,N(K)^\ell\cr
&\le\  \sfrac{2}{\al^{2}}\,N(f_{m;n})\,\sfrac{N(K)}{1-{2\over\al^2}N(K)}
}$$
Consequently
$$\eqalign{
N(f') &\le \sum_{m\ge 0,\ n\ge 1} N\big(f'_{m;n}\big) \cr
&\le \ \sfrac{2}{\al^2}\,\sfrac{N(K)}{1-{2\over\al^2}N(K)}
\sum_{m\ge0,\ n\ge 1} N(f_{m;n})\cr
&\le \ \sfrac{2}{\al^2}\,\sfrac{N(K)}{1-{2\over\al^2}N(K)}\,N(f) \cr
}$$
This also proves that the formal Taylor series expansion of $\cR_{K,C}$
converges to an analytic function.
\endproof

\corollary{\STM\cornormestcalR}{
Let $f(\xi)$ and $K(\xi,\xi',\eta)$ be Grassmann functions over $A$ with
$K$ even and $K(\xi,\xi',0)=0$. Assume that $N(K)_\0 < \sfrac{\al^2}{4}$.
If 
$$
\lw f'\rw\ =\ \frac{1}{\bbbone -\cR_{K,C}}(\lw f\rw ) - \lw f\rw 
$$
then
$$
N(f') \le   \sfrac{2}{\al^2}\,\sfrac{N(K)}{1-{4\over\al^2}N(K)}\,N(f)
$$
}

\prf Set $\be = \sfrac{2}{\al^2}\,\sfrac{N(K)}{1-{2\over\al^2}N(K)}$. 
Observe that
$\be_\0 = \sfrac{2}{\al^2}\,\sfrac{N(K)_\0}{1-{2\over\al^2}N(K)_\0} 
< \sfrac{{1\over 2}}{1-{1\over 2}} =1$.
For $\ell \ge 0$ set
$$
\cR_{K,C}^\ell(\lw f\rw ) \ =\  \lw f'_\ell\rw
$$
By Lemma \lemnormestcalR\  above
$$
N( f'_\ell )  \le  \be^\ell \, N(f)  
$$
Since $f'=\smsum_{\ell=1}^\infty f'_\ell$
$$
N(f')\ \le\ \smsum_{\ell=1}^\infty N(f'_\ell) 
\ \le\  N(f)\,\smsum_{\ell=1}^\infty \be^\ell 
\ =\ \sfrac{\be}{1-\be} \,N(f)\ =
\ \sfrac{2}{\al^2}\,\sfrac{N(K)}{1-{4\over\al^2}N(K)}\,N(f) 
$$
\endproof

\proposition{\STM\propnormestcalS}{ 
Let $f(\xi)\in\bigwedge_AV$. The formal Taylor series 
$\ \cS_{\lw U\rw,C}(\lw f\rw )$
converges to an analytic map on $\set{U(\xi)\in\bigwedge_AV}{U\hbox{ even},\  
N(U\cl 4\al)_\0 < \sfrac{\al^2}{4}}$.
Furthermore, if $U(\xi)$  is an even Grassmann function with coefficients 
in $A$ and $N(U\cl 4\al)_\0 < \sfrac{\al^2}{4}$, then
$$
N\Big( \cS_{\lw U\rw,C}(\lw f\rw )-f(0)\ \cl \al \Big)\ 
\le\  \sfrac{2}{\al^2}\,\sfrac{N(U\cl 4\al)}{1-{4\over\al^2}N(U\cl 4\al)}
\, N(f)
$$
}

\prf Set
$$
K(\xi,\xi',\eta) =  U(\xi+\xi'+\eta)- U(\xi+\xi')
$$
By Remark \remfunctnorm 
$$
N(K\cl \al) \le N\big(U(\xi+\xi'+\eta)\cl \al\big)\le N(U\cl 4\al)  
$$
By Proposition \propcalR, 
$R_{\lw U\rw,C} = \cR_{K,C}$. Therefore, by Theorem \theoremIII\ 
 and part (i) of Proposition \:\propBII,
$$
\cS_{\lw U\rw,C}(\lw f\rw )-f(0) 
= \int \Big(\sfrac{1}{\bbbone -\cR_{K,C}}(\lw f\rw )(\xi) - \lw f(\xi)\rw \Big) d\mu_C(\xi)
$$
Consequently, by Lemma \lemwicknorm\ and Corollary \cornormestcalR
$$
N\big( \cS_{\lw U\rw ,C}(\lw f\rw )-f(0) \big) \ 
\le\ \sfrac{2}{\al^2}\,\sfrac{N(K)}{1-{4\over\al^2}N(K)}\,N(f)
\ \le\ \sfrac{2}{\al^2}\,\sfrac{N(U\cl 4\al)}{1-{4\over\al^2}N(U\cl 4\al)}
\, N(f)
$$
\endproof

\sect{  Proof of Theorem \theorII}\PG\pgRIIIc

Recall that  the renormalization group map $\Om_C$ is defined by
$$
\Om_C(\lw W\rw)(\psi) = \log \sfrac{1}{Z}\int e^{\lw W\rw(\psi +\xi)} d\mu_C(\xi) 
\qquad{\rm where}\quad Z= Z\Big(\int e^{\lw W\rw (\xi)} d\mu_C(\xi)\Big)
$$
Let $A'=\bigwedge_A V'$ be the Grassmann algebra in the variables $\psi$ with coefficients in $A$. 
As $W(\psi +\xi)\in \bigwedge_A (V'\oplus V)\cong \bigwedge_{A'}V$
and $\int \ \cdot\ d\mu_C(\xi)$ maps $ \bigwedge_{A'}V$ to $A'$,
$\Om_C$ is map from (a subset of) $ \bigwedge_{A}V$ to $A'=\bigwedge_A V'
\cong \bigwedge_A V$ (since $V'$ is a copy of $V$).

Set 
$$\
U(\psi;\xi) = W(\psi+\xi) \in \bigwedge\nolimits_{A'}V
$$
Then $U(\psi,0)=W(\psi)$. By part (ii) of Proposition \:\propBII
$$
\lw U(\psi,\xi)\rw_\xi \ =\ \lw W\rw(\psi+\xi)
$$
and by Remark \remrenschw.i
$$
\Om_C(\lw W\rw)(\psi) - W(\psi)  
=\int_0^1 \big( \cS_{\lw t U\rw, C}(\lw U\rw ) 
  - U(\psi;0) \big)\,dt
 - \log Z
$$

We now wish to apply Proposition \propnormestcalS, with $(U,f,A)$ 
replaced by $(tU,U,A')$, to
bound $\Om_C(\lw W\rw)(\psi) - W$. We have been given, in the statement
of  Theorem \theorII, a system $\|\,\cdot\,\|$
of symmetric seminorms on the spaces $A_m\otimes V^{\otimes n}$.
Any $f\in A'_m\otimes V^{\otimes n}$ may be uniquely expressed as
 $$
f=\sum_{ m'+m''=m}f_{m',m''}
$$ 
with $f_{m',m''}\in A_{m'}\otimes \bigwedge^{m''}V\otimes V^{\otimes n}$. We define
$$
\|f\|' =\sum_{ m'+m''=m} \al^{m''}\ib^{m''}\,\|f_{m',m''}\|
$$
Then $\|\,\cdot\,\|'$ is a system of symmetric seminorms on the spaces
$A'_m\otimes V^{\otimes n}$ and the covariance $C$ has contraction bound
$\cb$ and integral bound $\ib$  with respect to these norms. For 
$f\in \bigwedge_{A'}V \cong  \bigwedge_A (V'\oplus V)$ and $\al>1$ let 
$N(f\cl \al)$ be the quantity of Definition \deffunctnorm, considering $f$ as an element of $\bigwedge_A (V'\oplus V)$ and using the seminorms 
$\|\,\cdot\,\|$; and let $N'(f\cl \al)$ be 
defined  viewing $f$ as an element of $\bigwedge_{A'}V$ and using 
the seminorms $\|\,\cdot\,\|'$. Then 
$N'(f\cl \al) = N(f\cl \al)$, while for $\al'>\al$,
$N'(f\cl \al') \le N(f\cl \al')$.

 By Remark \remfunctnorm
$$
N\big(U\cl \al\big) \le  N\big(W\cl 2\al\big)
$$
 Proposition \propnormestcalS\ now implies that
$$\deqalign{
N\big(\Om_C(\lw W\rw)(\psi) - W\,\cl \al)
&=N'\big(\Om_C(\lw W\rw)(\psi) - W\,\cl \al)  
& \le \sup_{0\le t \le 1}
    N'\big( \cS_{\lw t U\rw, C}(\lw U\rw )- U(0) \big) \cr
& \le \sfrac{2}{\al^2}\,
     \sfrac{N'( U\cl 4\al)}{1-{4\over\al^2}N'(U\cl 4\al)} 
     \, N'( U\cl \al )
& \le \sfrac{2}{\al^2}\,
     \sfrac{N( U\cl 4\al)}{1-{4\over\al^2}N(U\cl 4\al)} 
     \, N( U\cl \al ) \cr
& \le \sfrac{2}{\al^2}\,
     \sfrac{N( W\cl 8\al)}{1-{4\over\al^2}N(W\cl 8\al)} 
     \, N( W\cl 2\al )
& \le \sfrac{2}{\al^2}\,
     \sfrac{N( W\cl 8\al)^2}{1-{4\over\al^2}N(W\cl 8\al)} \cr
}$$
In the last step we used Remark \remfunctnorm, part (ii). The hypotheses of Proposition \propnormestcalS\ are fulfilled, since, by hypothesis
$$
N\big( U \cl 4\al \big)_\0
\le \ N\big(W\cl 8\al \big)_\0 <\sfrac{\al^2}{4} 
$$
\endproof

\remark{\STM\remjointanalyticity}{
Suppose that $\|\ \cdot\ \|_{\ib\cb}$ is a norm on the space of antisymmetric bilinear forms on $V$ and that there is a $\ka>0$ such that every $C$ 
with $\|C\|_{\ib\cb}<\ka$ has integral bound $\ib$ and contraction bound $\cb_\0+\sum_{\de\ne 0}\infty t^\de$. Then $\Om_C(\lw W\rw)$ is jointly
analytic in $C$ and $W$ on
$\set{(W,C)}{W\hbox{ even},\  
N\big(W\cl 8\al \big)_\0 <\sfrac{\al^2}{4},\ \|C\|_{\ib\cb}<\ka}$.

}

\vfill\eject

\chap{ More Estimates on the Renormalization Group Map}\PG\pgRIV

In the situation of [FKTf1--f3], the effective interaction is Wick ordered both with
respect to the covariance that is integrated out at the renormalization group step and a covariance that is approximately
the sum of the covariances for all future
renormalization group steps. In this Section we modify the construction of 
the previous two Sections to accommodate the second ``output'' Wick ordering.
Furthermore, we estimate the derivative of $\Om_C\big(\lw W\rw_{\psi,C+D}\big)$
with respect to the effective interaction $W$ and the covariances $C$ and $D$.

Let again $V$ be a complex vector space with generators $\{\xi_i\}$,  
let $A$ be a graded superalgebra, and let $\|\,\cdot\,\|$ be a system of symmetric seminorms. Furthermore let $C$ and $D$ be two covariances on $V$. 

\theorem{\STM\theoremIVa}{
Let $W(\psi)$ be an even Grassmann function with coefficients in $A$. Let
$$
\lw W'(\psi)\rw_{\psi,D}\ =\ \Om_C(\lw W\rw_{\psi,C+D})
$$
Let $\al\ge 1$ and assume that $\cb$ is a 
contraction bound for the covariance $C$ and $\ib$ is an integral bound 
for $C$ and for $D$. If $\,N(W\cl 32\al)_\0 <\al^2$, then
$$
N\big( W'-W\cl \al\big) \le \sfrac{1}{2\al^2}\,
     \sfrac{N( W\cl 32\al)^2}{1-{1\over\al^2}N( W\cl 32\al)} 
$$
}
\remark{\STM\remnormlaCD}{ By Remark \remnormal.ii, 
$Z\big(\lw W'(\psi)\rw_{\psi,D}\big)=0$.
In general $Z\big(W'\big)\ne 0$. If one defines
$$
\Om_{C,D}(W)(\psi) = \log \sfrac{1}{Z_{C,D}}\int e^{W(\psi +\xi)} d\mu_C(\xi) 
\ {\rm where}\  Z_{C,D}= Z\Big(\dblint e^{W(\psi+\xi)} d\mu_C(\xi)
d\mu_D(\psi)\Big)
$$
then $\lw W''(\psi)\rw_{\psi,D}\ =\ \Om_{C,D}(\lw W\rw_{\psi,C+D})$
obeys the normalization condition $Z(W'')=0$. Furthermore, $W'$ and $W''$
differ only by a constant, so that 
$$
N\big( W''-W\cl \al\big) \le \sfrac{1}{2\al^2}\,
     \sfrac{N( W\cl 32\al)^2}{1-{1\over\al^2}N( W\cl 32\al)}
$$

}
\proof{of Theorem \theoremIVa} Define $U$ and $U'$ by
$$
U(\psi) =\lw W(\psi)\rw_{\psi,D}\qquad 
U'(\psi) =\lw W'(\psi)\rw_{\psi,D}
$$
Then, by Lemma \:\lemBIV.i
$$
\lw U\rw_{\psi,C} \ =\ \lw W\rw_{\psi,C+D}\qquad
U' \ =\ \Om_C(\lw W\rw_{\psi,C+D})
=\ \Om_C(\lw U\rw_{\psi,C} )\qquad
U'-U=\lw W'-W\rw_{\psi,D}
$$
By Corollary \corwicknorm
$$
N\big( U\cl 16\al\big)_\0= N\big( \lw W\rw_{\psi,D}\cl 16\al\big)_\0
\le  N\big( W\cl 32\al\big)_\0<\al^2
$$
so that by Corollary \corwicknorm, followed by Theorem \theorII\ (with
$W=U$ and $2\al$ replacing $\al$)
$$\eqalign{
N\big( W'-W\cl \al\big)
&\le N\big( U'-U\cl 2\al\big)
\le\sfrac{1}{2\al^2}\,
     \sfrac{N( U\cl 16\al)^2}{1-{1\over\al^2}N( U\cl 16\al)} 
\le\sfrac{1}{2\al^2}\,
     \sfrac{N( W\cl 32\al)^2}{1-{1\over\al^2}N( W\cl 32\al)}
}$$
\endproof
\remark{\STM\remjointanalyticitywick}{
Suppose that $\|\ \cdot\ \|_{\ib}$ and $\|\ \cdot\ \|_{\ib\cb}$ are norms
 on the space of antisymmetric bilinear forms on $V$ and that there are 
$\ka,\ka'>0$ such that every $C$ 
with $\|C\|_{\ib\cb}<\ka$ has integral bound $\ib$ and contraction bound $\cb_\0+\sum_{\de\ne 0}\infty t^\de$ and every $D$ 
with $\|D\|_{\ib}<\ka'$ has integral bound $\ib$. 
Then $W'$ is jointly analytic in $C$, $D$ and $W$ on
$$
\set{(W,C,D)}{W\hbox{ even},\  
N\big(W\cl 32\al \big)_\0 <\al^2,\ \|C\|_{\ib\cb}<\ka,\ \|D\|_{\ib}<\ka'}
$$

}

The derivatives of $\Om_C(\lw W\rw_{\psi,C+D})$ with respect to
$W$, $C$ and $D$ are bounded in the following Theorem, which
is an amalgam of Lemmas \:\lemprftwoA, \:\lemprftwoB\ and 
\:\lemprftwoC\ below.

\theorem{\STM\theoremIVb}{
Let, for $\ka$ in a neighbourhood of $0$, $W_\ka(\psi)$ be an even Grassmann 
function and $C_\ka,D_\ka$ be covariances on $V$. Assume that $\al\ge 1$ and 
$$
N(W_0\cl 32\al)_\0<\al^2
$$
and that
$$\meqalign{
&C_0\ {\rm has\ contraction\ bound} \ \cb \quad&&\quad
&\sfrac{1}{2}\ib \ {\rm is\ an\ integral\ bound\ for\ } C_0,D_0 \cr
&\sfrac{d\hfill}{d\ka}C_\ka\big|_{\ka=0}\ {\rm has\ contraction\ bound} \ \cb' \quad&&\quad
&\sfrac{1}{2}\ib' \ {\rm is\ an\ integral\ bound\ for\ } \sfrac{d\hfill}{d\ka}D_\ka\big|_{\ka=0} \cr
}$$
and that $\cb\le\sfrac{1}{\mu}\cb^2$.
Set
$$\eqalign{
\lw \tilde W_\ka(\psi)\rw_{\psi,D_\ka}\ 
&=\ \Om_{C_\ka}(\lw W_\ka \rw_{\psi,C_\ka+D_\ka})\cr
}$$
Then
$$\eqalign{
N\big(\,\sfrac{d\hfill}{d\ka}[\tilde W_\ka-W_\ka]_{\ka=0}\,\cl \al\big)
&\le\ \sfrac{1}{2\al^2}\,
     \sfrac{N( W_0\cl 32\al)}{1-{1\over\al^2}N( W_0\cl 32\al)}
N\big(\,\sfrac{d\hfill}{d\ka}W_\ka\big|_{\ka=0}\,\cl 8\al\big) \cr
&\hskip1cm+\sfrac{1}{2\al^2}\,
     \sfrac{N( W_0\cl 32\al)^2}{1-{1\over\al^2}N( W_0\cl 32\al)}
\Big\{\sfrac{1}{4\mu}\cb'+\big(\sfrac{\ib'}{\ib}\big)^2\Big\}
}$$
}

\lemma{\STM\lemprftwoA}{ 
Let $C$ be a covariance on $V$ with contraction bound $\cb$ and integral bound
$\ib$. Let, for $\ka$ in a neighbourhood of $0$,
 $W_\ka(\psi) \in \bigwedge_A V'$ be an even Grassmann function.
\Item i)
Set
$$
\tilde W_\ka(\psi)\ =\ \Om_C(\lw W_\ka\rw_{\psi,C}) 
$$
If $\ 
N\big(W_0\cl 8\al \big)_\0 < \sfrac{\al^2}{4}
\ $, 
then
$$\eqalign{
N\big(\,\sfrac{d\hfill}{d\ka}[\tilde W_\ka-W_\ka]_{\ka=0}\,\cl \al\big)
\le\ \sfrac{2}{\al^2}\,
     \sfrac{N( W_0\cl 8\al)}{1-{4\over\al^2}N( W_0\cl 8\al)}
N\big(\,\sfrac{d\hfill}{d\ka}W_\ka\big|_{\ka=0}\,\cl 2\al\big) 
}$$
\Item ii) Let $D$ be a covariance on $V$ with integral bound $\ib$.
Set
$$
\lw \tilde W_\ka(\psi)\rw_{\psi,D}\ =\ \Om_C(\lw W_\ka\rw_{\psi,C+D})
$$
If $\ 
N\big(W_0\cl 32\al \big)_\0 < \al^2
\ $, 
then
$$\eqalign{
N\big(\,\sfrac{d\hfill}{d\ka}[\tilde W_\ka-W_\ka]_{\ka=0}\,\cl \al\big)
\le\ \sfrac{1}{2\al^2}\,
     \sfrac{N( W_0\cl 32\al)}{1-{1\over\al^2}N( W_0\cl 32\al)}
N\big(\,\sfrac{d\hfill}{d\ka}W_\ka\big|_{\ka=0}\,\cl 8\al\big) 
}$$
}
\prf
Set
$$\meqalign{
\null\hskip.75inU_\ka(\psi,\xi) &= W_\ka(\psi+\xi)  &&
U'_\ka(\psi,\xi) &= \sfrac{d\hfill}{d\ka}W_\ka(\psi+\xi)  \cr
\noalign{\noindent By Proposition \:\propBII.ii,}
\lw U_\ka(\psi,\xi)\rw_\xi &= \ \lw W_\ka\rw(\psi+\xi)\qquad &&
\lw U'_\ka(\psi,\xi)\rw_\xi &= \ \lW \sfrac{d\hfill}{d\ka}W_\ka\rW(\psi+\xi) \cr
\noalign{\noindent By Remark \remfunctnorm,}
N\big(U_\ka\cl \al\big) &\le N\big(W_\ka\cl 2\al\big) &&
N\big(U'_\ka\cl\al\big) &\le N\big(\sfrac{d\hfill}{d\ka}W_\ka\cl 2\al\big)\cr
}$$
Differentiating Definition \defrengroup,
$$
\sfrac{d\hfill}{d\ka}\Om_C\big(\lw W_\ka\rw\big) 
=\frac{\int \lW\sfrac{d\hfill}{d\ka} W_\ka\rW(\psi +\xi)\ e^{\lw W_\ka\rw(\psi +\xi)} d\mu_C(\xi) }
{\int e^{\lw W_\ka\rw(\psi +\xi)} d\mu_C(\xi) } 
=\cS_{\lw U_\ka\rw, C}(\lw U'_\ka\rw)
\qquad {\rm mod}\, A_0 
$$
so that
$$
\sfrac{d\hfill}{d\ka}\big[\Om_C\big(\lw W_\ka\rw\big) -W_\ka\big]
=\cS_{\lw U_\ka\rw, C}(\lw U'_\ka\rw)-U'_\ka(\psi,0)
\qquad {\rm mod}\, A_0 
$$
Define the system of symmetric seminorms $\|\ \cdot\ \|'$ and the norm
$N'\big( f\cl \al \big)$ as in the proof of Theorem  \theorII.
 Proposition \propnormestcalS\ now implies that
$$\eqalign{
N\big(\sfrac{d\hfill}{d\ka}[\tilde W_\ka-W_\ka]\,\cl \al)
&=N'\big(\sfrac{d\hfill}{d\ka}[\tilde W_\ka-W_\ka]\,\cl \al)  \cr
&= N'\big( \cS_{\lw  U_\ka\rw, C}(\lw U'_\ka\rw )- U'_\ka(\psi,0)\,\cl \al\big) 
              \cr
&\le \sfrac{2}{\al^2}\,
     \sfrac{N'( U_\ka\cl 4\al)}{1-{4\over\al^2}N'(U_\ka\cl 4\al)} 
     \, N'( U'_\ka\cl \al )\cr
& \le \sfrac{2}{\al^2}\,
     \sfrac{N( U_\ka\cl 4\al)}{1-{4\over\al^2}N(U_\ka\cl 4\al)} 
     \, N( U'_\ka\cl \al ) \cr
& \le \sfrac{2}{\al^2}\,
     \sfrac{N( W_\ka\cl 8\al)}
     {1-{4\over\al^2}N( W_\ka\cl 8\al)} 
     \, N\big(\sfrac{d\hfill}{d\ka}W_\ka\cl 2\al\big)\cr
}$$
 The hypotheses of Proposition \propnormestcalS\ are fulfilled, at $\ka=0$,
since, by hypothesis
$$
N\big( U_0 \cl 4\al \big)_\0 \
\le \ N\big(W_0\cl 8\al \big)_\0
 < \sfrac{\al^2}{4} 
$$
\Item ii)
Part (ii) follows from part (i) as Theorem \theoremIVa\  
follows from Theorem \theorII.
\endproof

\lemma{\STM\lemfcf}{
Let $\cb$ be a contraction bound for $C$ and $\cb'$ be a contraction bound 
for $C'$. If $\cb\le\sfrac{1}{\mu}\cb^2$, then
$$
N\big(\smsum_{i,j}\sfrac{\partial f}{\partial\xi_i}\,C'_{ij}
\,\sfrac{\partial g}{\partial\xi_j}\cl \al\big)
\le\sfrac{1}{\mu\al^2}\cb' \ N(f\cl 2\al)\,N(g\cl 2\al)
$$
}
\prf Write
$$
f(\xi) =\sum_{m, n\ge 0} f_{m;n}(\xi)\qquad
g(\xi) =\sum_{m', n'\ge 0} g_{m';n'}(\xi)
$$
with $f_{m;n} \in A_m[n]$ and $g_{m';n'} \in A_{m'}[n']$. Then, by Lemma
\lemGrasscontract\ and Definition \defGrasscontract,
$$
\smsum_{i,j}\sfrac{\partial f_{m;n}}{\partial\xi_i}\,C'_{ij}
\,\sfrac{\partial g_{m';n'}}{\partial\xi_j}
=-n\cont{\xi}{\ze}{C'}\big(f_{m;n}(\xi)g_{m';n'}(\ze)\big)\Big|_{\ze=\xi}
=-nn'\Cont{1}{n+1}{C'}\big(f_{m;n}\otimes g_{m';n'}\big)\Big|_{\ze=\xi}
$$
so that, by Lemma \lemGrasscompatnorm.ii and Definition \defcontractintbound,
$$\eqalign{
\Big\|\smsum_{i,j}\sfrac{\partial f_{m;n}}{\partial\xi_i}\,C'_{ij}
\,\sfrac{\partial g_{m';n'}}{\partial\xi_j}\Big\|
\le nn'\Big\|\Cont{1}{n+1}{C'}\big(f_{m;n}\otimes g_{m';n'}\big)\Big\|
\le nn'\cb'\ \big\|f_{m;n}\|\,\big\| g_{m';n'}\big\|
}$$
Hence, by Definition \deffunctnorm,
$$\eqalign{
N\big(\smsum_{i,j}\sfrac{\partial f}{\partial\xi_i}\,C'_{ij}
\,\sfrac{\partial g}{\partial\xi_j}\cl \al\big)
&\le \sfrac{1}{\ib^2}\cb\sum_{m,m',n,n'\ge 0}(\al\ib)^{n+n'-2}
\Big\|\smsum_{i,j}\sfrac{\partial f_{m;n}}{\partial\xi_i}\,C'_{ij}
\,\sfrac{\partial g_{m';n'}}{\partial\xi_j}\Big\|\cr
&\le \sfrac{1}{\mu\al^2}\cb'
\bigg[\sfrac{1}{\ib^2}\cb\smsum_{m,n,\ge 0}n(\al\ib)^{n}\big\|f_{m;n}\|\bigg]\,
\bigg[\sfrac{1}{\ib^2}\cb
           \smsum_{m',n',\ge 0}n'(\al\ib)^{n'}\big\|g_{m';n'}\|\bigg]\cr
&\le \sfrac{1}{\mu\al^2}\cb'
\bigg[\sfrac{1}{\ib^2}\cb\smsum_{m,n,\ge 0}(2\al\ib)^{n}\big\|f_{m;n}\|\bigg]\,
\bigg[\sfrac{1}{\ib^2}\cb
           \smsum_{m',n',\ge 0}(2\al\ib)^{n'}\big\|g_{m';n'}\|\bigg]\cr
&=\sfrac{1}{\mu\al^2}\cb' \ N(f\cl 2\al)\,N(g\cl 2\al)\cr
}$$
\endproof

\lemma{\STM\lemprftwoB}{ 
Let, for $\ka$ in a neighbourhood of $0$, $C_\ka$ be a covariance on $V$. Assume
that $C_0$ has contraction bound $\cb$ and integral bound $\ib$, that
$\sfrac{d\hfill}{d\ka}C_\ka\big|_{\ka=0}$  has contraction bound $\cb'$ and that
$\cb\le\sfrac{1}{\mu}\cb^2$. Let $W(\psi) \in \bigwedge_A V'$ be an even 
Grassmann function.
\Item i)
Set
$$
\tilde W_\ka(\psi)\ =\ \Om_{C_\ka}(\lw W\rw_{\psi,C_\ka}) 
$$
If $\ 
N\big(W\cl 8\al \big)_\0 < \sfrac{\al^2}{4}
\ $, 
then
$$\eqalign{
N\big(\,\sfrac{d\hfill}{d\ka}\tilde W_\ka\big|_{\ka=0}\,\cl \al\big)
\le \sfrac{1}{2\al^2}\,
     \sfrac{N( W\cl 8\al)^2}{1-{4\over\al^2}N( W\cl 8\al)}\ \sfrac{1}{\mu}\cb' 
}$$
\Item ii) Let $D$ be a covariance on $V$ with integral bound $\ib$.
Set
$$
\lw \tilde W_\ka(\psi)\rw_{\psi,D}\ =\ \Om_{C_\ka}(\lw W\rw_{\psi,C_\ka+D})
$$
If $\ 
N\big(W\cl 32\al \big)_\0 < \al^2
\ $, 
then
$$\eqalign{
N\big(\,\sfrac{d\hfill}{d\ka}\tilde W_\ka\big|_{\ka=0}\,\cl \al\big)
\le\sfrac{1}{8\al^2}\,
     \sfrac{N( W\cl 32\al)^2}{1-{1\over\al^2}N( W\cl 32\al)}\ \sfrac{1}{\mu}\cb'
}$$
}
\prf
Set
$\ 
U(\psi,\xi) = W(\psi+\xi)
\ $. By Proposition \:\propBII.ii,
$\ 
\lw U(\psi,\xi)\rw_{\xi,C_\ka} = \ \lw W\rw_{C_\ka}(\psi+\xi)
\ $ and by Remark \remfunctnorm,
$\ 
N\big(U\cl \al\big) \le N\big(W\cl 2\al\big)\ 
$.
By Lemma \:\lemCovderiv.iv, setting $C'(\ka)=\sfrac{d\hfill}{d\ka}C_\ka$,
$$\eqalign{
\sfrac{d\hfill}{d\ka}\tilde W_\ka 
&=\frac{\sfrac{d\hfill}{d\ka}\int e^{\lw W\rw_{C_\ka}(\psi +\xi)} d\mu_{C_\ka}(\xi) }
{\int e^{\lw W\rw_{C_\ka}(\psi +\xi)} d\mu_{C_\ka}\xi) } \qquad {\rm mod}\, A_0 \cr
&=\frac{-{1\over 2}  \int\big[\sum_{i,j}
\lW\sfrac{\partial U}{\partial\xi_i}\rW_{\xi,C_t}C'_{ij}(\ka)
\ \lW\sfrac{\partial U}{\partial\xi_j}\rW_{\xi,C_\ka}\big]
e^{\lw U(\psi,\xi)\rw_{\xi,C_\ka}}\  d\mu_{C_\ka}(\xi) }
{\int e^{\lw U(\psi,\xi)\rw_{\xi,C_\ka}} d\mu_{C_\ka}\xi) } \cr
&=-\half \cS_{\lw U\rw_{\xi,C_\ka}, C_\ka}\Big(\smsum_{i,j}
\lW\sfrac{\partial U}{\partial\xi_i}\rW_{\xi,C_\ka}C'_{ij}(\ka)
\ \lW\sfrac{\partial U}{\partial\xi_j}\rW_{\xi,C_\ka}\Big)\cr
&= \cS_{\lw U\rw_{\xi,C_\ka}, C_\ka}(\lw V_\ka\rw_{\xi,C_\ka})
}$$
where
$$
\lw V_\ka\rw_{\xi,C_\ka}
=-\half\smsum_{i,j}
\lW\sfrac{\partial U}{\partial\xi_i}\rW_{\xi,C_\ka}C'_{ij}(\ka)
\ \lW\sfrac{\partial U}{\partial\xi_j}\rW_{\xi,C_\ka}
$$
As in Lemma \lemprftwoB,
 Proposition \propnormestcalS\ now implies that
$$\eqalign{
N\big(\sfrac{d\hfill}{d\ka}\tilde W_\ka\big|_{\ka=0}\,\cl \al)
&=N'\big(\sfrac{d\hfill}{d\ka}\tilde W_\ka\big|_{\ka=0}\,\cl \al)  \cr
&\le N'\big(\sfrac{d\hfill}{d\ka}\tilde W_\ka\big|_{\ka=0}-V_0(\psi,0)\,\cl \al)
                 +N'\big(V_0(\psi,0)\,\cl \al)  \cr
& =
    N'\big( \cS_{\lw  U\rw_{\xi,C_0}, C_0}(\lw V_0\rw_{\xi,C_0} )- V_0(\psi,0) \big)
     +N'\big(V_0(\psi,0)\,\cl \al) \cr
&\le \sfrac{2}{\al^2}\,
     \sfrac{N'( U\cl 4\al)}{1-{4\over\al^2}N'(U\cl 4\al)}\, N'( V_0\cl \al )
      +N'\big(V_0\,\cl \al)\cr
& \le \sfrac{4}{\al^2}\,
     \sfrac{N( U\cl 4\al)}{1-{4\over\al^2}N(U\cl 4\al)} \, N( V_0\cl \al )
     +N( V_0\cl \al ) \cr
& \le  \sfrac{1}{1-{4\over\al^2}N(U\cl 4\al)} \, N( V_0\cl \al ) \cr
& \le \sfrac{1} {1-{4\over\al^2}N( W\cl 8\al)} \, N\big(V_0\cl \al\big)\cr
}$$
By Proposition \propBII.i (three times)
$$\eqalign{
&V_0(\psi,\xi)
=\int \lw V_0\rw_{\xi,C_0}(\psi,\xi+\xi')\ d\mu_{C_0}(\xi')\cr
&\hskip.2cm=-\half\sum_{i,j}\int
\sfrac{\partial U}{\partial\xi_i}(\psi,\xi+\xi'+\ze)\ C'_{ij}(0)
\ \sfrac{\partial U}{\partial\xi_j}(\psi,\xi+\xi'+\ze')\ 
d\mu_{-C_0}(\ze)\ d\mu_{-C_0}(\ze')\ d\mu_{C_0}(\xi')\cr
&\hskip.2cm=-\half\sum_{i,j}\int
\sfrac{\partial W}{\partial\xi_i}(\psi+\xi+\xi'+\ze)\ C'_{ij}(0)
\ \sfrac{\partial W}{\partial\xi_j}(\psi+\xi+\xi'+\ze')\ 
d\mu_{-C_0}(\ze)\ d\mu_{-C_0}(\ze')\ d\mu_{C_0}(\xi')\cr
}$$
By Lemma \lemwicknorm, with $s=0$, followed by Lemma \lemfcf\ and then
Remark \remfunctnorm,
$$\eqalign{
N\big(V_0\cl \al\big)
&\le\half\sum_{i,j}  
N\Big(\sfrac{\partial W}{\partial\xi_i}(\psi+\xi+\xi'+\ze)\ C'_{ij}(0)
\ \sfrac{\partial W}{\partial\xi_j}(\psi+\xi+\xi'+\ze')\ \cl \al\Big)\cr
&\le\sfrac{1}{2\mu\al^2}\cb'N\big(W(\psi+\xi+\xi'+\ze)\,\cl 2\al\big)
N\big(W(\psi+\xi+\xi'+\ze')\, \cl 2\al\big)\cr
&\le\sfrac{1}{2\mu\al^2}\cb'N\big(W \cl 8\al\big)^2\cr
}$$ 
\Item ii)
Part (ii) follows from part (i) as Theorem \theoremIVa\  
follows from Theorem \theorII.
\endproof

\lemma{\STM\lemprftwoC}{ 
Let $C$ and, for $\ka$ in a neighbourhood of $0$, $D_\ka$ be covariances on $V$. 
Assume that $C$ has contraction bound $\cb$ and integral bound $\half\ib$, that
$D_0$ has integral bound $\half\ib$ and that $\sfrac{d\hfill}{d\ka}D_\ka\big|_{\ka=0}$
has integral bound $\half\ib'$. Let $W(\psi) \in \bigwedge_A V'$ be an even 
Grassmann function.
Set
$$
\lw \tilde W_\ka(\psi)\rw_{\psi,D_\ka}\ =\ \Om_{C}(\lw W\rw_{\psi,C+D_\ka})
$$
If $\ 
N\big(W\cl 32\al \big)_\0 < \al^2
\ $, 
then
$$\eqalign{
N\big(\,\sfrac{d\hfill}{d\ka}\tilde W_\ka\big|_{\ka=0}\,\cl \al\big)
\le\ \sfrac{1}{2\al^2}\,\sfrac{N( W\cl 32\al)^2}
     {1-{1\over\al^2}N( W\cl 32\al)}\big(\sfrac{\ib'}{\ib}\big)^2
}$$
}
\prf
Set
$\ 
E_z=D_0+z\sfrac{d\hfill}{d\ka}D_\ka\big|_{\ka=0}
\ $
and define $V_z$ by
$$
\lw V_z(\psi)\rw_{\psi,E_z}\ =\ \Om_{C}(\lw W\rw_{\psi,C+E_z})
$$
As $D_\ka$ and $E_\ka$ agree to order $\ka$, 
$\sfrac{d\hfill}{d\ka}\tilde W_\ka\big|_{\ka=0}=\sfrac{d\hfill}{dz}V_z\big|_{z=0}$.
Set $\veps_d=\big(\sfrac{\ib'}{\ib}\big)^2$. Then, by Remark \scaleIntConsts, 
$$
\sfrac{1}{2}\ib+\sfrac{1}{2}\sqrt{|z|}\,\ib'\le \ib
$$
is an integral bound for  $E_z$  for all $|z|\le\sfrac{1}{\veps_d}$.
By Theorem \theoremIVa 
$$\eqalign{
N\big( V_z-W\cl \al\big) 
&\le \sfrac{1}{2\al^2}\,
     \sfrac{N( W\cl 32\al)^2}
     {1-{1\over\al^2}N( W\cl 32\al)}\cr
}$$
for all  $|z|\le\sfrac{1}{\veps_d}$.
By the Cauchy integral formula, if $f(z)$ is analytic and bounded in absolute
value by $Q$ on $|z|\le r$, then
$$
f'(z) =\sfrac{1}{2\pi \imath}\int_{|\ze|=r} \sfrac{f(\ze)}{(\ze-z)^2} d\ze
$$
and
$$
\big|f'(0)\big| \le \sfrac{1}{2\pi}\sfrac{Q}{r^2} 2\pi r= Q\sfrac{1}{r}
$$
Hence
$$
N\big(\,\sfrac{d\hfill}{d\ka}\tilde W_\ka\big|_{\ka=0}\,\cl \al\big)
=N\big(\sfrac{d\hfill}{d z}\big[ V_z-W\big]_{z=0}\cl \al\big) 
\le \sfrac{\veps_d}{2\al^2}\,\sfrac{N( W\cl 32\al)^2}
     {1-{1\over\al^2}N( W\cl 32\al)}
= \sfrac{1}{2\al^2}\,\sfrac{N( W\cl 32\al)^2}
     {1-{1\over\al^2}N( W\cl 32\al)}\big(\sfrac{\ib'}{\ib}\big)^2
$$
\endproof

\vfill\eject

\appendix{A}{ Wick--Ordering}\PG\pgRA

Let $A$ be a superalgebra, $V$ be a complex vector space and $C$ an
antisymmetric bilinear form (covariance) on $V$.
Wick ordering with respect to a covariance
$C$ is the $A$--linear map on $\bigwedge\nolimits_A V$,
$$
f(\xi) \mapsto \ \lw f(\xi)\rw_{\xi,C} 
$$
characterized by
$$
 \lw e^{\Sigma\, \xi_i\ze_i}\rw_{\xi,C}\ = \ 
e^{1/2\,\Sigma\, \ze_i C_{ij} \ze_j}\,e^{\Sigma\, \xi_i\ze_i}
$$
for any set $\{\ze_i\}$ of odd Grassmann variables.
If the context admits, we delete the Wick ordering covariance $C$ or the variable $\xi$ (or both) from the symbol $\lw \,\cdot\,\rw_{\xi,C}$ for Wick 
ordering. Also recall that the Grassmann Gaussian integral with covariance $C$
is characterized by
$$
\int e^{\Sigma\, \xi_i\ze_i}\, d\mu_C(\xi) = e^{-1/2\,\Sigma\, \ze_i C_{ij} 
\ze_j}
$$

\lemma{\STM\lemBI}{
For $n,m\ge 0$
$$
\int (\lw \xi_{i_1}\xi_{i_2}\cdots \xi_{i_n}\rw) \, 
(\lw \xi_{j_m}\xi_{j_{m-1}}\cdots \xi_{j_1}\rw)\,d\mu_C(\xi)
\ = \cases{  \det \big(C_{i_kj_\ell}\big) & if $m=n$ \cr
            \   0 & if $m\ne n$ \cr}
$$
}
\prf This follows easily from the definitions by induction. See, for example,
Proposition I.31 of [FKTffi].
\endproof

Choose a copy of $V$ with generating system $(\xi_i')$ corresponding to the 
generating system $(\xi_i)$ of $V$.

\proposition{\STM\propBII}{  
\Item{(i)} Let $f(\xi) = \lw \hat f(\xi)\rw_\xi\ \in \bigwedge\nolimits_A V\ $. Then
$$
f(\xi) = \int \hat f(\xi+i\xi')\,d\mu_C(\xi')
= \int \hat f(\xi+\xi')\,d\mu_{-C}(\xi') \quad,\quad
\hat f(\xi) = \int  f(\xi+\xi')\,d\mu_C(\xi')
$$
In particular,  $f(0) = \int \hat f(i\xi)\,d\mu_C(\xi)= \int \hat f(\xi)\,d\mu_{-C}(\xi)$.
\Item{(ii)}
Let $f(\xi) = \lw \hat f(\xi)\rw_\xi\ \in \bigwedge\nolimits_A V\ $. Then 
$$
f(\xi+\xi') =\  \lw \hat f(\xi+\xi') \rw_\xi\ =\ \lw \hat f(\xi+\xi') \rw_{\xi'}
$$
\Item{(iii)}
For $f_1(\xi),\cdots,f_\ell(\xi)\in\bigwedge_AV$
$$
\lw f_1(\xi)\rw_\xi \cdots\lw f_\ell(\xi)\rw_\xi\ = \
\lW \int \lw f_1(\xi+\xi')\rw_{\xi'} \cdots \lw f_\ell(\xi+\xi')\rw_{\xi'} \,d\mu_C(\xi')\rW_\xi
$$
\Item{(iv)}
For $f(\xi) \in\bigwedge_AV$
$$
e^{\lw f(\xi)\rw_\xi} =\ \lW \int e^{\lw f(\xi+\xi')\rw_{\xi'} } d\mu_C(\xi') \rW_\xi
$$
}

\prf 
\Item{(i)} It suffices to prove the identities for 
$\hat f(\xi)=e^{\Sigma\, \xi_j\ze_j}$.
In this case
$$
\int \hat f(\xi+i\xi')\,d\mu_C(\xi') 
= \int e^{\Sigma\, \xi_j\ze_j}\,e^{i\Sigma\, \xi'_j\ze_j}\,d\mu_C(\xi') 
=  e^{\Sigma\, \xi_j\ze_j}\,e^{1/2\,\Sigma\, \ze_iC_{ij}\ze_j} 
= \ \lw \hat f(\xi)\rw_C\ =\ f(\xi)
$$
and
$$
\int \hat f(\xi+\xi')\,d\mu_{-C}(\xi') 
= \int e^{\Sigma\, \xi_j\ze_j}\,e^{\Sigma\, \xi'_j\ze_j}\,d\mu_{-C}(\xi') 
=  e^{\Sigma\, \xi_j\ze_j}\,e^{1/2\,\Sigma\, \ze_iC_{ij}\ze_j} 
= \ \lw \hat f(\xi)\rw_C\ =\ f(\xi)
$$
The statement about $\hat f(\xi)$ is proven in the same way, and the formula for
$f(0)$ follows immediately.
\Item{(ii)} Again we may assume that $\hat f(\xi) = e^{\Sigma\, \xi_i \ze_i}$. Then
$$\eqalign{
f(\xi) &= e^{1/2\, \Sigma\, \ze_i C_{ij} \ze_j } 
      \, e^{\Sigma \xi_i \ze_i} \cr
\hat f(\xi+\xi') &= e^{\Sigma\, \xi_i \ze_i}\, e^{\Sigma \,\xi'_i \ze_i}\cr
}$$
and
$$\eqalign{
\lw \hat f(\xi+\xi')\rw_\xi\  
&= e^{\half \Sigma\, \ze_i C_{ij}\ze_j +\Sigma\, \xi_i \ze_i}\, 
   e^{\Sigma\, \xi'_i \ze_i} \cr
&= e^{1/2\, \Sigma\, \ze_i C_{ij}\ze_j +\Sigma\, (\xi_i+\xi'_i) \ze_i} \cr
&= f(\xi+\xi')
}$$
Similarly one shows that $\lw \hat f(\xi+\xi')\rw_{\xi'}\ =f(\xi+\xi')$.
\Item{(iii)}
Let
$$
\lw g(\xi)\rw \ =\ \lw f_1(\xi)\rw \cdots\lw f_\ell(\xi)\rw
$$
By part (ii)
$$
\lw g(\xi+\xi')\rw_{\xi'}\ =\ \lw f_1(\xi+\xi')\rw_{\xi'} \cdots\lw f_\ell(\xi+\xi')\rw_{\xi'}
$$
Therefore
$$\eqalign{
g(\xi) &= \int \lw g(\xi+\xi')\rw_{\xi'} \,d\mu_C(\xi') \cr
&= \int \lw f_1(\xi+\xi')\rw_{\xi'} \cdots\lw f_\ell(\xi+\xi')\rw_{\xi'} \,d\mu_C(\xi')
}$$
\Item{(iv)} By part (iii)
$$\eqalign{
e^{\lw f(\xi)\rw} 
&= \smsum_{\ell=0}^\infty \sfrac{1}{\ell!}\,\big( \lw  f(\xi)\rw\big)^\ell \cr
&= \smsum_{\ell=0}^\infty \sfrac{1}{\ell!}\,
    \lW \int \big( \lw  f(\xi+\xi')\rw_{\xi'}\,\big)^\ell d\mu_C(\xi') \rW_\xi \cr
&=  \ \lW  \int e^{\lw   f(\xi+\xi')\rw_{\xi'} } d\mu_C(\xi') \rW_\xi \cr
}$$
\endproof

\corollary{\STM\corBIII}{
For $f_1(\xi),\cdots,f_\ell(\xi),g(\xi)\in\bigwedge_AV$
$$
\lw f_1(\xi)\rw_\xi \cdots\lw f_\ell(\xi)\rw_\xi\ \lw g(\xi)\rw_\xi\ = \
\lww \int \hskip-6pt \int \hskip -4pt
\lW  \big( \smprod_{i=1}^\ell \lw f_i(\xi+\xi'+\eta)\rw_{\xi'} \big)\rW_\eta\ 
\lw g(\xi+\eta)\rw_\eta  \,d\mu_C(\xi')\,d\mu_C(\eta)\rww_\xi
$$
}

\prf By iterated application of part (iii) of Proposition \propBII
$$\eqalign{
\lw f_1(\xi)\rw_\xi \cdots\lw &f_\ell(\xi)\rw_\xi\ \lw g(\xi)\rw_\xi\ 
 =\ \lW  \int \smprod_{i=1}^\ell \lw f_i(\xi+\xi')\rw_{\xi'}\,d\mu_C(\xi')\rW_\xi\ 
\lw g(\xi)\rw_\xi  \cr
&= \ \lww \int \lW  \int 
\smprod_{i=1}^\ell \lw f_i(\xi+\xi'+\eta)\rw_{\xi'}\,d\mu_C(\xi') \rW_\eta\ 
\lw g(\xi+\eta)\rw_\eta  \ d\mu_C(\eta) \rww_\xi
}$$
\endproof

\lemma{\STM\lemBIV}{Let $D$ be a second covariance on $V$, and let 
$f(\xi) \in\bigwedge_AV$.
$$\leqalignno{
\lw f(\xi)\rw_{\xi,C+D} \ &=\ \lW\lw f(\xi)\rw_{\xi,C}\rW_{\xi,D}&(i)\cr
\lw f\rw_{\,C+D}(\xi+\xi') \ 
     &= \ \lW \,\lw  f(\xi+\xi')\rw_{\xi,C}\,\rW_{\xi',D}&(ii)\cr
\lw f(\xi)\rw_{\xi,C+D} \ &=\ \lw f'(\xi)\rw_{\xi,C}&(iii)\cr
}$$
where
$$
f'(\xi) = \int f(\xi+i\xi')\,d\mu_D(\xi')= \int f(\xi+\xi')\,d\mu_{-D}(\xi')
$$
}

\prf Again we may assume that $f(\xi)=e^{\Sigma\, \xi_i\ze_i}$.
Set $\hat f(\xi)=\lw f(\xi)\rw_{\xi,C}
=e^{1/2\, \Sigma\, \ze_i C_{ij} \ze_j } \, e^{\Sigma\, \xi_i \ze_i}$.
\Item{(i)}
 $$
\ \lw \hat f(\xi)\rw_{\xi,D} 
\ = \  e^{1/2\, \Sigma\, \ze_i C_{ij} \ze_j } 
\, \lw e^{\Sigma\, \xi_i \ze_i}\rw_{\xi,D}
\ =\  e^{1/2\, \Sigma\, \ze_i (C_{ij}+D_{ij}) \ze_j } 
\, e^{\Sigma\, \xi_i \ze_i}
\ =\ \lw f(\xi)\rw_{\xi,C+D}
$$
\Item{(ii)}
$$\eqalign{
\lW \,\lw  f(\xi+\xi')\rw_{\xi,C}\,\rW_{\xi',D}\
&=\ e^{1/2\, \Sigma\, \ze_i C_{ij} \ze_j }\,
   \lw e^{\Sigma\, (\xi_i+\xi_i')\ze_i}\rw_{\xi',D} \cr
&=\ e^{1/2\, \Sigma\, \ze_i (C_{ij}+D_{ij}) \ze_j }\,
   e^{\Sigma\, (\xi_i+\xi_i')\ze_i} \cr
&=\ \lw  f\rw_{\,C+D}(\xi+\xi')
}$$
\Item{(iii)}
By part (i) and parts (i),(ii) of Proposition \propBII
$$\eqalign{
\lw f(\xi)\rw_{\xi,C+D}\ &=\ \lw \hat f(\xi)\rw_{\xi,D} 
   \ =\ \int \hat f(\xi+i\xi')\,d\mu_D(\xi')
   \ =\ \int \lw f(\xi+i\xi')\rw_{\xi,C}\,d\mu_D(\xi') \cr
&=\ \lW \int f(\xi+i\xi')\,d\mu_D(\xi')\,\rW_{\xi,C}
}$$
\endproof


\lemma{\STM\lemBV}{
Let $f(\xi),g(\xi),h(\xi) \in \bigwedge\nolimits_A V\ $. Then
$$
\int \lw f(\xi)g(\xi)\rw_\xi h(\xi)\,d\mu_C(\xi)
= \int \lw f(\xi)\rw_\xi 
\int \lw g(\xi')\rw_{\xi'}\ \lw h(\xi+\xi')\rw_\xi \,d\mu_C(\xi')
\,d\mu_C(\xi)
$$
}

\prf We may assume that
$$
f(\xi)=e^{\Sigma\, \xi_i\ze_i} \qquad\qquad
g(\xi)=e^{\Sigma\, \xi_i\ze_i'}  \qquad\qquad
h(\xi)=e^{\Sigma\, \xi_i\eta_i}
$$
with additional Grassmann variables $\ze_i,\ze_i',\eta_i$. Then
$$\eqalign{
\int \lw f(\xi)g(\xi)\rw_\xi h(\xi)\,d\mu_C(\xi)
&= \int e^{1/2\,\Si\,(\ze_i+\ze_i')C_{ij}(\ze_j+\ze_j')}\,
  e^{\Si\,\xi_i(\ze_i+\ze_i'+\eta_i)}\,d\mu_C(\xi) \cr
&= e^{ 1/2\,\Si\,[(\ze_i+\ze_i')C_{ij}(\ze_j+\ze_j')  
     - (\ze_i+\ze_i'+\eta_i)C_{ij}(\ze_j+\ze_j'+\eta_j)] } \cr
&= e^{ -1/2\,\Si\,\eta_i C_{ij} \eta_j }\,
   e^{ -\Si\,(\ze_i+\ze_i')C_{ij} \eta_j }\cr
}$$
On the other hand
$$\eqalign{
\int \lw g(\xi')\rw_{\xi'}\ \lw h(\xi+\xi')\rw_\xi \,d\mu_C(\xi')\ 
&=\ \lW  \int  e^{1/2\,\Si\,\ze_i'C_{ij}\ze_j'}\,
 e^{\Si\,\xi'_i(\ze_i'+\eta_i)}\,e^{\Sigma\, \xi_i\eta_i}\,d\mu_C(\xi') \rW_\xi\cr
&=\ \lW e^{ 1/2\,\Si\,[\ze_i'C_{ij}\ze_j' - (\ze_i'+\eta_i)C_{ij}(\ze_j'+\eta_j) }
 \,e^{\Sigma\, \xi_i\eta_i} \rW_\xi\cr
&=\ e^{ 1/2\,\Si\,[\ze_i'C_{ij}\ze_j'  + \eta_i C_{ij}\eta_j
   - (\ze_i'+\eta_i)C_{ij}(\ze_j'+\eta_j) }
 \,e^{\Sigma\, \xi_i\eta_i} \cr
&=\ e^{ -\Si\,\ze_i'C_{ij}\eta_j }\,e^{\Sigma\, \xi_i\eta_i} \cr
}$$
Therefore, also
$$\eqalign{
\int \lw f(\xi)\rw_\xi \int \lw g(\xi')\rw_{\xi'} \lw h(\xi+\xi')\rw_\xi &\,d\mu_C(\xi')\,d\mu_C(\xi) \cr
&= \int  e^{1/2\,\Si\,\ze_i C_{ij} \ze_j}\,
  e^{ -\Si\,\ze_i'C_{ij}\eta_j }\,e^{\Sigma\, \xi_i(\ze_i+\eta_i)}
  \,d\mu_C(\xi) \cr
&=e^{ 1/2\,[\Si\,\ze_i C_{ij} \ze_j - (\ze_i+\eta_i)C_{ij}(\ze_j+\eta_j)]}\,
  e^{ -\Si\,\ze_i'C_{ij}\eta_j } \cr
&= e^{ -1/2\,\Si\,\eta_i C_{ij} \eta_j }\,
   e^{ -\Si\,(\ze_i+\ze_i')C_{ij} \eta_j }\cr
}$$
\endproof
\corollary{\STM\corBVIa}{
Let $f(\xi),h(\xi) \in \bigwedge\nolimits_A V\ $. Then
$$
\int f(\xi) h(\xi)\,d\mu_C(\xi)
= \int  f(\xi)   \lww\int h(\xi+\xi') \,d\mu_C(\xi')\rww_\xi
\,d\mu_C(\xi)
$$
}
\prf Set $g(\xi)=1$ and replace $\lw f(\xi)\rw_\xi$ by $f(\xi)$ in
Lemma \lemBV.
\endproof

\lemma{\STM\lemBVI}{ 
Let $f(\xi)$ be a homogeneous Grassmann polynomial of degree two. Write
$$
f(\xi'+\xi^{\prime\prime}) = f(\xi')+f(\xi^{\prime\prime}) + 
  f_{\rm mix}(\xi',\xi^{\prime\prime})
$$
Furthermore let $g(\xi),h(\xi)\in \bigwedge_A V$. Then
$$\eqalign{
\int \lw f(\xi)\rw_\xi\,\lw g(\xi)h(\xi)&\rw_\xi \,d\mu_C(\xi)
=\int \hskip -5pt \int f_{\rm mix}(\xi',\xi^{\prime\prime})\,
    \lw g(\xi')\rw_{\xi'} \,\lw h(\xi^{\prime\prime})\rw_{\xi^{\prime\prime}}   \,d\mu_C(\xi')\,d\mu_C(\xi'') \cr
& + h(0)\int \lw f(\xi)\rw_\xi\,\lw g(\xi)\rw_\xi \,d\mu_C(\xi) 
   + g(0)\int \lw f(\xi)\rw_\xi\,\lw h(\xi)\rw_\xi \,d\mu_C(\xi) \cr
}$$
}
\prf 
By Lemma \lemBV\ and part (ii) of Proposition \propBII
$$\eqalign{
\int \lw f(\xi)\rw_\xi\,\lw &g(\xi)h(\xi)\rw_\xi \,d\mu_C(\xi)
= \int \hskip -5pt \int \lW \lw f(\xi'+\xi^{\prime\prime})\rw_{\xi'}\rW_{\xi''}
\ \lw g(\xi')\rw_{\xi'}\ \lw h(\xi^{\prime\prime})\rw_{\xi^{\prime\prime}} \,d\mu_C(\xi')\,d\mu_C(\xi'')  \cr
&= \int \hskip -5pt \int 
\lw \big[ f(\xi')+f(\xi^{\prime\prime}) + f_{\rm mix}(\xi',\xi^{\prime\prime})\big]
\rw_{\xi',\xi^{\prime\prime}}
\ \lw g(\xi')\rw_{\xi'}\ \lw h(\xi^{\prime\prime})\rw_{\xi^{\prime\prime}} \,d\mu_C(\xi')\,d\mu_C(\xi'')  \cr 
}$$
Since $f_{\rm mix}$ has degree one in $\xi'$ and in $\xi^{\prime\prime}$
$$\eqalign{
\int \hskip -5pt \int 
\lw  f_{\rm mix}(\xi',\xi^{\prime\prime})\rw_{\xi',\xi^{\prime\prime}}
\ \lw g(\xi')\rw_{\xi'}\ &\lw h(\xi^{\prime\prime})\rw_{\xi^{\prime\prime}} \,d\mu_C(\xi')\,d\mu_C(\xi'') \cr
&=\int \hskip -5pt \int f_{\rm mix}(\xi',\xi^{\prime\prime})\,
    \lw g(\xi')\rw_{\xi'} \,\lw h(\xi^{\prime\prime})\rw_{\xi^{\prime\prime}}
\,d\mu_C(\xi')\,d\mu_C(\xi'')    
}$$
Clearly
$$\eqalign{
\int \hskip -5pt \int 
\lw  f(\xi')\rw_{\xi',\xi^{\prime\prime}}
\ \lw g(\xi')\rw_{\xi'}\ \lw h(\xi^{\prime\prime})\rw_{\xi^{\prime\prime}} \,d\mu_C(\xi')\,d\mu_C(\xi'') 
&= h(0)\int \lw f(\xi)\rw_\xi\,\lw g(\xi)\rw_\xi \,d\mu_C(\xi) \cr
\int \hskip -5pt \int 
\lw  f(\xi^{\prime\prime})\rw_{\xi',\xi^{\prime\prime}}
\ \lw g(\xi')\rw_{\xi'}\ \lw h(\xi^{\prime\prime})\rw_{\xi^{\prime\prime}} \,d\mu_C(\xi')\,d\mu_C(\xi'') 
&= g(0)\int \lw f(\xi)\rw_\xi\,\lw h(\xi)\rw_\xi \,d\mu_C(\xi) \cr
}$$
\endproof

\lemma{\STM\lemCovderiv}{ 
Let $C(\ka)$ be a $C^1$ family of antisymmetric bilinear forms 
(covariances). Then
\Item{i)} 
$$\eqalign{
\sfrac{d\hfill}{d\ka}\int f(\xi)\  d\mu_{C(\ka)}(\xi) 
&=\half \sfrac{d^2\hfill}{dt^2} \dblint f(\xi+t\xi')\  d\mu_{C(\ka)}(\xi)
\  d\mu_{C'(\ka)}(\xi')\Big|_{t=0}\cr
&=-\half \sum_{i,j}C'_{ij}(\ka) \int\sfrac{\partial\hfill}{\partial\xi_i}
\sfrac{\partial\hfill}{\partial\xi_j}\,f(\xi)\  d\mu_{C(\ka)}(\xi)\cr
}$$
\Item{ii)}
$$\eqalign{
\sfrac{d\hfill}{d\ka}\lw f(\xi)\rw_{C(\ka)}
&=\half \sfrac{d^2\hfill}{dt^2} \lww\int f(\xi+t\xi')\  d\mu_{-C'(\ka)}(\xi)
\rww_{\xi,C(\ka)}\Big|_{t=0}\cr
&=\half \sum_{i,j}C'_{ij}(\ka) \lW\sfrac{\partial\hfill}{\partial\xi_i}
\sfrac{\partial\hfill}{\partial\xi_j}\,f(\xi)\rW_{C(\ka)}\cr
}$$
\Item{iii)} If $f$ is even,
$$\eqalign{
\sfrac{d\hfill}{d\ka}\int e^{f(\xi)}\  d\mu_{C(\ka)}(\xi) 
&=-\half \sum_{i,j}C'_{ij}(\ka) \int\Big[
\sfrac{\partial f}{\partial\xi_i}\sfrac{\partial f}{\partial\xi_j}+
\sfrac{\partial\hfill}{\partial\xi_i}\sfrac{\partial \hfill}{\partial\xi_j}
            f(\xi)\Big]
e^{f(\xi)}\  d\mu_{C(\ka)}(\xi)\cr
}$$
\Item{iv)} If $f$ is even,
$$
\sfrac{d\hfill}{d\ka}\int e^{\lw f(\xi)\rw_{C(\ka)}}\  d\mu_{C(\ka)}(\xi) 
=-\half \sum_{i,j}C'_{ij}(\ka) \int\Big[
\lW\sfrac{\partial f}{\partial\xi_i}\rW_{C(\ka)}
\ \lW\sfrac{\partial f}{\partial\xi_j}\rW_{C(\ka)}\Big]
e^{\lw f(\xi)\rw_{C(\ka)}}\  d\mu_{C(\ka)}(\xi)
$$

}

\prf 
\Item{i)} 
$$\eqalignno{
\sfrac{d\hfill}{d\ka}\int f(\xi)\  d\mu_{C(\ka)}(\xi)
&=\half\sfrac{d^2\hfill}{dt^2}\int f(\xi)\  d\mu_{C(\ka+t^2)}(\xi)\Big|_{t=0}\cr
&=\half\sfrac{d^2\hfill}{dt^2}\int f(\xi)\  d\mu_{C(\ka)+t^2C'(\ka)}(\xi)\Big|_{t=0}
\cr
&=\half\sfrac{d^2\hfill}{dt^2}\dblint f(\xi+\xi')\  d\mu_{C(\ka)}(\xi)\,d\mu_{t^2C'(\ka)}(\xi')\Big|_{t=0}
\cr
&=\half\sfrac{d^2\hfill}{dt^2}\dblint f(\xi+t\xi')\  d\mu_{C(\ka)}(\xi)\,d\mu_{C'(\ka)}(\xi')\Big|_{t=0}
\cr
&=-\half \sum_{i,j} \dblint\xi'_i\xi'_j\sfrac{\partial\hfill}{\partial\xi_i}
\sfrac{\partial\hfill}{\partial\xi_j}\,f(\xi)
\ d\mu_{C(\ka)}(\xi)\,d\mu_{C'(\ka)}(\xi')\cr
&=-\half \sum_{i,j}C'_{ij}(\ka) \int\sfrac{\partial\hfill}{\partial\xi_i}
\sfrac{\partial\hfill}{\partial\xi_j}\,f(\xi)\  d\mu_{C(\ka)}(\xi)\cr
}$$

\Item{ii)} 
$$\eqalignno{
\sfrac{d\hfill}{d\ka}\lW f(\xi)\rW_{C(\ka)}
&=\sfrac{d\hfill}{d\ka}\int f(\xi+\xi'')\,d\mu_{-C(\ka)}(\xi'')\cr
&=\half\sfrac{d^2\hfill}{dt^2}\dblint f(\xi+\xi''+t\xi')\  d\mu_{-C(\ka)}(\xi'')\,d\mu_{-C'(\ka)}(\xi')\Big|_{t=0}
\cr
&=\int\Big\{\half\sfrac{d^2\hfill}{dt^2}\int f(\xi+t\xi'+\xi'')\  d\mu_{-C'(\ka)}(\xi')\Big|_{t=0}\Big\}\
d\mu_{-C(\ka)}(\xi'')\cr
&=\lww\half\sfrac{d^2\hfill}{dt^2}\int f(\xi+t\xi')\  d\mu_{-C'(\ka)}(\xi')\Big|_{t=0}\rww_{C(\ka)}\cr
&=\half \sum_{i,j}C'_{ij}(\ka) \lW\sfrac{\partial\hfill}{\partial\xi_i}
\sfrac{\partial\hfill}{\partial\xi_j}\,f(\xi)\rW_{C(\ka)}\cr
}$$

\Item iii)
By the first part
$$\eqalign{
\sfrac{d\hfill}{d\ka}\int e^{f(\xi)}\  d\mu_{C(\ka)}(\xi) 
&=-\half \sum_{i,j}C'_{ij}(\ka) \int\sfrac{\partial\hfill}{\partial\xi_i}
\sfrac{\partial\hfill}{\partial\xi_j}\,e^{f(\xi)}\  d\mu_{C(\ka)}(\xi)\cr
&=-\half \sum_{i,j}C'_{ij}(\ka) \int\sfrac{\partial\hfill}{\partial\xi_i}\Big[
e^{f(\xi)}\sfrac{\partial\hfill}{\partial\xi_j}f(\xi)\Big]\  d\mu_{C(\ka)}(\xi)\cr
&=-\half \sum_{i,j}C'_{ij}(\ka) \int e^{f(\xi)}\Big[
\sfrac{\partial f}{\partial\xi_i}(\xi)
\,\sfrac{\partial f}{\partial\xi_j}(\xi)
+\sfrac{\partial\hfill}{\partial\xi_i}\sfrac{\partial\hfill}{\partial\xi_j}
        f(\xi)\Big]\  d\mu_{C(\ka)}(\xi)\cr
}$$

\Item iv)
By parts (iii) and (ii)
$$\eqalignno{
\sfrac{d\hfill}{d\ka}\int e^{\lw f(\xi)\rw_{C(\ka)}}\  d\mu_{C(\ka)}(\xi) 
&=-\half \sum_{i,j}C'_{ij}(\ka) \int\Big[
\sfrac{\partial \lw f\rw}{\partial\xi_i}
\sfrac{\partial \lw f\rw}{\partial\xi_j}+
\sfrac{\partial\hfill}{\partial\xi_i}\sfrac{\partial \hfill}{\partial\xi_j}
            \lw f(\xi)\rw\Big]
e^{\lw f(\xi)\rw }\  d\mu_{C(\ka)}(\xi)\cr
&\hskip1cm + \int\big[ \sfrac{d\hfill}{d\ka}\lw f(\xi)\rw_{C(\ka)} \big]
e^{\lw f(\xi)\rw_{C(\ka)}}\  d\mu_{C(\ka)}(\xi)\cr
&=-\half \sum_{i,j}C'_{ij}(\ka) \int\Big[
\sfrac{\partial \lw f\rw}{\partial\xi_i}
\sfrac{\partial \lw f\rw}{\partial\xi_j}+
\sfrac{\partial\hfill}{\partial\xi_i}\sfrac{\partial \hfill}{\partial\xi_j}
            \lw f(\xi)\rw\Big]
e^{\lw f(\xi)\rw }\  d\mu_{C(\ka)}(\xi)\cr
&\hskip1cm + \half \sum_{i,j}C'_{ij}(\ka)
\int\big[  \sfrac{\partial\hfill}{\partial\xi_i}
\sfrac{\partial\hfill}{\partial\xi_j}\,\lW f(\xi)\rW_{C(\ka)} \big]
e^{\lw f(\xi)\rw_{C(\ka)}}\  d\mu_{C(\ka)}(\xi)\cr
&=-\half \sum_{i,j}C'_{ij}(\ka) \int\Big[
\lW\sfrac{\partial f}{\partial\xi_i}\rW_{C(\ka)}
\ \lW\sfrac{\partial f}{\partial\xi_j}\rW_{C(\ka)}\Big]
e^{\lw f(\xi)\rw_{C(\ka)}}\  d\mu_{C(\ka)}(\xi)\cr
}$$
\endproof

\vfill\eject

\appendix{B}{ Gram Bounds}\PG\pgRB

Let $C$ be an antisymmetric bilinear form (covariance) on the vector space $V$, 
and let $\{\xi_i\}$ be a system of generators for $V$.

\proposition{\STM\propGII}{
Suppose that $V$ is the direct sum of two subspaces $V_a$ and $V_c$ such that
each of the generators $\xi_i$ lies either in $V_a$ or $V_c$ and
$$
C(\xi,\xi')=0 \qquad {\rm if\ both}\ \xi,\xi'\in V_a\ \   {\rm or\ both}\ 
   \xi,\xi'\in V_c
$$
Assume furthermore that there is a Hilbert space $\cH$, and that there
is associated to each generator $\xi_i$ a vector $w_i\in \cH$. Set
$$
S=\sup_i \|w_i\|
$$
\Item{i)}
If 
$$
C(\xi_i,\xi_j) = \<w_i,w_j\>_\cH \qquad {\rm if\ \ } \xi_i\in V_a, \,\xi_j\in V_c
$$
for all $i,j$, then
$$
\Big| \int \xi_{i_1} \cdots \xi_{i_m}\, d\mu_C(\xi)  \Big| 
\le S^m
$$
for all $i_1,\cdots,i_m$.
\Item{ii)}
Assume that, for each generator $\xi_i$, there exists a real number $\tau_i$ 
such that
$$
C(\xi_i,\xi_j) = \cases{e^{-(\tau_i-\tau_j)}\<w_i,w_j\>_\cH & if $\tau_i>\tau_j$ \cr
                        0 & if $\tau_i\le \tau_j$
}$$
for all $i,j$ with $\xi_i\in V_a, \,\xi_j\in V_c$. Then again
$$
\Big| \int \xi_{i_1} \cdots \xi_{i_m}\, d\mu_C(\xi)  \Big| 
\le S^m
$$
for all $i_1,\cdots,i_m$.

\noindent
In both cases, $2S$ is an integral bound for the 
covariance $C$ with respect to the norms of Example \egcompatnorm.
}

\proof{of part (i)} If the integral does not vanish, we may reorder the
factors in the integrand $\xi_{i_1} \cdots \xi_{i_m}=\pm \xi_{j_1}\xi_{\ell_1}
\cdots\xi_{j_n}\xi_{\ell_n}$ so that $\xi_{j_p}\in V_a$ and $\xi_{\ell_p}\in
V_c$ for all $1\le p\le n=\sfrac{m}{2}$. Then
$$
 \int \xi_{i_1} \cdots \xi_{i_m}\, d\mu_C(\xi) 
= \pm \det\big[C(\xi_{j_p},\xi_{\ell_q})\big]_{1\le p,q\le n}\
$$
Part (i) now follows by Gram's inequality. 
Part (ii) shall be proven following Lemma \:\lemGIV.
\endproof
\example{\STM\exGIII}{
In [FKTo3], part (i) of Proposition \propGII\ is applied to covariances of the 
form
$$
C(\xi_i,\xi_j)
= \de_{\si_i,\si_j} \int \frac{d^{d+1}k}{(2\pi)^{d+1}} e^{\imath<k,x_i-x_j>_-}
\frac{\chi(k)}{\imath k_0 - e(\k)}
$$
where $\si_i,\si_j\in\{\uparrow,\downarrow\}$ are spins, 
$x_i=(\tau_i,\x_i),\ x_j=(\tau_j,\x_j)\in\bbbr^{d+1}$ 
are points in space--(imaginary)time, $k=(k_0,\k)\in\bbbr\times\bbbr^d$
and
$<k,x_i-x_j>_- = -k_0(\tau_i-\tau_j) + \k\cdot(\x_i-\x_j)$.
The function $e(\k)$ is the dispersion relation, minus the chemical potential,
and $\chi(k)$ is a nonnegative cutoff function. In this case, we may take
$\cH= L^2(\bbbr^{d+1}\times\bbbc^2)$,
$$
w_i(k,\si)=\de_{\si,\si_i}e^{-\imath<k,x_i>_-}
\cases{\sqrt{ \sfrac{1}{(2\pi)^{d+1}}\sfrac{\chi(k)}{\imath k_0 - e(\k)}}
                & if $\xi_i\in V_c$\cr
\noalign{\vskip.1in}
  \overline{\sqrt{ \sfrac{1}{(2\pi)^{d+1}}\sfrac{\chi(k)}{\imath k_0 - e(\k)}}}
                & if $\xi_i\in V_a$\cr}
$$
for any single--valued square root, and
$$
S=\sqrt{\int \sfrac{d^{d+1}k}{(2\pi)^{d+1}}\Big|
\sfrac{\chi(k)}{\imath k_0 - e(\k)}\Big|}
$$

Part (ii) of Proposition \propGII\ will be used in 
[FKTo2]. It is designed to deal with covariances
of the form
$$
C(\xi_i,\xi_j)
= \de_{\si_i,\si_j} \int \frac{d^{d+1}k}{(2\pi)^{d+1}} e^{\imath<k,x_i-x_j>_-}
\frac{U(\k)}{\imath k_0 - 1}
$$
in which the nonnegative cutoff function $U(\k)$ is independent of $k_0$.
In this case $\int \sfrac{d^{d+1}k}{(2\pi)^{d+1}}\Big|
\sfrac{U(\k)}{\imath k_0 - 1}\Big|$ diverges.
As
$$
\int \frac{d k_0}{2\pi} e^{-\imath k_0(\tau_i-\tau_j)}
\frac{U(\k)}{\imath k_0 - 1}
=\cases{-U(\k)e^{-(\tau_i-\tau_j)}&if $\tau_i>\tau_j$\cr
                   0 & if $\tau_i\le\tau_j$\cr}
$$
(actually, the case $\tau_i=\tau_j$ is defined by the limit $\tau_j\rightarrow
\tau_i+$), we may take $\cH= L^2(\bbbr^{d}\times\bbbc^2)$,
$$
w_i(\k,\si)=\de_{\si,\si_i}e^{-\imath\k\cdot\x_i}\sqrt{ 
\sfrac{1}{(2\pi)^{d}}U(\k)}
\cases{-1 & if $\xi_i\in V_a$\cr  1 & if $\xi_i\in V_c$\cr}
$$
and then
$$
S=\sqrt{\int \sfrac{d^d\k}{(2\pi)^{d}}\big|U(\k)\big|}
$$

}

To prepare for the proof of part (ii) of Proposition \propGII,
let $\bigwedge \cH = \bigoplus\limits_{n\ge 0} \bigwedge^n \cH$ be the Grassmann
algebra over $\cH$. The element $1\in\bbbc = \bigwedge^0 \cH$ is also denoted
by $\Omega_0$ (the ground state).
On each of the summands $\bigwedge^n \cH$ there is an inner
product such that 
$$
\<v_1\cdots v_n,\,v'_1\cdots v'_n\> 
= \det \big( \< v_i,v'_j \> \big)_{i,j=1,\cdots,n}
$$
for all $v_1,\cdots,v_n,\,v'_1,\cdots,v'_n \in \cH$. 
On $\cH$ there is an inner product determined by the requirement that
$\bigoplus\limits_{n\ge 0} \bigwedge^n \cH$ be an orthogonal direct sum.
For $v\in\cH$, let
$a^\dagger(v)$ be the operator on $\bigwedge \cH$ that maps 
$f\in \bigwedge^n \cH\,$ to $\,vf\in \bigwedge^{n+1} \cH$, and $a(v)$ its
adjoint. We have the standard

\lemma{\STM\lemGIII}{ For all $v,w\in \cH$
\Item{i)} 
$$\eqalign{
\{a(v),a(w)\} & = \{a^\dagger(v),a^\dagger(w)\} =0 \cr
\{a(v),a^\dagger(w)\} & =\<v,\,w\>
}$$
\Item{ii)}
The operator norms $\|a(v)\|, \|a^\dagger(v)\|$  are bounded by $\|v\|_\cH$.
}
\prf
Let $\big( e_j \big)_{j\in \cJ}$ be an orthonormal basis for $\cH$, 
indexed by a totally ordered set $\cJ$. For each finite subset $J$ 
of $\cJ$ set
$$
e_J = e_{j_1}\cdots e_{j_n}  \qquad\qquad {\rm when\ } J=\{j_1,\cdots,j_n\}\ 
   {\rm with\ } j_1 \prec \cdots \prec j_n   
$$
The elements $\ e_J,\,|J|=n\ $ are an orthonormal basis for $\bigwedge^n\cH$.
Then
$$\eqalign{
a^\dagger(e_j)e_J &= \cases{ \ep_{J,j}\ e_{J\cup \{j\}}
                               & if $j\notin J$ \cr
                             0 & if $j\in J$   } \cr
a(e_j)e_J &= \cases{ \ep_{J\setminus \{j\},j}\ e_{J\setminus \{j\}} 
                               & if $j\in J$ \cr
                             0 & if $j\notin J$   } \cr
}$$
where, for $j\notin J$, $\ep_{J,j}$ is the sign of the permutation that brings
the sequence $j,J$ to standard order.

In the case $v=e_j,\,w=e_{j'}$, part (i) of the Lemma follows directly from this
description. The general case follows from this special case, since all terms
are bilinear in $v$ and $w$.

To prove part (ii) of the Lemma observe that, for all $u\in\cH$ and $f\in\bigwedge \cH$
$$
\|a(u)f\|^2_{\wedge\cH}+\|a^\dagger(u)f\|^2_{\wedge\cH}
=\big<\{a(u),a^\dagger(u)\}f,f\big>
=\|u\|_\cH^2\ \|f\|^2_{\wedge\cH}
$$
so that
$$
\|a(u)f\|_{\wedge\cH}\le \|u\|_\cH\ \|f\|_{\wedge\cH}\qquad
\|a^\dagger(u)f\|_{\wedge\cH}\le \|u\|_\cH\ \|f\|_{\wedge\cH}
$$
\endproof

Let $N$ be the number operator on $\bigwedge \cH$. By definition, 
its restriction to 
$\bigwedge^n \cH$ is multiplication by $n$.
For each index $i$, labeling a generator $\xi_i$, set
$$
a_i = \cases{ e^{\tau_i N}a(w_i)e^{-\tau_i N} = e^{-\tau_i} a(w_i)
                      & if $\xi_i\in V_a$ \cr
\noalign{\vskip.1in}
              e^{\tau_i N}a^\dagger(w_i)e^{-\tau_i N} = e^{\tau_i} a^\dagger(w_i)
                      & if $\xi_i\in V_c$ \cr
}$$
A sequence $i_1,\cdots,i_m$ is called time ordered if
$ \tau_{i_1} \ge \cdots \ge \tau_{i_m}$ and for $1\le k<\ell \le m$
$$
\tau_{i_k} = \tau_{i_\ell},\ \ \xi_{i_k} \in V_a  \qquad
{\rm implies} \ \ \xi_{i_\ell} \in V_a
$$

\lemma{\STM\lemGIV}{
Let $i_1,\cdots,i_m$ be a time ordered sequence. Then, under the assumptions of 
part (ii) of Proposition \propGII
$$
\int \xi_{i_1} \cdots \xi_{i_m}\, d\mu_C(\xi) 
= \<\Omega_0,\,a_{i_1}\cdots a_{i_m} \Omega_0 \>
$$
}

\prf The proof is by induction on $m$. The cases $m=0$ and $m=1$ are trivial.
 We perform the induction step $m-2 \rightarrow m$.

Assume first that $\xi_{i_1} \in V_a$. Then
$$
\{a_{i_1},\,a_{i_k}\} 
 = \cases{ e^{-\tau_{i_1}+\tau_{i_k}}\, \{a(w_{i_1}),\,a^\dagger(w_{i_k})\}
         = e^{-(\tau_{i_1}-\tau_{i_k})}\, \<w_{i_1},\,w_{i_k}\> & 
                                     if $\xi_{i_k} \in V_c$ \cr
           0 & if $\xi_{i_k} \in V_a$ \cr
}$$
and $a_{i_1} \Omega_0 = 0$. Therefore
$$\eqalign{
\<\Omega_0,\,a_{i_1}\cdots a_{i_m} \Omega_0 \>
&= \smsum_{k=2 \atop \xi_{i_k} \in V_c}^m (-1)^k\,e^{-(\tau_{i_1}-\tau_{i_k})}\,
  \<w_{i_1},\,w_{i_k} \>\,
\<\Omega_0,\,a_{i_2}\cdots a_{i_{k-1}} a_{i_{k+1}}\cdots a_{i_m}\,\Omega_0\> \cr
&= \smsum_{k=2}^m (-1)^k C(\xi_{i_1},\xi_{i_k})\,
 \int \xi_{i_2} \cdots \xi_{i_{k-1}} \xi_{i_{k+1}}\cdots \xi_{i_m}\, d\mu_C(\xi)\cr
&=\int \xi_{i_1} \cdots \xi_{i_m}\, d\mu_C(\xi) \cr
}$$
For the second equality we used the assumption on $C$, the fact that the sequence
$i_1,\cdots,i_m$ is time ordered, and the induction hypothesis. 
The third equality is the integration by parts formula.

Now assume that $\xi_{i_1} \in V_c$. Then
$$
\<\Omega_0,\,a_{i_1}\cdots a_{i_m} \Omega_0 \>
= e^{\tau_{i_1}} \<a(w_{i_1})\Omega_0,\,a_{i_2}\cdots a_{i_m} \Omega_0 \> =0
$$
and, by the integration by parts formula
$$
\int \xi_{i_1} \cdots \xi_{i_m}\, d\mu_C(\xi)
=\smsum_{k=2}^m (-1)^k C(\xi_{i_1},\xi_{i_k})\,
 \int \xi_{i_2} \cdots \xi_{i_{k-1}} \xi_{i_{k+1}}\cdots \xi_{i_m}\, d\mu_C(\xi)
=0
$$
since $C(\xi_{i_1}, \xi_{i_k}) = - C(\xi_{i_k}, \xi_{i_1}) = 0$ for $k=2,\cdots,m$.
\endproof

\proof{ of part (ii) of Proposition  \propGII}
We may assume that the sequence $i_1,\cdots,i_m$ is time ordered. If 
$\#\{ \nu\,\big|\, \xi_{i_\nu} \in V_c\} 
 \ne \#\{ \nu\,\big|\, \xi_{i_\nu} \in V_a\}$ then 
$\int \xi_{i_1} \cdots \xi_{i_m}\, d\mu_C(\xi) =0$. 
Otherwise, by Lemma \lemGIV\ and Lemma \lemGIII
$$\eqalign{
\Big| \int \xi_{i_1} \cdots \xi_{i_m}\,& d\mu_C(\xi)  \Big|
 = \big|\<\Omega_0,\,a_{i_1}\cdots a_{i_m} \Omega_0 \> \big| \cr
& = \Big|\<\Omega_0,\,a^{(\dagger)}(w_{i_1})e^{-(\tau_{i_1}-\tau_{i_2})N}
     a^{(\dagger)}(w_{i_2})
     \cdots e^{-(\tau_{i_{m-1}}-\tau_{i_m})N} a^{(\dagger)}(w_{i_m}) \Omega_0 \> \Big| \cr
& \le \big\|a^{(\dagger)}(w_{i_1})\big\|\, \big\|e^{-(\tau_{i_1}-\tau_{i_2})N}\big\|\,\big\|a^{(\dagger)}(w_{i_2})\big\|
     \cdots \big\|e^{-(\tau_{i_{m-1}}-\tau_{i_m})N}\big\|\,
     \big\| a^{(\dagger)}(w_{i_m})\big\| \cr
& \le \smprod_{k=1}^m \big\|a^{(\dagger)}(w_{i_k})\big\|\ \le\ S^m
}$$
Here, we used that $a_{i_1}\cdots a_{i_m} \Omega_0 \in \bigwedge^0\cH$ and that 
the restriction of the number operator $N$ to $\bigwedge^0\cH$ is identically zero.
\endproof

\vfill\eject

\titlea{ References}\PG\pgRIref

\item{[BS]} F. Berezin, M. Shubin: {\it The Schr\"odinger Equation}, Kluwer (1991). 
Supplement 3: D.Le\u ites, {\bf Quantization and supermanifolds}.
\smallskip%
\item{[DR]}  M. Disertori and V. Rivasseau,
{\bf Continuous constructive fermionic renormalization}, Annales Henri Poincar\'e, 
{\bf 1} (2000).
\smallskip%
\item{[FKLT1]} J. Feldman, H. Kn\"orrer, D. Lehmann, E. Trubowitz, {\bf
Fermi Liquids in Two-Space Dimensions}, in {\it Constructive Physics} 
V. Rivasseau ed. 
Springer Lecture Notes in Physics 446, 267-300 (1995).
\smallskip%
\item{[FKLT2]} J. Feldman, H. Kn\"orrer, D. Lehmann, E. Trubowitz, 
{\bf Are There Two Dimensional Fermi Liquids?}, in {\it Proceedings of the XIth
International Congress of Mathematical Physics}, D. Iagolnitzer ed., 
440-444 (1995).
\smallskip%
\item{[FKT1]} J. Feldman, H. Kn\"orrer, E. Trubowitz, 
{\bf A Representation for Fermionic Correlation Functions},
Communications in Mathematical Physics {\bf 195},  465--493 (1998).
\smallskip%
\item{[FKTf1]} J. Feldman, H. Kn\"orrer, E. Trubowitz, 
{\bf A Two Dimensional Fermi Liquid, Part 1: Overview}, preprint.
\smallskip%
\item{[FKTf2]} J. Feldman, H. Kn\"orrer, E. Trubowitz, 
{\bf A Two Dimensional Fermi Liquid, Part 2: Convergence}, preprint.
\smallskip%
\item{[FKTf3]} J. Feldman, H. Kn\"orrer, E. Trubowitz, 
{\bf A Two Dimensional Fermi Liquid, Part 3: The Fermi Surface}, preprint.
\smallskip%
\item{[FKTffi]} J. Feldman, H. Kn\"orrer, E. Trubowitz, 
     {\bf Fermionic Functional Integrals and the Renormalization Group},
     Andr\'e Aisenstadt Monograph Series, to appear.
\smallskip%
\item{[FKTo1]} J. Feldman, H. Kn\"orrer, E. Trubowitz, 
{\bf Single Scale Analysis of Many Fermion Systems, Part 1: Insulators}, preprint.
\smallskip%
\item{[FKTo2]} J. Feldman, H. Kn\"orrer, E. Trubowitz, 
{\bf Single Scale Analysis of Many Fermion Systems, Part 2: The First Scale}, preprint.
\smallskip%
\item{[FKTo3]} J. Feldman, H. Kn\"orrer, E. Trubowitz, 
{\bf Single Scale Analysis of Many Fermion Systems, Part 3: Sectorized Norms}, preprint.
\smallskip%
\item{[FKTo4]} J. Feldman, H. Kn\"orrer, E. Trubowitz, 
{\bf Single Scale Analysis of Many Fermion Systems, Part 4: Sector Counting}, preprint.
\smallskip%
\item{[FT]} J. Feldman, E. Trubowitz, 
{\bf Perturbation Theory for Many Fermion Systems}, Helvetica Physica
Acta {\bf 63} (1990) 156-260.
\smallskip%
\item{[PT]}  J. P\"oschel, E. Trubowitz, {\it Inverse Spectral Theory}, 
Academic Press (1987).
\smallskip%
\item{[S]}  M. Salmhofer,
{\bf Improved Power Counting and Fermi Surface Renormalization}, Reviews in 
Mathematical Physics, {\bf 10} (1998) 553-578.
\smallskip%
\item{[SW]}  M. Salmhofer and C. Wieczerkowski,
{\bf Positivity and Convergence in Fermionic Quantum Field Theory}, Journal of 
Statistical Physics, {\bf 99} (2000) 557-586.

\vfill\eject

\hoffset=-0.1in
\titlea{Notation}\PG\pgRInot

\vskip1in

\centerline{
\vbox{\offinterlineskip
\hrule
\halign{\vrule#&
         \strut\hskip0.05in\hfil#\hfil&
         \hskip0.05in\vrule#\hskip0.05in&
          #\hfil\hfil&
         \hskip0.05in\vrule#\hskip0.05in&
          #\hfil\hfil&
           \hskip0.05in\vrule#\cr
height2pt&\omit&&\omit&&\omit&\cr
&Not'n&&Description&&Reference&\cr
height2pt&\omit&&\omit&&\omit&\cr
\noalign{\hrule}
height2pt&\omit&&\omit&&\omit&\cr
&$Z(f)$&&degree zero component of $f$&&Definition \defsuperalgebra.iii&\cr
height2pt&\omit&&\omit&&\omit&\cr
&$\bigwedge V$&&Grassmann algebra over $V$&&Example \exsuper&\cr
height2pt&\omit&&\omit&&\omit&\cr
&$\bigwedge\nolimits_A V $&&Grassmann algebra over $V$ with coefficients
in $A$&&Example \exsuper&\cr
height4pt&\omit&&\omit&&\omit&\cr
&$A_m[n_1,\cdots,n_r]$&&partially antisymmetric elements of $A_m\otimes V^{\otimes (n_1+\cdots+n_r)}$
&&Definition \defAmnonetonr&\cr
height4pt&\omit&&\omit&&\omit&\cr
&$\int e^{\Sigma\, \xi_i\ze_i}\, d\mu_C(\xi)$
&&$=e^{-1/2\,\Sigma\, \ze_i C_{ij} \ze_j}$
Grassmann Gaussian  integral&&before Definition \defrengroup&\cr
height4pt&\omit&&\omit&&\omit&\cr
&$\Om_C(W)(\psi)$&&$=\log \sfrac{1}{Z}\int e^{W(\psi +\xi)} d\mu_C(\xi)$
renormalization group map&&Definitions \defrengroup, \defrengroupInfDim&\cr
height4pt&\omit&&\omit&&\omit&\cr
&$\cS(f)$&&$=\sfrac{1}{Z(U,C)}\, \int f(\xi)\, e^{U(\xi)}\, d\mu_C(\xi)$
Schwinger functional&&before Remark \remrenschw&\cr
height4pt&\omit&&\omit&&\omit&\cr
&$R$&&$R$--operator&&before Theorem \theoremIII&\cr
height2pt&\omit&&\omit&&\omit&\cr
&$R_C(K_1,\cdots,K_\ell)$&&$\ell^{\rm th}$ Taylor coefficient of $R$&&
(\eqRCdef)&\cr
height2pt&\omit&&\omit&&\omit&\cr
&$\cR_{K,C}(f)$&&
$\lww  \int\hskip-4pt\int \lW  e^{\lw K(\xi,\xi',\eta)\rw_{\xi'}} -1 \rW_\eta\,f(\eta)\,
d\mu_C(\xi')\,d\mu_C(\eta) \rww_\xi$&&Definition \defcalR&\cr
height2pt&\omit&&\omit&&\omit&\cr
&$\lw e^{\Sigma\, \xi_i\ze_i}\rw_{\xi,C}$&&$=e^{1/2\,\Sigma\, \ze_i C_{ij} \ze_j}\,e^{\Sigma\, \xi_i\ze_i}$
Wick ordering&&after Remark \remnormal&\cr
height2pt&\omit&&\omit&&\omit&\cr
&$\Cont{i}{j}{C},\ \cont{\xi}{\xi'}{C}$&&contractions&&Definitions \defContract,
\defGrasscontract&\cr
height2pt&\omit&&\omit&&\omit&\cr
&$\fN_d$&&norm domain&&Definition \defFancynormdomain&\cr
height2pt&\omit&&\omit&&\omit&\cr
&$\cb$&&contraction bound&&Definition \defcontractintbound.i&\cr
height2pt&\omit&&\omit&&\omit&\cr
&$\ib$&&integral bound&&Definition \defcontractintbound.ii&\cr
height2pt&\omit&&\omit&&\omit&\cr
&$N(f\cl \al)$&&$\sfrac{1}{\ib^2}\,\cb\!\sum_{m,n_1,\cdots, n_r\ge 0}\,
\al^{|n|}\,\ib^{|n|} \,\|f_{m; n_1,\cdots, n_r}\|$&&Definition \deffunctnorm&\cr
height2pt&\omit&&\omit&&\omit&\cr
}\hrule}}

\end